\documentclass{article}
\usepackage[utf8]{inputenc}
\usepackage[table]{xcolor}
\usepackage{amsmath}
\usepackage{setspace}
\usepackage{geometry}
\usepackage{graphicx}
\usepackage{subfigure}
\usepackage{indentfirst}
\usepackage{commath}
\usepackage{amssymb}
\usepackage{float}
\usepackage{bbm}
\usepackage{algorithm2e}
\usepackage{gensymb}
\usepackage{authblk}
\usepackage{tabularray}
\usepackage{xcolor}

\usepackage{amsthm}

\title{Modeling Transient Changes in Circadian Rhythms}
\author[1,2]{Ziyu Zhao}
\author[3,4]{Dae-Sung Hwangbo} 
\author[5]{Sumit Saurabh}
\author[2,4]{Clark Rosensweig}
\author[2,4]{Ravi Allada}
\author[1,2,4,8]{William L. Kath}
\author[1,2,6,7,8]{Rosemary Braun}

\affil[1]{{\small Department of Engineering Sciences and Applied Mathematics, Northwestern University, Evanston, IL 60208, USA}}
\affil[2]{{\small NSF-Simons Center for Quantitative Biology, Northwestern University, Evanston, IL 60208, USA}}
\affil[3]{{\small Department of Biology, University of Louisville, Louisville, KY 40292, USA}}
\affil[4]{{\small Department of Neurobiology, Northwestern University, Evanston, IL 60208, USA}}
\affil[5]{{\small Department of Biology, Loyola University, Chicago, IL 60660, USA}}
\affil[6]{{\small Department of Molecular Biosciences, Northwestern University, Evanston, IL 60208, USA}}
\affil[7]{{\small Department of Physics and Astronomy, Northwestern University, Evanston, IL 60208, USA}}
\affil[8]{{\small Northwestern Institute on Complex Systems, Northwestern University, Evanston, IL 60208, USA}}

\date{April 14, 2023}

\usepackage{titlesec}
\titlespacing{\section}{0pt}{\parskip}{-\parskip}
\titlespacing{\subsection}{0pt}{\parskip}{-\parskip}
\titlespacing{\subsubsection}{0pt}{\parskip}{-\parskip}

\def\todo#1{{\color{red} **** #1 ****\typeout{ }\typeout{LaTeX Warning: **** To do: #1 ****}\typeout{ }}}
\def\todo#1{\typeout{ }\typeout{LaTeX Warning: **** To do: #1 ****}\typeout{ }}

\begin{document}

\maketitle
\doublespacing
\large

\begin{abstract}
\setlength{\parindent}{0pt}
\noindent
The circadian clock can adapt itself to external cues, but the molecular mechanisms and regulatory networks governing circadian oscillations' transient adjustments are still largely unknown. Here we consider the specific case of circadian oscillations transiently responding to a temperature change. Using a framework motivated by Floquet theory, we model the mRNA expression level of the fat body from \textit{Drosophila melanogaster} following a step change from 25\degree{}C to 18\degree{}C. Using the method we infer the adaptation rates of individual genes as they adapt to the new temperature. To deal with heteroskedastic noise and outliers present in the expression data we employ quantile regression and wild bootstrap for significance testing. Model selection with finite-sample corrected Akaike Information Criterion (AICc) is performed additionally for robust inference. We identify several genes with fast transition rates as potential sources of temperature-mediated responses in the circadian system of fruit flies, and the constructed network suggests that the proteasome may play important roles in governing these responses.

\todo{WLK: should we name the method?}


\end{abstract}

\newpage
\section{Introduction}
Circadian rhythms are internal clocks within organisms regulating physical and behavioral changes that follow an approximately 24 hr cycle \cite{young2001time}. The molecular oscillator in the core clock system drives the rhythmic expression of hundreds of other genes \cite{BX}. This oscillation is stable under steady environmental conditions, but it can also adapt to changes in light and temperature in the environment. Circadian rhythms can be entrained by light/dark cycles \cite{plautz1997independent}, but can be phase-shifted by a light pulse \cite{myers1996light}. In \textit{Drosophila Melanogaster}, light is involved in the degradation of \textit{tim} \cite{hunter1996regulation}, and this been widely employed in models of the core clock \cite{leloup1999limit,leise2007mathematical}.

Analogously to light/dark cycles, studies have also shown that the circadian rhythm can be entrained by daily temperature cycles \cite{yoshii2005temperature}. Short-duration heat pulses can mediate rapid degradation of the core clock proteins \textit{per} and \textit{tim} and induce phase shifts in the internal clock \cite{sidote1998differential}. Another property of the circadian rhythm is temperature compensation: the free (undriven) period of the core clock is largely independent of temperature. Several models have been proposed to describe the behavior of the core clock circuit \cite{kurosawa2017temperature}.  In such models it is common for reaction rates to increase with temperature according to the Arrhenius equation \cite{ruoff1996temperature} or by simply specifying $Q_{10}$ values for the reaction rates \cite{smolen2004simulation} and to fine-tune parameters so that cyclic patterns and the oscillation period are maintained at different temperatures.

It has been observed that temperature affects the identity and phase of genes under circadian regulation ~\cite{BX}. While the above models capture the behavior of the clock itself, they do not describe the response of non-clock genes to temperature. Moreover, they cannot provide specific answers as to how fast each component reacts and adapts to a new temperature. This is not a trivial question, in that we cannot expect every component in the regulatory network, whether they are genes or proteins, to adapt at the same rate given the complexity of gene regulation.  Changes in the core clock components (e.g., \textit{per} and \textit{tim}) might be due to changes in their individual rates or mediated by other components in the network that can be regarded as `sources of temperature-mediated response'. To identify those potential sources, it is necessary to first accurately estimate the adaptation rate for each component as it adapts to the new temperature via experimental data.

In this paper, we utilize a time series model based upon Floquet theory \cite{floquet1883equations} (see Methods section) to simultaneously describe the cycling patterns at two different constant temperatures and the gene-by-gene exponential transitions from one temperature to another. These transition rates provide the rate of adaption and, moreover, inform the chronological sequence of reactions or regulations that take place in a gene regulatory network as the system adapts to a new temperature. For example, transitions from a gene with a more rapid rate cannot be induced by other genes with slower rates. 

When using statistical tools to analyze gene expression patterns across different environmental conditions, care must be taken to avoid violating methodological assumptions.  Regression methods are often based upon the assumption of independent, identically-distributed (\textit{i.i.d.})\ normal residuals, and this can easily be violated when environmental changes take place.  In particular, gene expression can have different variance under different temperatures \cite{huang2020genotype}, possibly due to the different chemical reaction rates, leading to non-\textit{i.i.d.}\ and/or non-Gaussian residuals. For cycling genes with large amplitudes, it is reasonable to expect that molecular noise will vary with the average expression level at different phases in an oscillation.  This is also true for the transition or adaptation phase when variance may change gradually because mean levels gradually adjust to new values.  Under this scenario, it is therefore important to consider residual heteroskedasticity when conducting statistical tests and inferences. 

We adopt the wild bootstrap \cite{wu1986jackknife}, a method that deals with heteroskedastic residuals, as the basis for our significance tests. We also incorporate quantile regression \cite{koenker1978regression} instead of the traditional least-square regression to account for outliers observed in our experimental data, such that regression results will be robust to these outliers and provide more accurate estimates on transition rates.   Correlations between our model parameters are taken into account.  Model selection is also performed to further validate the significance test results.  We take the conservative approach of only claiming confidence about the estimated transition rates about genes that have concordant inferences from both of the above methods.  The goal is to provide more accurate estimates of the transition rates, and we believe the cost of being able to provide concordant and conclusive results on only a smaller set of genes is a reasonable one. Given the known challenges associated with detecting circadian oscillations \cite{ness2020timetrial,BX}, we also take advantage of a separate circadian time-series experiment \cite{Hwangbo2023.01.04.522718} to enhance reproducibility.

Previous works \cite{sidote1998differential, yoshii2007induction} collect behavioral and molecular data from locomotor activity recording, immunoblotting or RT-PCR, but they are only able to report dynamics of several specific genes or proteins and present a small part of the large set of expression changes. This paper is the first study to probe transients following temperature perturbation with high-throughput RNA-seq such that mRNA expression is measured not only under steady temperature, but also immediately after the temperature step change takes place. Thus, it is possible for us to analyze gene expression data during the adaptation stage, and identify potential genes that may play an important role in mediating the adaptation to the new temperature. Although only gene expression data are analyzed under this framework, our method can be extend to other components such as protein or gene isoform expression as well. 

In summary, we propose a generalizable modeling framework that can characterize transient behavior from empirical observations of noisy oscillatory dynamics, and use this framework to analyze high-throughput RNA-seq data to identify key molecular elements involved in the response of the fly's circadian system to temperature changes.

\newpage
\section{Methods}
\subsection{Experimental methods}
\subsubsection{Study design}
For the transient experiment, two groups of wild-type \textit{Drosophila Melanogaster} flies are grown in constant 25\degree{}C entrained with 12:12 LD cycles and high-calorie diet. After entrainment, one group of fruit flies are moved into an 18\degree{}C environment with other conditions unchanged, and the other group remains in 25\degree{}C. One replicate of fatbody samples is collected on Day4/5 (Day4 ZT6 to Day5 ZT18) under both temperature every 2hrs.  In addition, a single replicate is collected during the adaptation stage during Day1/2 under 18\degree{}C every 2hrs (Day1 ZT0 to Day2 ZT12).  A second replicate is collected on Day4/5 (Day4 ZT6 to Day5 ZT18) under both temperatures every 6hrs, as well as on Day1 (Day1 ZT0 to Day1 ZT24) under 25\degree{}C every 6hrs [Figure~\ref{Figure.1}].  Data from 25\degree{}C are treated as being collected at negative times (from -120hr to 0hr) during the analysis phase in order to fit the model curve. 

In a previous experiment \cite{Hwangbo2023.01.04.522718} (refer to their paper for details), flies were placed in 25\degree{}C (growth conditions) or 18\degree{}C for five days prior to sampling. All flies are entrained with 12:12 LD cycles. Fatbody samples are collected in LD conditions on Day5 from ZT02 to ZT24 every 2 hours with 2 replicates at each ZT for all three groups. 

\subsubsection{Fly rearing and fat body dissection}

All experiments were carried out with \textit{Iso31} flies (an isogenic $w^{1118}$ strain). Age-matched female flies were collected and aged for 3 days with males before being transferred to an entrainment incubator. Following three days of entrainment, the experimental group was shifted to 18\degree{}C at seven days old while the control group was kept at 25\degree{}C. At each time point, females were collected from the incubator, anesthetized on a CO2 pad, and then dissected in phosphate buffered saline (137 mM NaCl, 2.7 mM KCl, 8 mM N$\text{a}_2$HP$\text{O}_4$, and 2 mM K$\text{H}_2$P$\text{O}_4$). Abdominal fat body dissections were performed by first removing the anal plate and guts, then cutting a slit in the ventral abdominal cuticle and removing the remaining ovaries, malphigian tubes, and other internal organs. Finally, the abdominal cuticle with attached fat body was severed from the thorax and collected for RNA isolation. Ten abdominal fat bodies were collected and combined to generate data for each timepoint.

\subsubsection{RNA-Seq}

RNA was isolated from the abdominal fat bodies using Trizol LS (ThermoFisher, \#10296028). Briefly, 300 µL of Trizol LS was added to fat bodies in 100 µL of PBS. The tissue was homogenized with a motorized pestle for 2 minutes before adding another 600 µL of Trizol LS (3:1 mixture of Trizol LS:PBS). The resulting solution was centrifuged at 12,000 x g for 10 minutes at 4\degree{}C. The aqueous supernatant layer was collected in a new tube, while carefully avoiding disturbing the other layers of the phase separated solution. RNA was extracted from the aqueous supernatant layer by vigorously shaking with 240 µL of chloroform (Fisher Scientific, \#C298), again carefully avoiding other layers following phase separation. The aqueous phase was transferred to a new tube and the RNA was precipitated by incubating at room temperature with 500 µL of 100\% isopropanol (Sigma-Aldrich, \#I9516). Following centrifugation at 12,000 x g for 10 minutes at 4\degree{}C, the supernatant was removed leaving only the RNA pellet. The pellet was washed with 1 mL of 75\% ethanol (Sigma-Aldrich, \#E7023), then air dried for 5-10 minutes before resuspension in RNase-free water. 

Purified RNA was sent to Novogene (Sacramento, CA) for library preparation. Libraries were prepared from mRNA purified from total RNA using poly-T oligo-attached magnetic beads (NEBNext Ultra II RNA Library Prep kit for Illumina, New England Biolabs, E7775). Non-stranded library preparation was carried out using the NEBNext Ultra II RNA Library Prep kit for Illumina according to manufacturer protocol. Libraries were subsequently sequenced on a Novaseq 6000 S4 flow cell. 20 million paired-end reads (PE150) were generated for each sample.  
  
\subsubsection{Read alignment, quantitation, and quality control assessment}  
Basic quality checking of sequence files was performed with FastQC \cite{FastQC}.   Paired-end reads were first trimmed using Atropos version 1.1.31 \cite{Didion2017} using the options
\begin{verbatim}
 atropos trim --aligner insert -a AGATCGGAAGAGCACACGTCTGAACTCCAGTCA \
        -A AGATCGGAAGAGCGTCGTGTAGGGAAAGAGTGT --minimum-length 50
\end{verbatim}
Reads were then aligned and quantified using STAR \cite{BIOINFO29p15} (version 2.7.10a\textunderscore{}alpha\textunderscore{}220818) and RSEM \cite{BMCBIOINFO2011v12p323} (version 1.3.1). STAR and RSEM indexes were first built using the Ensembl \emph{Drosophila melanogaster} BDGP6.32 reference (release 107) using standard parameters. STAR was used with the options
\begin{verbatim}
 --outFilterType BySJout --alignIntronMax 1000000 \ 
 --quantMode GeneCounts TranscriptomeSAM
\end{verbatim}
to produce raw counts and also a BAM file with reads aligned to transcriptome.    RSEM was then used with options {\tt --paired-end --strandedness none} to produce tags-per-million (TPM) counts for each gene from transcriptome alignments.   Postprocessing of the count data into table
form was performed with custom Perl, Python and shell scripts.

Because FastQC reported significant sequence duplication in the samples, we also performed the same analysis as above after deduplicating the reads.   First, a BAM file was created using STAR with the options
\begin{verbatim}
 --outFilterType BySJout --alignIntronMax 1000000 \
 --outSAMmultNmax 1 --outSAMtype BAM SortedByCoordinate 
\end{verbatim}
to produce only uniquely mapped reads, and then duplicate paired-end reads were removed using bamUtil \cite{bamutil} with the options
\begin{verbatim}
 bam dedup --rmDups --excludeFlags 0xB04 --oneChrom  
\end{verbatim}
The resulting reads were then re-aligned with STAR to produce a BAM file with reads aligned to the transcriptome, followed by RSEM with the same options as above to produce TPM counts.  De-duplication significantly decreased the number of reads with large TPM values (e.g., $>$ 100), and as the result the TPM values of genes with smaller TPM values were increased by approximately a factor of 1.6-1.7, as seen in Fig.~\ref{SIFigure.2}.   

Single-end reads were quality-assessed with FastQC and then trimmed with Atropos using the options
\begin{verbatim}
 atropos -a AGATCGGAAGAGCACACGTCTGAACTCCAGTCAC --minimum-length 16
\end{verbatim}
Reads were then aligned and quantified using STAR with the same options as before,
\begin{verbatim}
 --outFilterType BySJout --alignIntronMax 1000000 \ 
 --quantMode GeneCounts TranscriptomeSAM
\end{verbatim}
followed by RSEM with the options {\tt --strandedness none}.  De-duplication could not be performed on the single-ended reads, of course. 

\subsection{Analysis Method}
\subsubsection{Model Setup}
\noindent
We assume mRNA expression levels (TPM values) of oscillating genes can be described as components of a limit-cycle oscillator driven by external periodic forcing (light/dark cycle entrainment).  After a temperature step change in the external environment a new limit cycle forms due to temperature compensation.  In general, mRNA levels are not expected to have values consistent with the new limit cycle when the temperature change takes place, and as a result they should experience a transient and asymptotically approach the new limit cycle as time advances.  Floquet theory \cite{Chicone2000,Hale1971,Stoker1950} shows that when expression values are close to those associated with the new limit cycle the genes' dynamical evolution will be described by a sum of terms where each is the product of an exponential decay and a periodic function.   (See the SI for additional details.)    Thus, for each gene the approach to new limit cycle will be dominated by the product of an exponential and a periodic function.  In addition, the gene expression dynamics on the final limit cycle itself will be periodic.  The simplest choice for a periodic function is a sinusoid, of course, and we will adopt this form. 

We therefore model the transient dynamics of the mRNA expression levels due to the temperature step change as an exponential transition from one sinusoidal oscillation to another, with all periods equal to 24 hrs ($\omega=2\pi/24$),
\begin{align}
    y_i & = \hat{y_i} + \epsilon_i =  F(t_i;A_1,A_2,A_3,A_4,A_5,A_6,A_7,A_8,\lambda) + \epsilon_i\\
        & = A_1 + A_2 \cos{\omega t_i} + A_3 \sin{\omega t_i}\\
    & \mbox{\ \ } + \mathbbm{1}_{t_i \geq 0} \cdot \big[ A_4 (e^{-\lambda t_i} \cos{\omega t_i} - 1) + A_5 (e^{-\lambda t_i} \sin{\omega t_i} - 0) + A_6 (e^{-\lambda t_i} - 1) \big]\\
    & \mbox{\ \ } + \mathbbm{1}_{t_i \geq 0} \cdot \big[A_7 (\cos{\omega t_i} - 1) + A_8 (\sin{\omega t_i} - 0)\big] + \epsilon_i
\end{align}
For each gene, $y_i$ is the mRNA tags-per-million (TPM) expression value collected at time $t_i$. The temperature step change is assumed to take place at $t=0$, so $t_i\leq0$ corresponds to samples collected at 25\degree{}C while $t_i>0$ corresponds to samples collected at 18\degree{}C.  For $t\leq0$, our model function $F$ describes a limit cycle composed of first order Fourier series terms but for $t>0$ $F$ includes transient terms (an exponential function multiplying two Fourier terms) in addition to the new limit cycle. This model forces $\lim_{t_i \rightarrow 0^+} y_i = \lim_{t_i \rightarrow 0^-} y_i$ because the mRNA expression level should be continuous (but not necessary differentiable) at the time of the temperature step. By allowing a few parameters to be equal to zero, our model also accounts for the scenario of one gene cycling only at one temperature (e.g. $A_2{=}A_3{=}0$ if cycling at 18\degree{}C only) or not cycling at all ($A_2{=}A_3{=}A_7{=}A_8{=}0$). 

The parameter $\lambda$ in the model is the estimated rate at which expression values adapt to the new temperature. The characteristic time constant $T=\lambda^{-1}$ represents the time needed to complete $e^{-1}\sim63\%$ of this adaptation.  In some cases individual parameters describe some feature of interest associated with the mRNA expression (e.g., $A_7$ is the difference between the cosine components of the limit cycles under two different temperatures), while in others it is some combinations of parameters that represent key features of interest (e.g., the median baseline expression at low temperature is $A_1-A_4-A_6-A_7$, and the phase at low temperature is $\arctan[(A_3+A_8)/(A_2+A_7)]$).  We will perform significance tests on several features of interest. 

\subsubsection{Quantile Regression and Nonlinear Model Fitting}
\noindent
We choose to fit our model to experimental data via quantile regression with $\tau=0.5$ (median regression), which aims to find the conditional median via minimizing the sum of the absolute deviances between the data and the model curve. Compared to least-square regression, quantile regression is more robust to outliers \cite{john2015robustness}. It does not make any assumptions about the residuals' structure, which allows us to account for potential outliers and deal with heteroskedastic noise in the experimental data. An advantage of quantile regression is that it is invariant under any monotonic transformations.

The regression model is nonlinear due to the transition rate $\lambda$, and thus there is no guarantee that our objective function is convex. A poorly chosen initial parameter guess in an iterative optimization method may result in convergence to only a local minimum.  Our model is linear for every fixed value of $\lambda$, however. Thus, in our fitting algorithm, we sweep the value of $\lambda$ on a grid, and choose the parameter with the smallest objective function value as the initial guess for further iteration.  An additional finer grid is used if the minimum lands at an end of the allowed $\lambda$ range to ensure a global minimum has not been missed.   We employ the \textit{quantreg} package in R \cite{quantreg} for linear quantile regression in the sweep, and the L-BFGS-B \cite{byrd1995limited} method as the nonlinear optimization method.  We use the coarse grid $\Lambda_1 = \{0.04,0.07,...,1\}$ and the fine grid $\Lambda_2 = \{0.04,0.05,...,1\}$ and the allowed $\lambda$ range to be $\Lambda_3 = [0.03,1.5]$, corresponding to transition times in the range [0.67,33.3] hours.   The experimental observations cannot be expected to accurately detect transition times shorter or longer than these values. 

\subsubsection{Significance Tests Using Wild Bootstrap}

We use ``Wild Bootstrap'' \cite{feng2011wild}, a bootstrapping method for heteroskedastic noise, to compute the standard errors on the parameter estimates. $p$-values are calculated by fraction of $n=10^3$ bootstrapped parameters (or their combinations in some characteristics of interest) that have larger distance to the mean of their empirical bootstrap distribution than the origin. We choose the significance level $\alpha = 5\%$. When considering the oscillation pattern, we compute the joint distribution of coefficients of the cosine and sine terms at each temperature. In this case the empirical distribution is 2D and we use Mahalanobis distance as the distance metric, which is a generalized distance metric for a multi-dimensional distribution. 

Significance tests are performed in two ways: 1) to check if fitted parameters (or their combinations) are significant, by using their empirical distributions; 2) to check if characteristics of mRNA expression exhibit a significant difference between the two temperatures by using point-wise differences between the empirical distributions of their corresponding parameters (or their combinations), in analogy to the paired sample $t$-test. We are interested in the median baseline expression, oscillation pattern (and oscillation phase and amplitude if the oscillation is significant), and whether or not they exhibit differences between temperatures [Table~\ref{Table.1}]. These tests are performed in parallel with each other but with the same empirical bootstrap distributions. Genes are classified into one of 16 categories according to the significance test results [Table~\ref{Table.2}].

\subsubsection{Model Selection for Accurate Transition Rates}
\noindent
These significance tests do not include tests of the transition rates. Since in our algorithm every fitted $\lambda$ is larger than 0 because of the assumed bounds, the bootstrapped $\lambda$ is always positive no matter how many bootstrap iterations are performed.  Thus, 0 will always be excluded from the confidence interval of the fitted $\lambda$.   As an alternative, model selection can help to determine whether these transitions terms are necessary in order to describe the dynamics of gene expression.   In addition, since an unnecessarily complicated model with redundant parameters can make estimates of $\lambda$ inaccurate, we add a penalty on the number of parameters and search for the most parsimonious model that is sufficient to describe the gene expression dynamics.

\begin{figure}[H]
    \centering
    \includegraphics[scale=0.6]{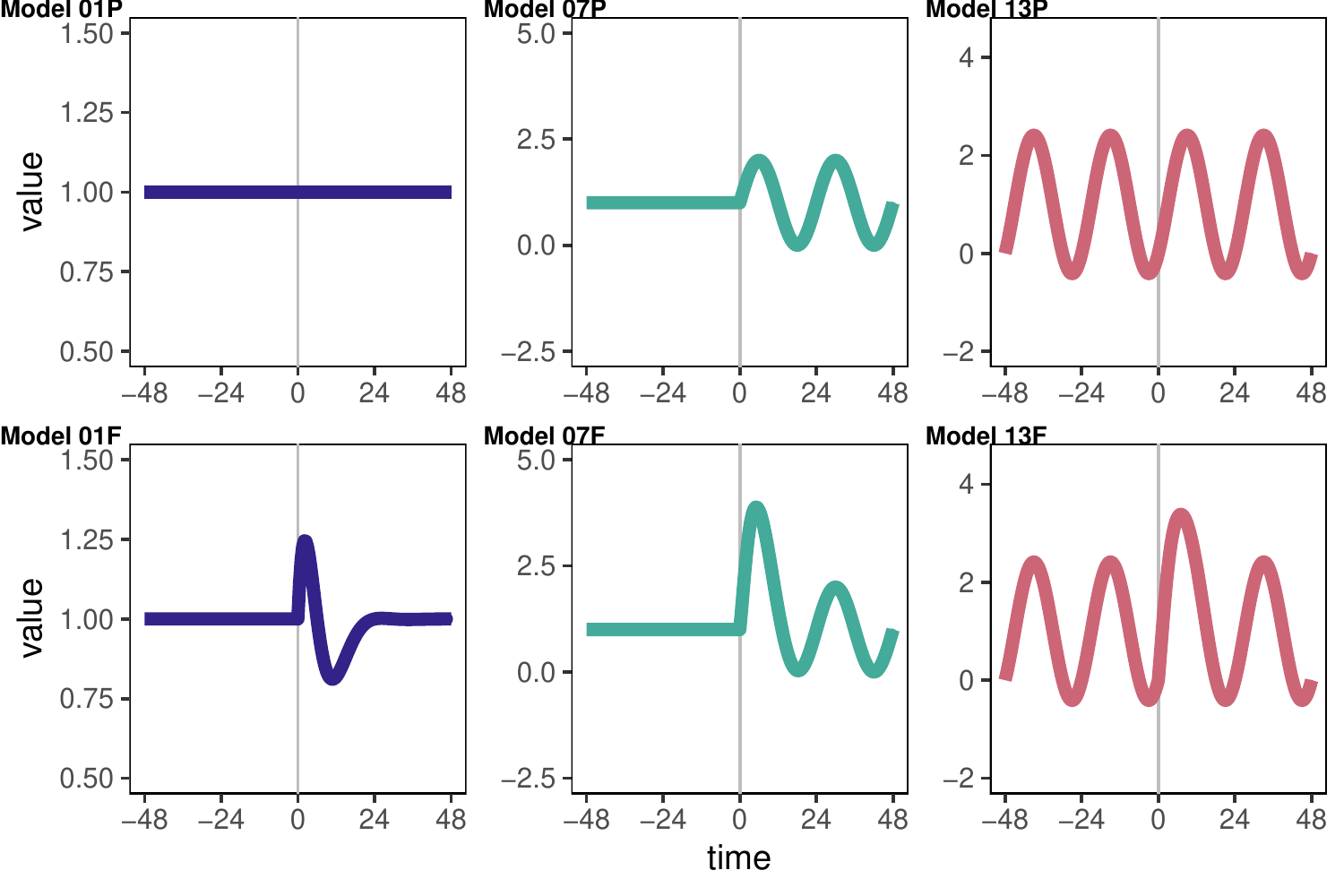}
    \caption{\textbf{Sketches of full and partial models for three types.} Models from Type 01, Type 07 and Type 13 are shown for visualization purposes. For all detailed models and their sketches, refer to SI Table~\ref{SITable.1}.}
    \label{Figure.2}
\end{figure}

In particular, we utilize the finite-sample-corrected AIC (AICc) to perform model selection on a set of 20 candidate models classified according to whether or not the gene expression is cycling at the two temperatures, is differentially cycling, and/or has different mean expression  [Figure~\ref{Figure.2}, SI Table~\ref{SITable.1}].  In addition, a candidate partial model will be defined as one without a transient and thus has only cycling or constant expression (i.e., only two limit cycles or equilibrium points), while a full model will be defined as one containing an exponential function governing a transition between different cycling or constant expression dynamics. Possible explanations for a partial model being a more parsimonious description of the dynamics are: 1) the characteristics of mRNA expression exhibit no changes with a temperature step ($\lambda=0$ by inference, as in Model 01P and Model 13P), or 2) the transient changes happen too quickly to detect with our 2hr sampling frequency ($\lambda=\infty$ by inference, as in other partial models).

The design of candidate models follows three principles: 1) the classified model type should be theoretically valid, in the sense that inferences about cycling and differential cycling are internally consistent; 2) candidate models within each type should share the same characteristics as their model type (i.e., have the same constraints on parameter values); 3) candidate models should be continuous at $t{=}0$ if possible. There are only 10 theoretically valid (internally consistent) categories (e.g., for Type 09, if one gene is only cycling at high temperature but not at low temperature, it is impossible to have the same oscillation pattern at both temperatures.) The only exceptions for the third principle are the Type 02 partial model (Model 02P) and the Type 14 partial model (Model 14P) because they would violate the second design principle without a step change at $t{=}0$ (via an indicator function).

We apply the same proposed nonlinear model fitting procedure [SI Algorithm~\ref{Algorithm.1}] to fit all the candidate full models. The partial models are linear and therefore using linear quantile regression in the R \textit{quantreg} package is sufficient. The asymmetric Laplace distribution (symmetric for median regression, $\tau=0.5$) is used to compute the likelihoods in AICc \cite{koenker1999goodness}, and only genes with large enough differences in AICc between the best model and the second best model are accepted ($\Delta$AICc $\geq2$) \cite{burnham2004multimodel}. We focus only on genes with concordant classifications from both significance test and model selection (i.e., both methods produce the same gene expression dynamics), and infer transition rates from the model selection.  For full models, the transition rate $\lambda$ is inferred from the fitted model with the lowest AICc, and for partial models $\lambda=0$ or $\infty$ depending on the inferred type.

\subsubsection{Gene Selection Criteria}
While the transient response to the temperature change was not measured in the previous aforementioned experiment \cite{Hwangbo2023.01.04.522718}, these data provide an opportunity to focus on genes whose steady-state dynamics (constant expression or rhythmic cycling in the final 24 hours in the transient study) are reproduced in both experiments. Under each temperature, we perform differential cycling detection using limorhyde \cite{singer2019limorhyde} on $Z$-scored TPM data in the transient study collected at Day5 only and the $Z$-scored TPM data in the previous experiment at all time points. Genes detected as significantly differentially cycling ($p\leq0.05$) between the two experiments under either temperature are considered irreproducible cyclers and are excluded from subsequent functional and gene expression analysis. These genes are still used to evaluate the performance of the transient detection method and demonstrate residual heteroskedasticity, however.

\newpage
\section{Results}
\subsection{Effect of Noise on Significance Test and Model Selection}
We take the conservative approach of estimating adaptation rates $\lambda$ only for genes that yield concordant results from the two different fitting methods.  As a result, we are more confident about these genes' characteristics (e.g., cycling status and adaptation rates). 

\begin{figure}[H]
    \centering
    \subfigure[]{
    \includegraphics[scale=0.5]{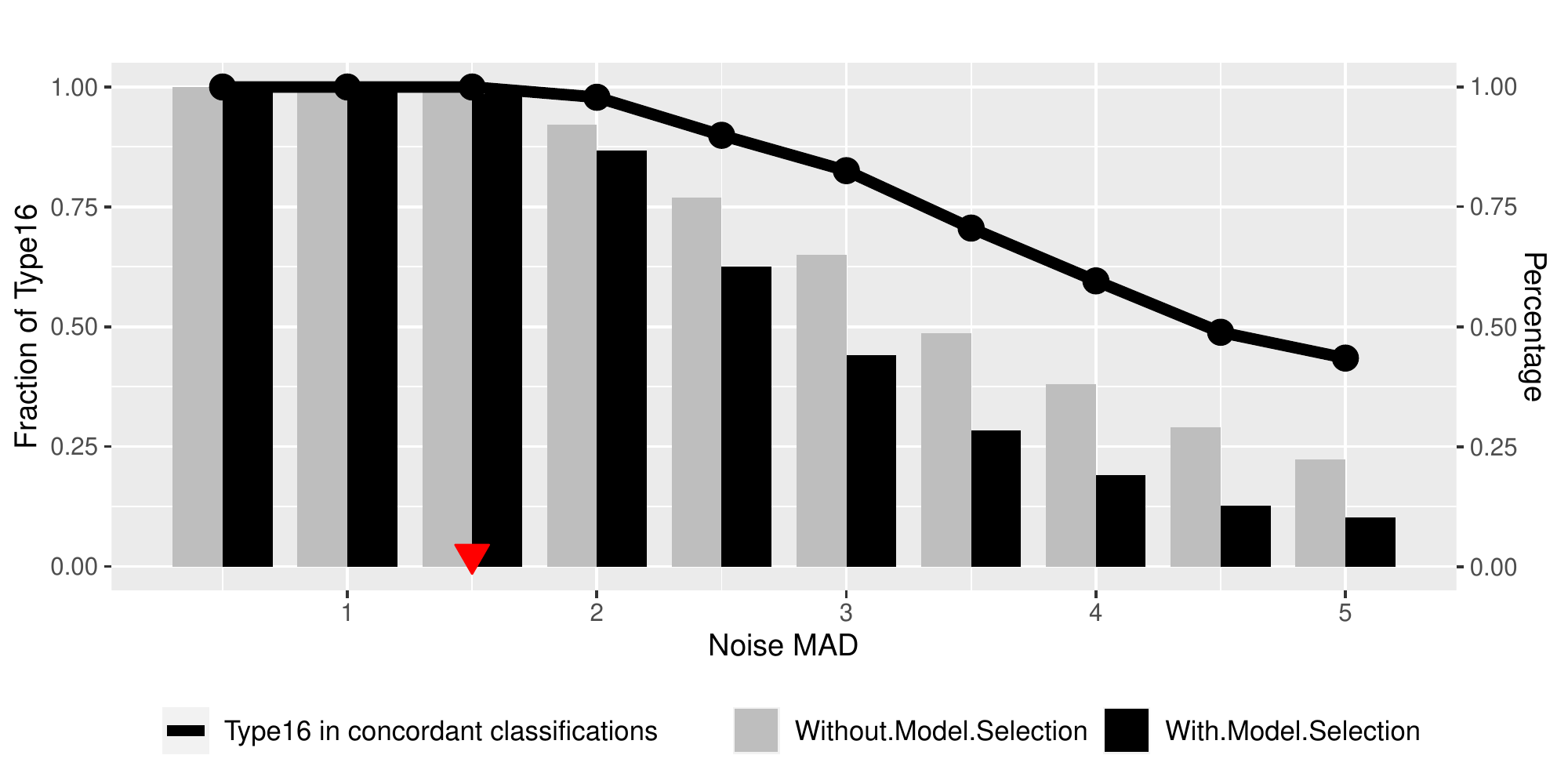}
    }
    \subfigure[]{
    \includegraphics[scale=0.5]{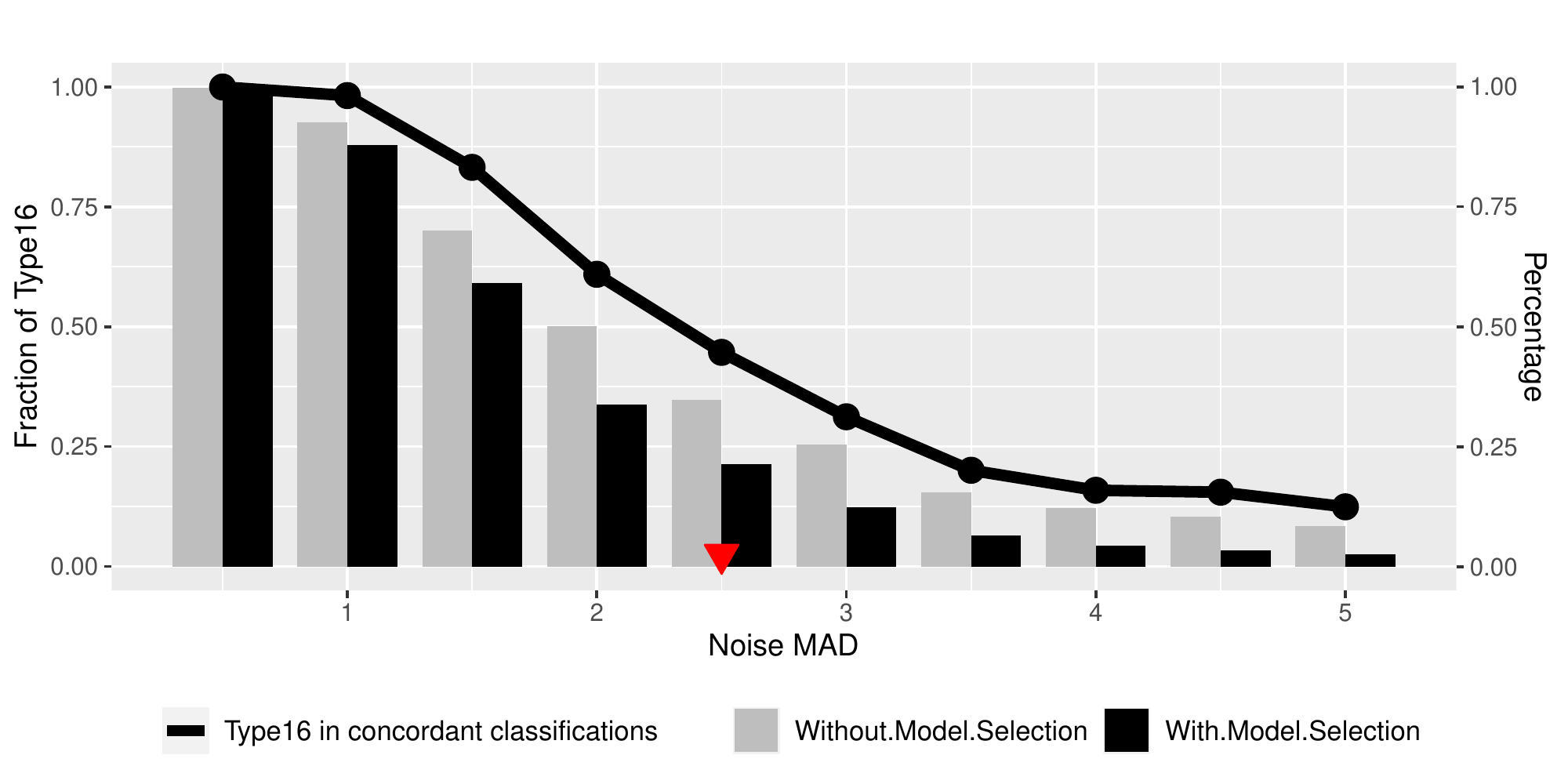}
    }
    \caption{\textbf{Simulation results with model parameters taken from two gene fits (fitted full model curves from \textit{per} and \textit{Hsf}).} Detailed parameter values used in the simulation are shown in SI Figure~\ref{Figure.16}. Residuals are drawn from the Laplace distribution with different mean average deviations (MAD), with $n=10^3$ trials for each MAD. Red triangle indicates the closest bar of the residual MAD to the experimental data of \textit{per} and \textit{Hsf}. Bar plots show the fractions of Type 16 from the significance test only, and concordant Type 16 from the two methods combined.  The black line shows the percentage of Type 16 among all concordant classifications. At the noise level of the experimental data, \textit{per} is classified as concordant while \textit{Hsf} is not. (a) Simulations starting with fitted curve from \textit{per}. (b) Simulations starting with fitted curve from \textit{Hsf}.}
    \label{Figure.3}
\end{figure}

To investigate the performance of our method on classifying genes' dynamics into different categories, we perform random simulation trials starting with parameters obtained from the fits of full model curves to the collected experimental data from two genes, \textit{per} and \textit{Hsf}.  We start from actual fits so that the simulated data are not far from realistic gene expression dynamics [Figure~\ref{Figure.3}, SI Figure~\ref{SIFigure.1}].  Synthetic data are created by adding residuals to the two fitted curves. Residuals are drawn from a Laplace distribution to be consistent with the likelihood function used for model selection, but we vary the simulated residuals' mean absolute deviance (MAD) by a factor that can be up to twice that associated with actual experimental data. 

Combining the significance test and model selection and requiring them to be concordant is a stricter requirement than for each method separately, and thus the numbers of correctly and concordantly classified trials (black bars) are smaller than correctly classified trials from the significance test alone (grey bars).  However, the fraction of trials that are correctly classified within only concordant results (black lines) is higher than the fraction of correctly classified trials using just the significance test (grey bars) for all noise levels [Figure~\ref{Figure.3}].   This is a trade-off, of course: we improve the classification accuracy at the cost of assigning inconclusive results to a fraction of trials. 

As the noise level increases, the fraction of incorrect trials increases in both cases, although the categories to which they are mis-assigned are often different [SI Figure~\ref{SIFigure.1}]. Furthermore, as can be seen from the different slopes of the true positive rates as functions of noise level [Figure~\ref{Figure.3}], the two starting fits have different sensitivities to noise. Thus, \textit{per} is classified as concordant at the experimental noise level while \textit{Hsf} is not. An advantage of requiring the two methods to be concordant, however, is that for both \textit{per} and \textit{Hsf} as the noise level increases it is more likely to conclude that one trial is non-concordant and be excluded from subsequent analysis rather than being misclassified. [SI Figure~\ref{SIFigure.1}]. 

\subsection{Comparing Sampling Schemes using Simulated Data from Existing Model}
\begin{figure}[H]
    \centering
    \subfigure[Simulated data from existing model.]{
    \includegraphics[scale=0.4]{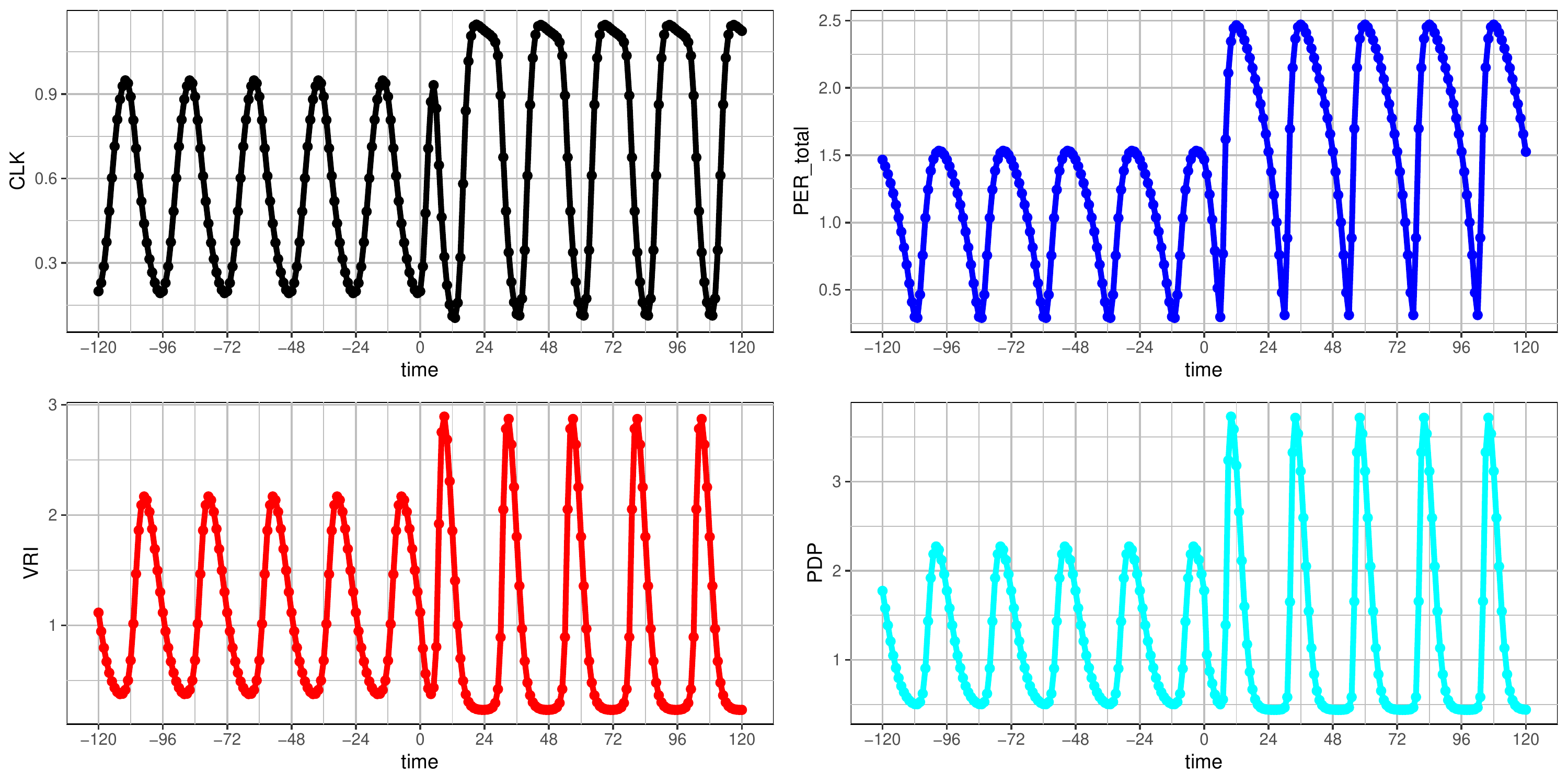}
    }
    \subfigure[Distribution of $\lambda$s.]{
    \includegraphics[scale=0.37]{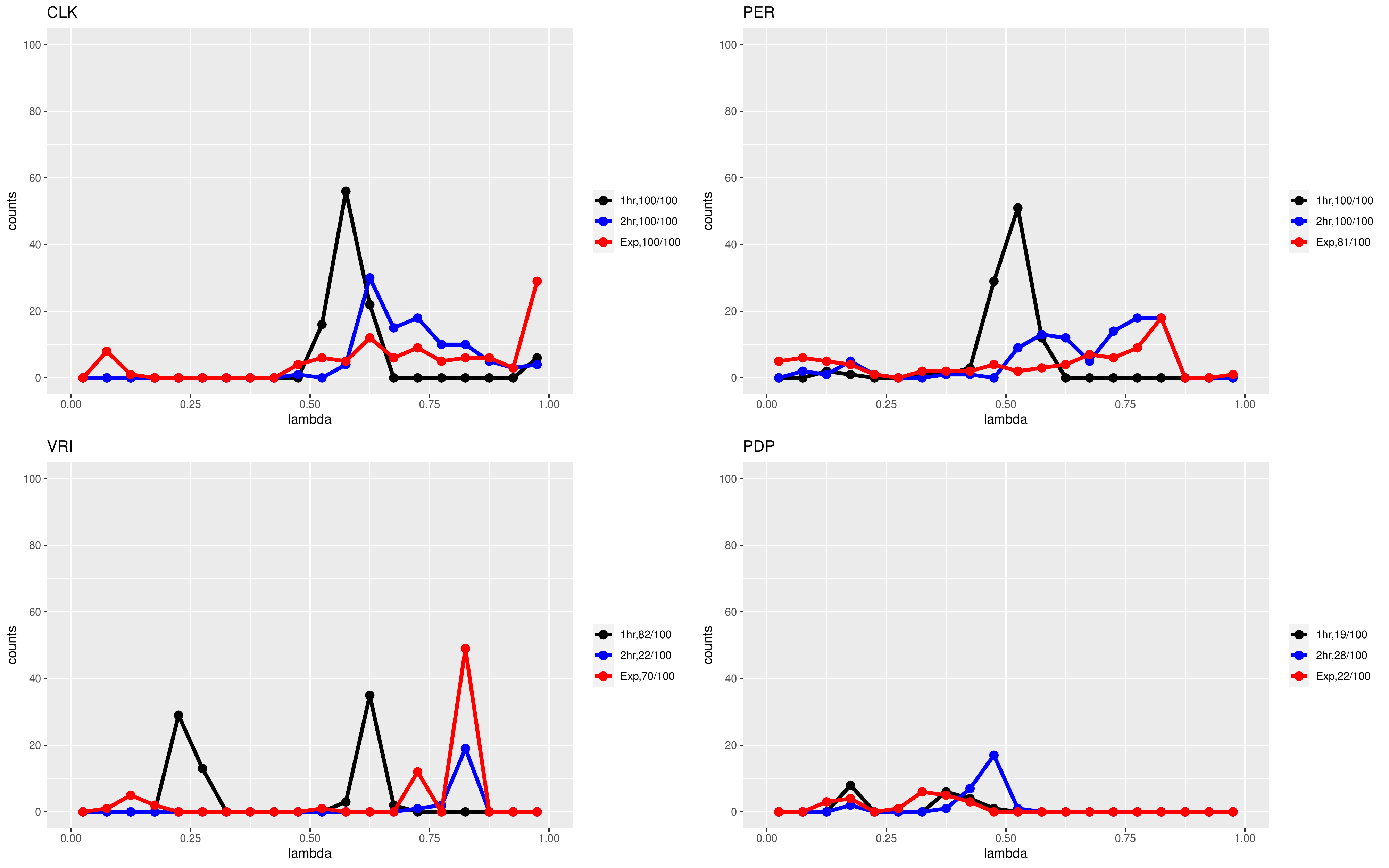}
    }
    \caption{\textbf{Distribution of estimated $\lambda$ for all concordant trials from simulated data.} Parameters change immediately to new values in response to temperature changes at t=0. Black: sample every 1hr; blue: sample every 2hr; red: our sampling scheme.}
    \label{Figure.4}
\end{figure}

We also compared the performance of our method using different sampling schemes. We simulated an existing model \cite{smolen2004simulation} that includes detailed parameter values after a temperature increase. Simulations are performed for 480hrs. At t=240hr there is a step temperature increase, when all parameter values are instantaneously changed to their new values at the high temperature. We take the data in [120hr,  360hr], shift it to $t\in$ [-120hr, 120hr] and compare three sampling schemes: 1) sample every 1hr, 2 replicates; 2) sample every 2hrs, 2 replicates; 3) sample as our experimental sampling scheme. Heteroskedastic residuals are drawn from Laplacian distribution in order to be consistent with the likelihood function used in model selection, with MAD equal to $15\%$ of the expression level. For each sampling scheme we perform n=$10^3$ trials, and the distribution of estimated $\lambda$ from concordant trials is plotted [Figure~\ref{Figure.4}]. The performance of our method is poor for \textit{vri} and \textit{Pdp-1} regardless of which sampling scheme is used, mainly because their expressions have a sharp peak that cannot be properly described by only first order Fourier terms in our model. It should be noted, however, that the model peaks are sharper than those present in the actual data. For \textit{Clk} and \textit{per}, sampling every 2hrs is comparable with our current sampling scheme, but sampling every 1hr outperforms the other sampling schemes as it provides more accurate estimates for $\lambda$. This is not difficult to understand, in that a dense sampling scheme can better capture the dynamics when the adaptation rate $\lambda$ is large.

We also consider the scenario when the adaptation rate is controlled by external inputs. This is modeled by making one parameter, $v_{pyct}$ (maximum cytosolic \textit{PER} phosphorylation rate) approach its new value exponentially at rate $\lambda=0.05$ [Figure~\ref{Figure.5}]. From both the simulated data and distribution of fitted $\lambda$ values, it is clear that the adaptation rate of the whole system is dominated by the driving transient with rate $\lambda=0.05$, and this can be detected with our method. In the context of a slow adaptation rate, sampling every 1hr can yield slightly better performance but of course this comes at the cost of collecting and processing more samples.

\subsection{Heteroskedastic Residuals from the Model Fits}
\begin{figure}[H]
    \centering
    \subfigure[]{
    \includegraphics[scale=0.28]{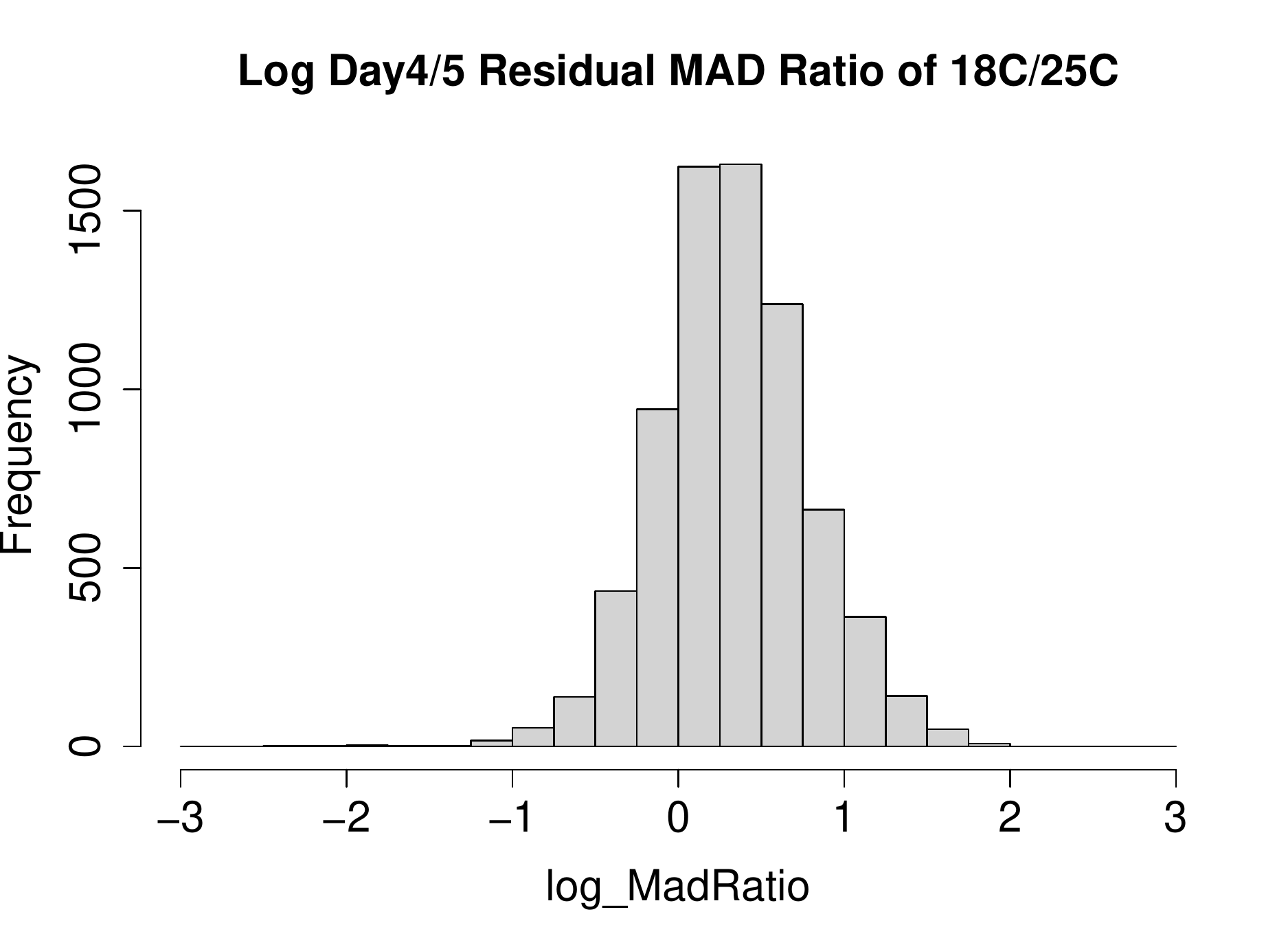}
    }
    \subfigure[]{
    \includegraphics[scale=0.28]{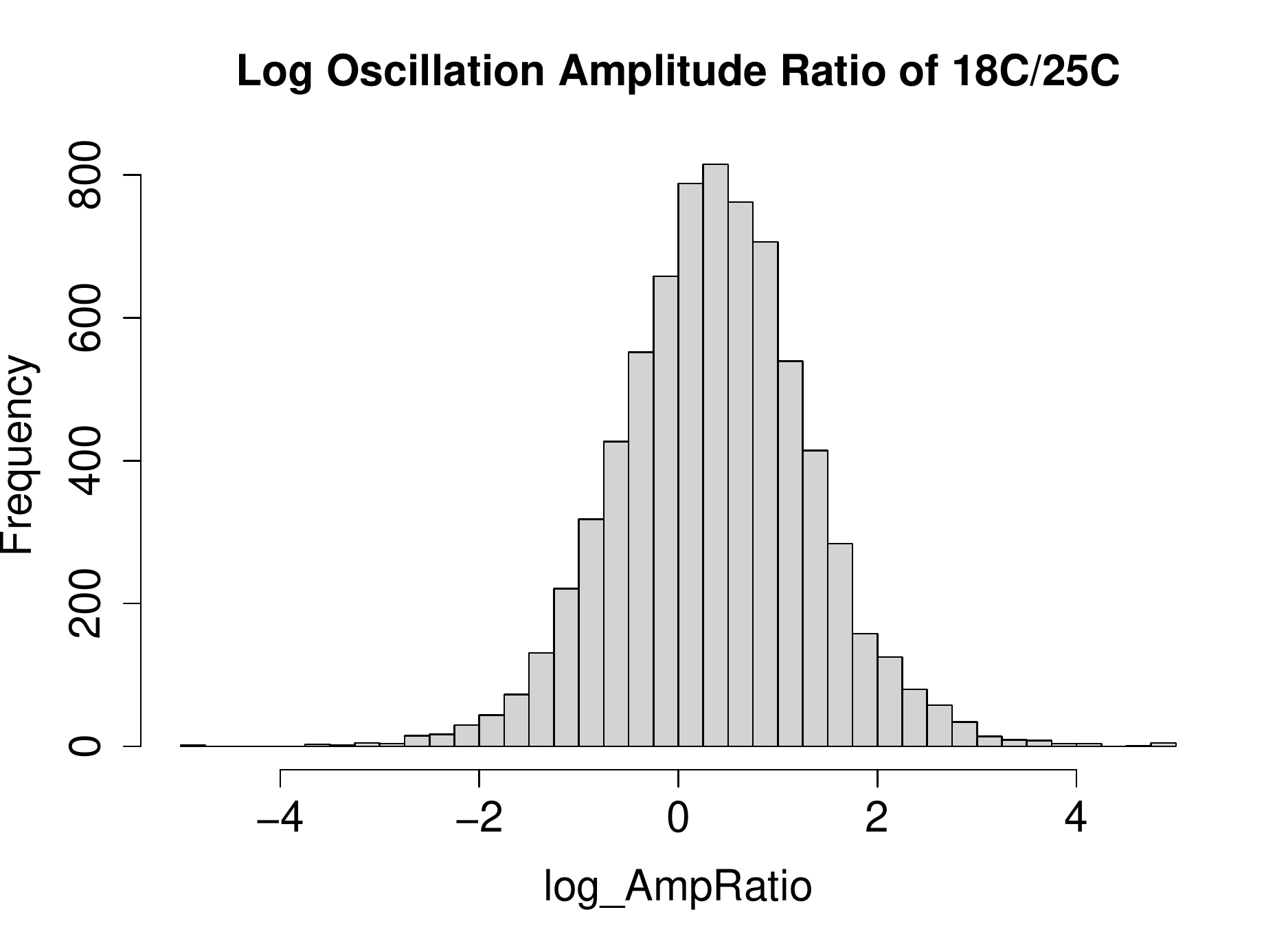}
    }
    \subfigure[]{
    \includegraphics[scale=0.28]{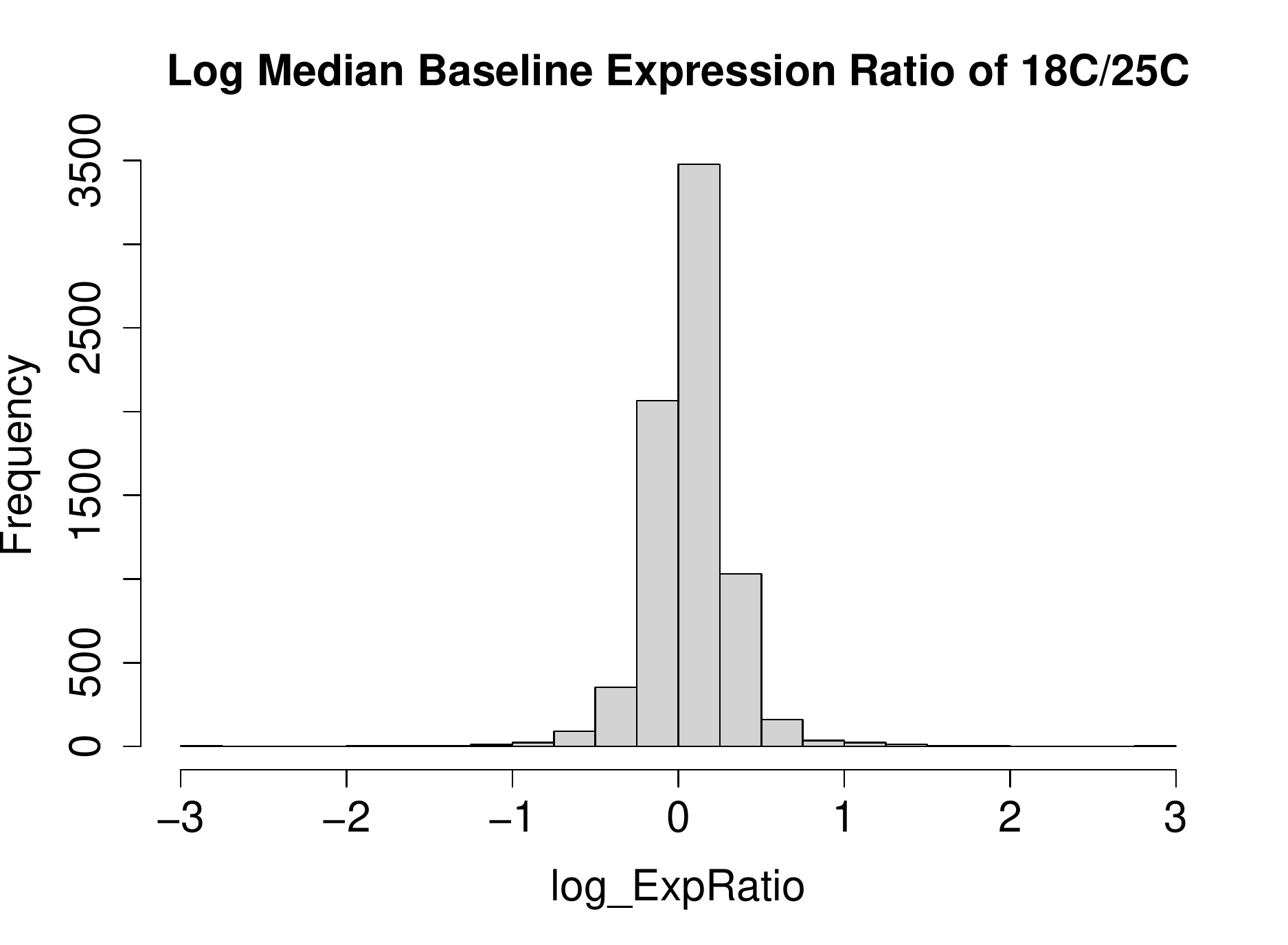}
    }
    \subfigure[]{
    \includegraphics[scale=0.28]{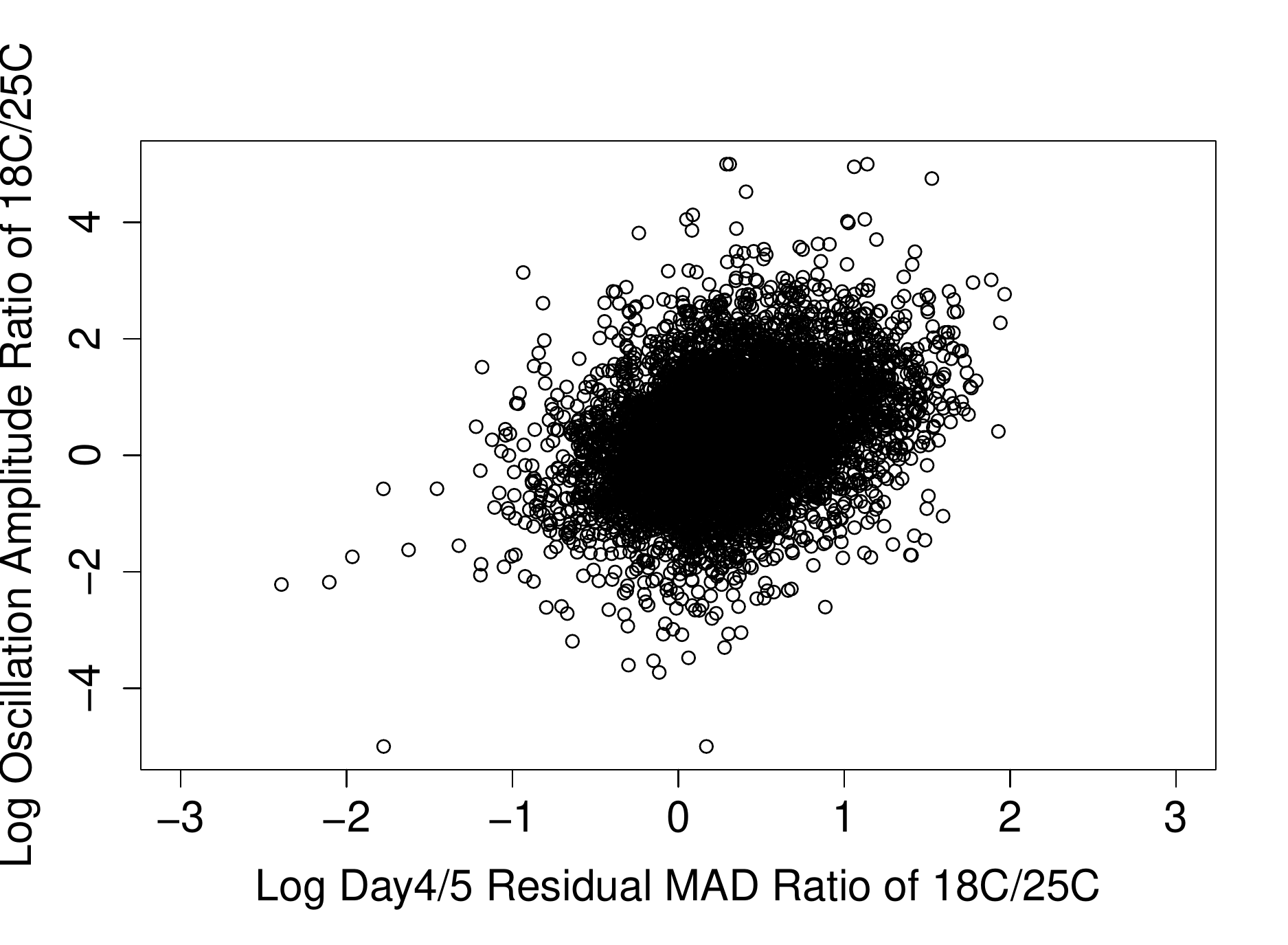}
    }
    \subfigure[]{
    \includegraphics[scale=0.28]{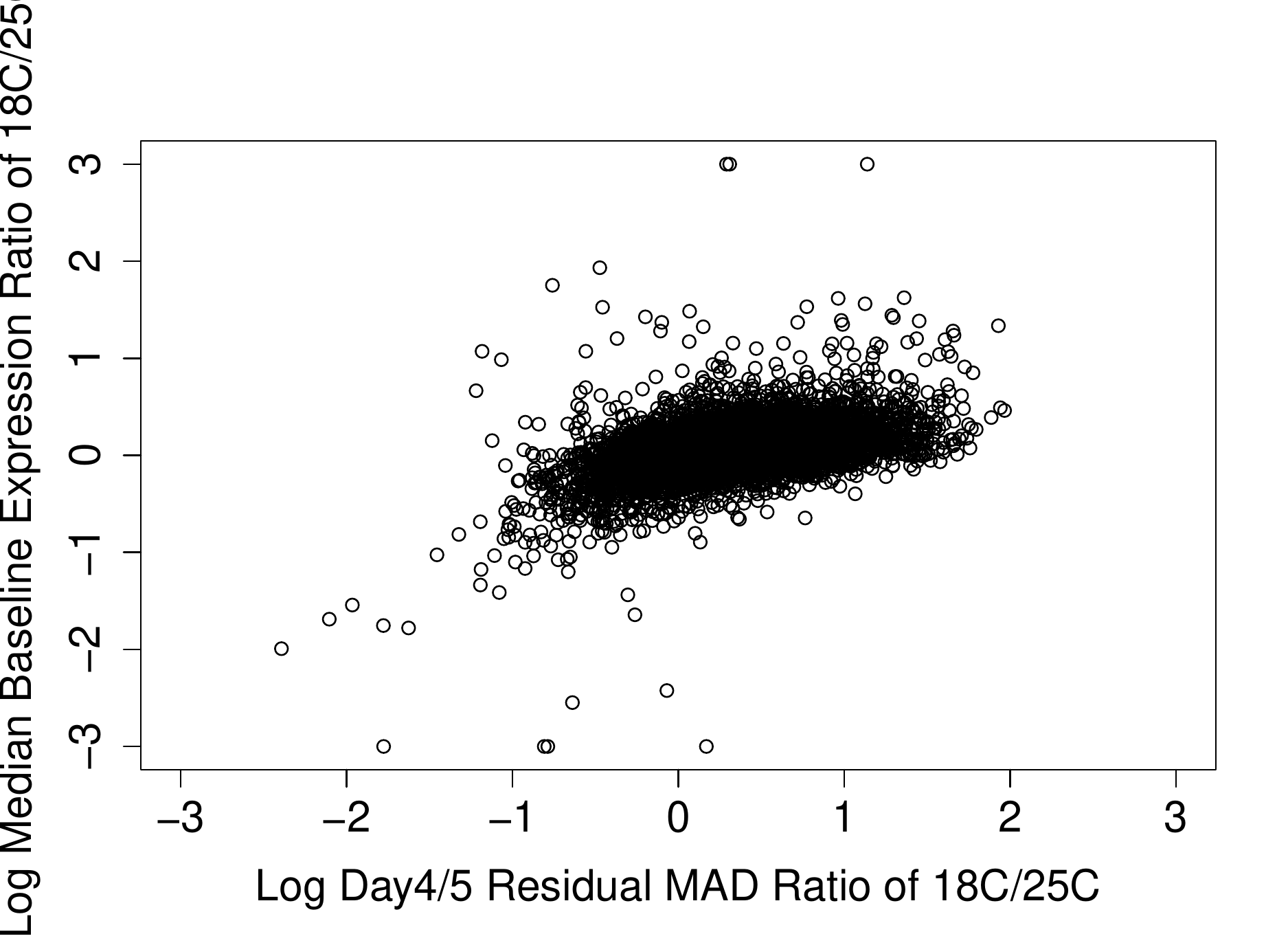}
    }
    \caption{\textbf{Residuals exhibit heteroskedasticity between different temperatures.} Residual MAD are calculated by dividing MAD of residuals at 18\degree{}C Day4/Day5 over 25\degree{}C Day4/Day5. Oscillation amplitude ratio and median baseline expression ratio are calculated in the same manner with fitted curve. (a) Histogram of log residual MAD ratio between different temperatures. $\mu=0.329$, $p<2.2\times10^{-16}$. (b) Histogram of log oscillation amplitude ratio between different temperatures. $\mu=0.362$, $p<2.2\times10^{-16}$. (c) Histogram of log median baseline expression ratio between different temperatures. $\mu=0.072$, $p<2.2\times10^{-16}$. (d) Scatter plot between log residual MAD ratio and log oscillation amplitude ratio (Pearson correlation = 0.34). (e) Scatter plot between log residual MAD ratio and log median baseline expression ratio (Pearson correlation = 0.45).}
    \label{Figure.6}
\end{figure}

\begin{figure}[H]
    \centering
    \subfigure[\textit{betaTub97EF}]{
    \includegraphics[scale=0.4]{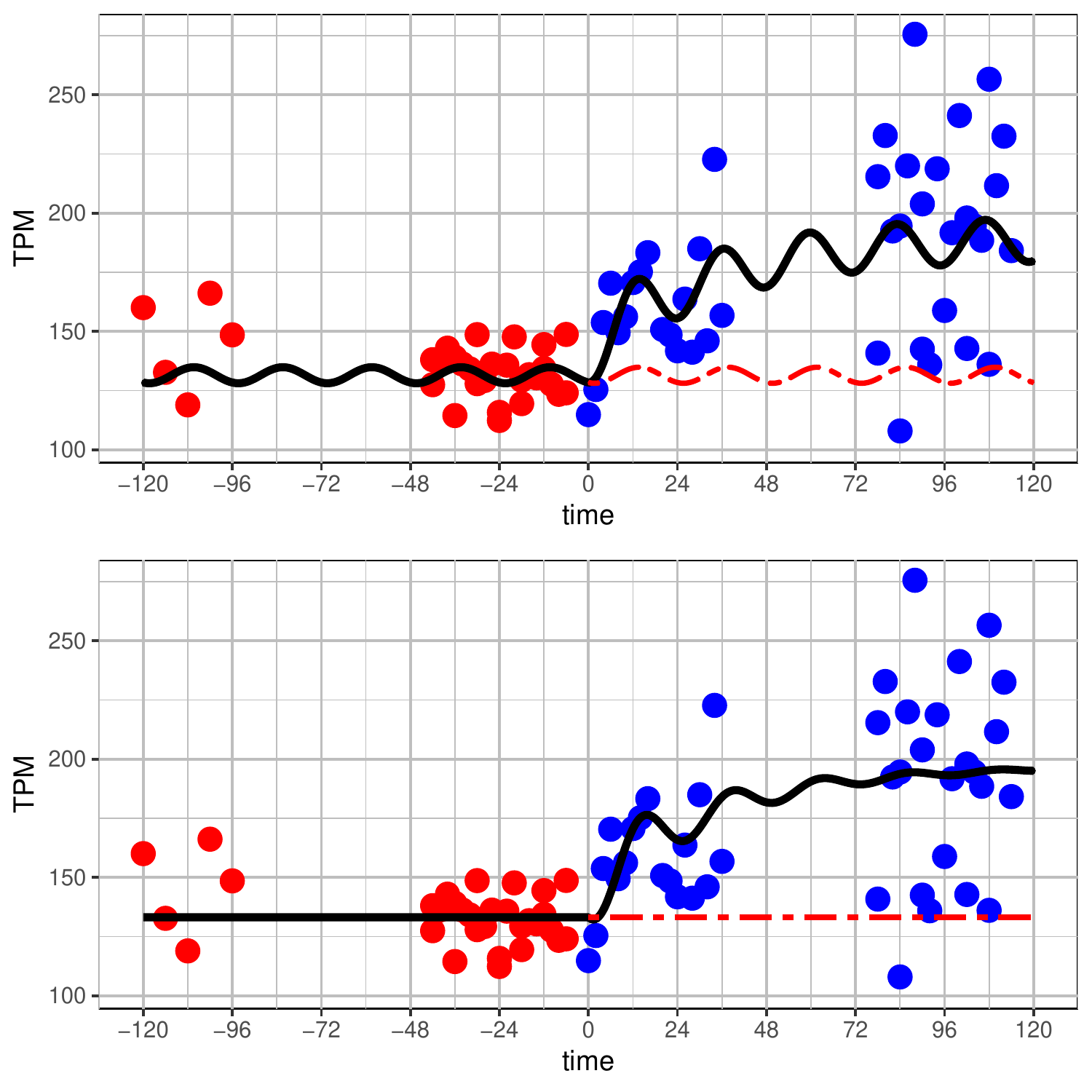}
    }
    \subfigure[\textit{sei}]{
    \includegraphics[scale=0.4]{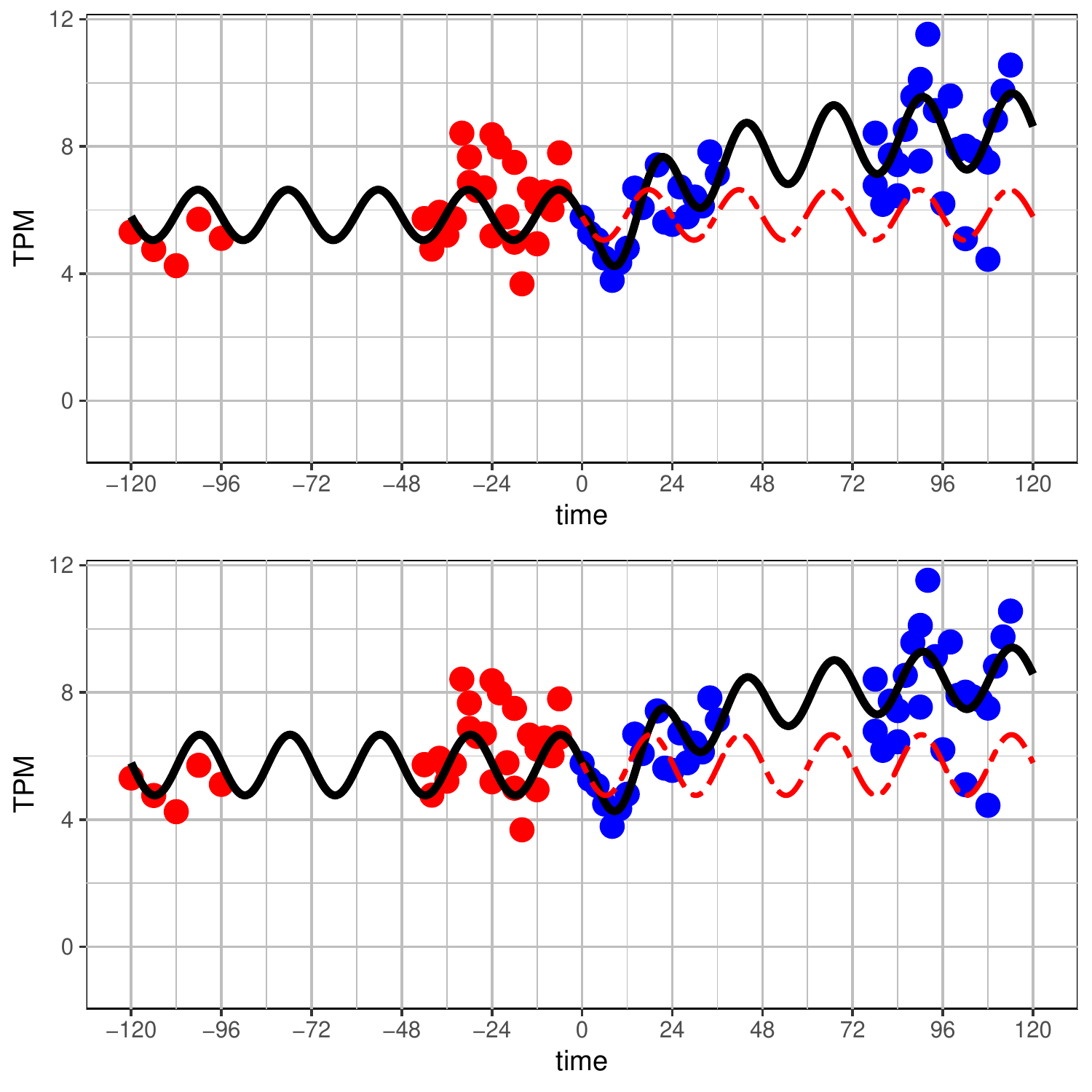}
    }
    \caption{\textbf{Example fits of residual heteroskedasticity under different temperatures.} Top figure shows the fit of the full model, and the bottom figure shows the best model fit with minimum AICc. Black curve is the fitted model curve, and red dashed line is the extension of fitted curve under 25\degree{}C. (a) $\beta$-Tubulin at 97EF (\textit{betaTub97EF}). This gene is concordantly classified as Type 02 (Model 02F). Residuals have larger MAD in 18\degree{}C Day4/5 than in 25\degree{}C. (b) seizure (\textit{sei}). This gene is concordantly classified as Type 14 (Model 14F). Residuals are larger in 18\degree{}C and 25\degree{}C Day4/5 than 18\degree{}C Day1/2.}
    \label{Figure.7}
\end{figure}

For each gene, we plot the histogram of log ratios between mean absolute deviance (MAD) of residuals at 18\degree{}C Day4/5 and 25\degree{}C Day4/5 [Figure~\ref{Figure.6}(a)], times sufficiently long for the gene expression dynamics to have mostly completed the adaptation stage and settled into steady expression/oscillation. The distribution of log ratios is not centered at zero ($\mu=0.329$, $p<2.2\times10^{-16}$), indicating larger fitted residuals at low temperature.  Inspecting plots of individual fitted curves also supports this claim [Figure~\ref{Figure.7}(a)].  Furthermore, there are other types of residual heteroskedasticity observed, for example, the residual MAD in the transition stage (18\degree{}C Day1/2) can be different from that after adaptation (18\degree{}C/25\degree{}C Day4/5) [Figure~\ref{Figure.7}(b)]. 

It is also observed that fitted oscillation amplitudes and median expression levels behave similarly to the residuals [Figure~\ref{Figure.6}(b),Figure~\ref{Figure.6}(c)], i.e., these fitted characteristics are larger in low temperature than in high temperature. Moderate correlations between log residual MAD ratio and log oscillation amplitude ratio (0.34, Pearson correlation) and log median baseline expression ratio (0.45, Pearson correlation) are observed [Figure~\ref{Figure.6}(d), Figure~\ref{Figure.6}(e)]. This suggests that mutual effects of larger median baseline expression level and oscillation amplitude result in larger mRNA expression variations, and thus larger levels of noise may be introduced at low temperature (e.g. intrinsic molecular noise and experimental measurement noise). These results illustrate the importance of taking residual heteroskedasticity into consideration when performing significance tests, especially across different environmental conditions. 

\subsection{Dynamics of Concordant Cycling Genes and Core Clock Genes}
\begin{figure}[H]
    \centering
    \includegraphics[scale=0.5]{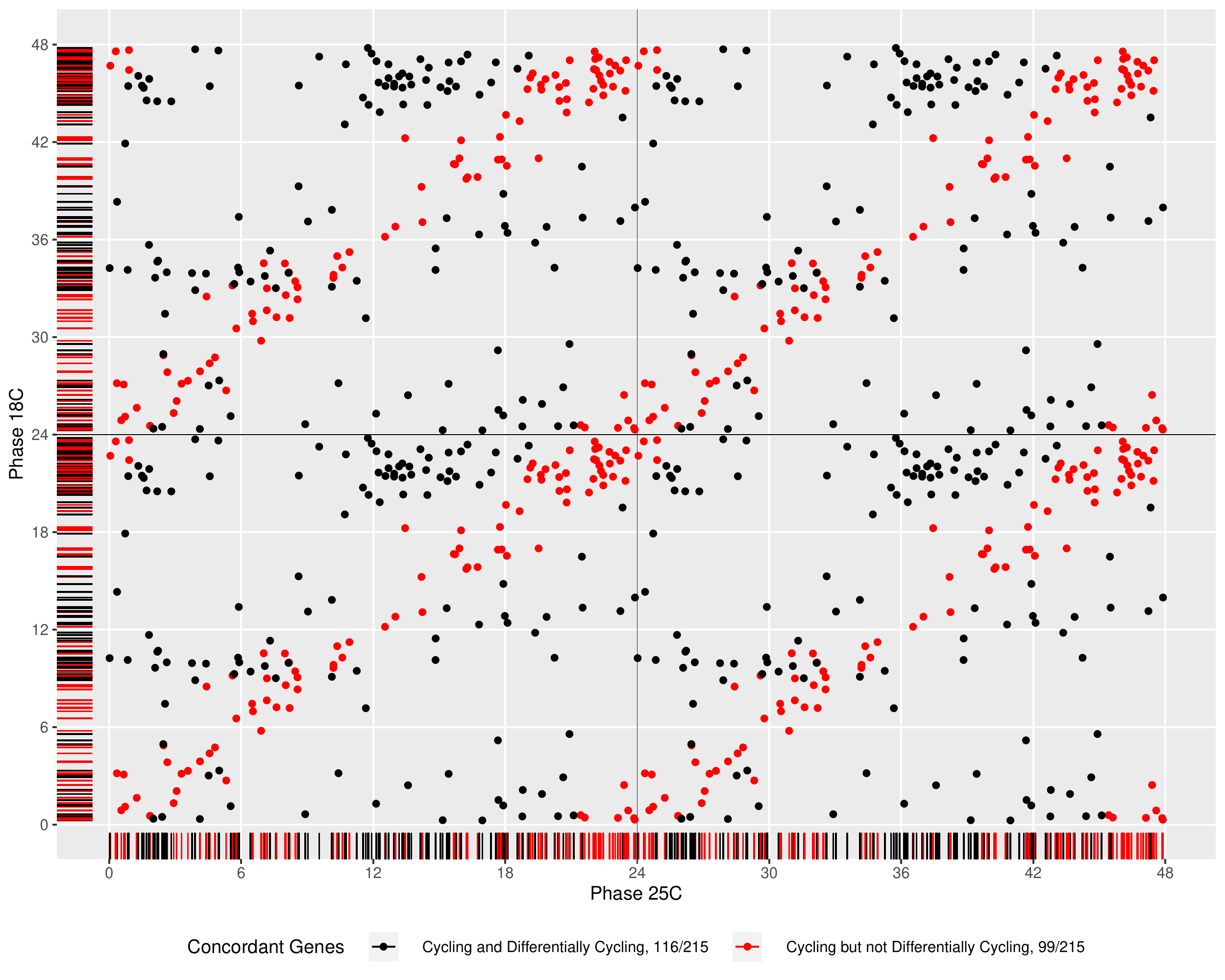}
    \caption{\textbf{Empirical distribution of estimated phases at two temperatures for concordant cycling genes.} The scatter plot is double-plotted for better visualization. Rug plots next to the axes show the density of phases under each temperature. Non differentially-cycling genes have similar estimated phases under two temperatures, while differentially-cycling genes usually do not. Red: Cycling but not differentially cycling genes; black: cycling and differentially cycling genes.}
    \label{Figure.8}
\end{figure}

We identified 1,082 concordant genes out of all 7,310 genes ($\sim$14.8\%) with median TPM value $>5$ under either temperature. 708 genes pass the cycling detection reproducibility criteria with respect to the previous dataset, and 215 are detected as cycling under both temperatures.  99 genes are detected as not differentially cycling with estimated phases at two temperatures to be almost the same, while the other 116 genes are detected as differentially cycling [Figure~\ref{Figure.8}]. 

\begin{figure}[H]
    \centering
    \subfigure[18\degree{}C Day1/2]{
    \includegraphics[scale=0.3]{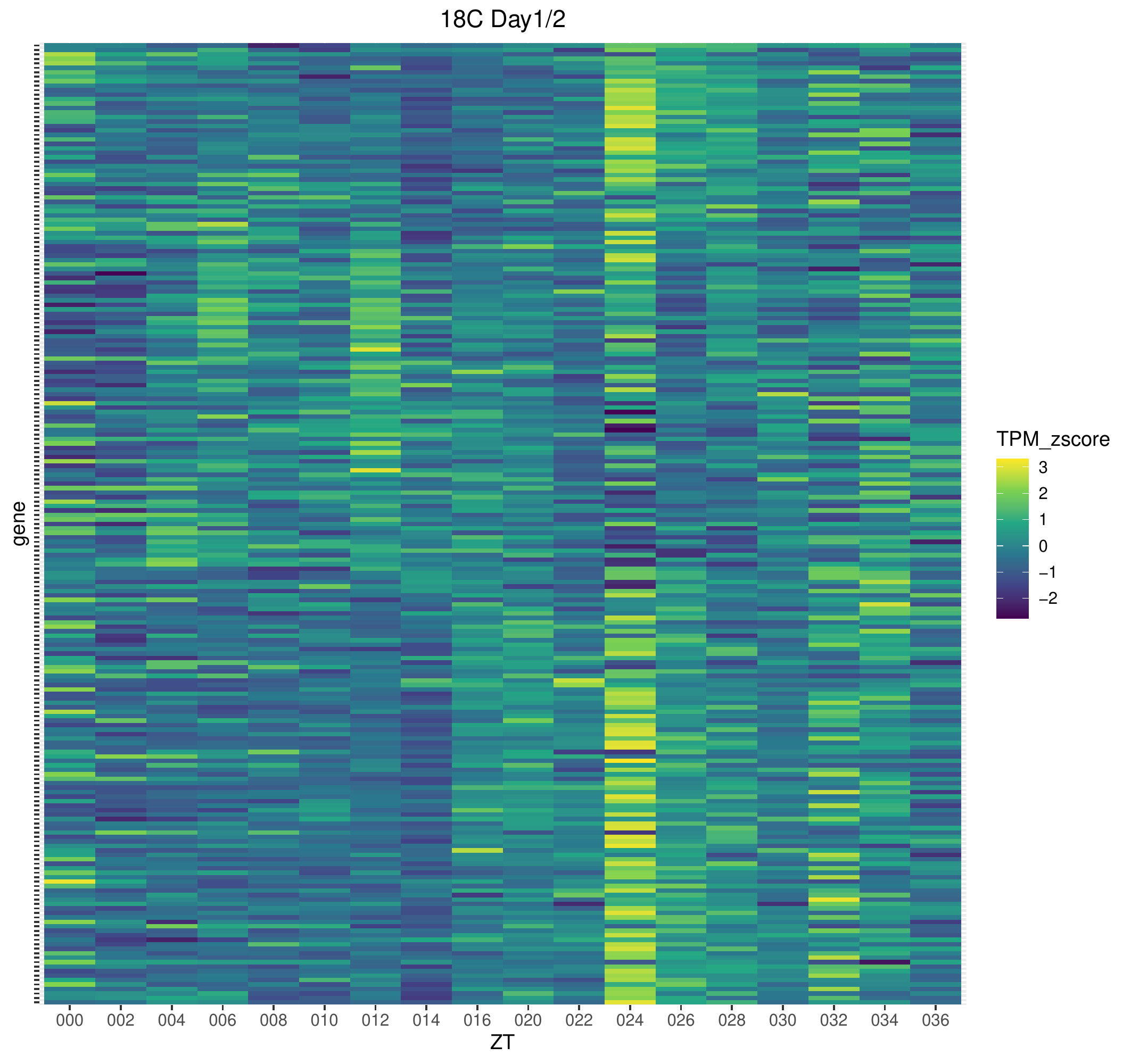}
    }
    \subfigure[18\degree{}C Day4/5]{
    \includegraphics[scale=0.3]{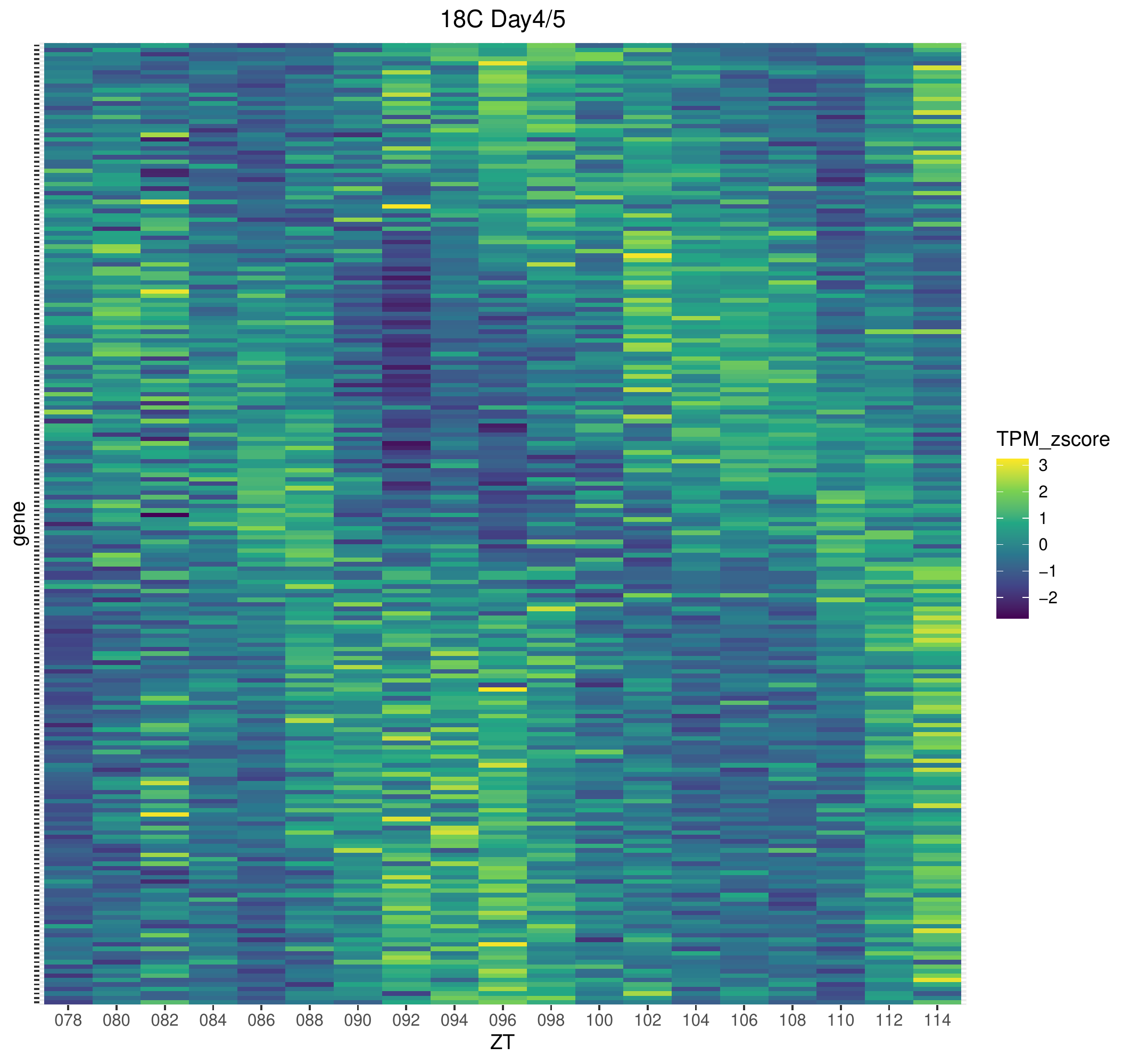}
    }
    \subfigure[25\degree{}C]{
    \includegraphics[scale=0.3]{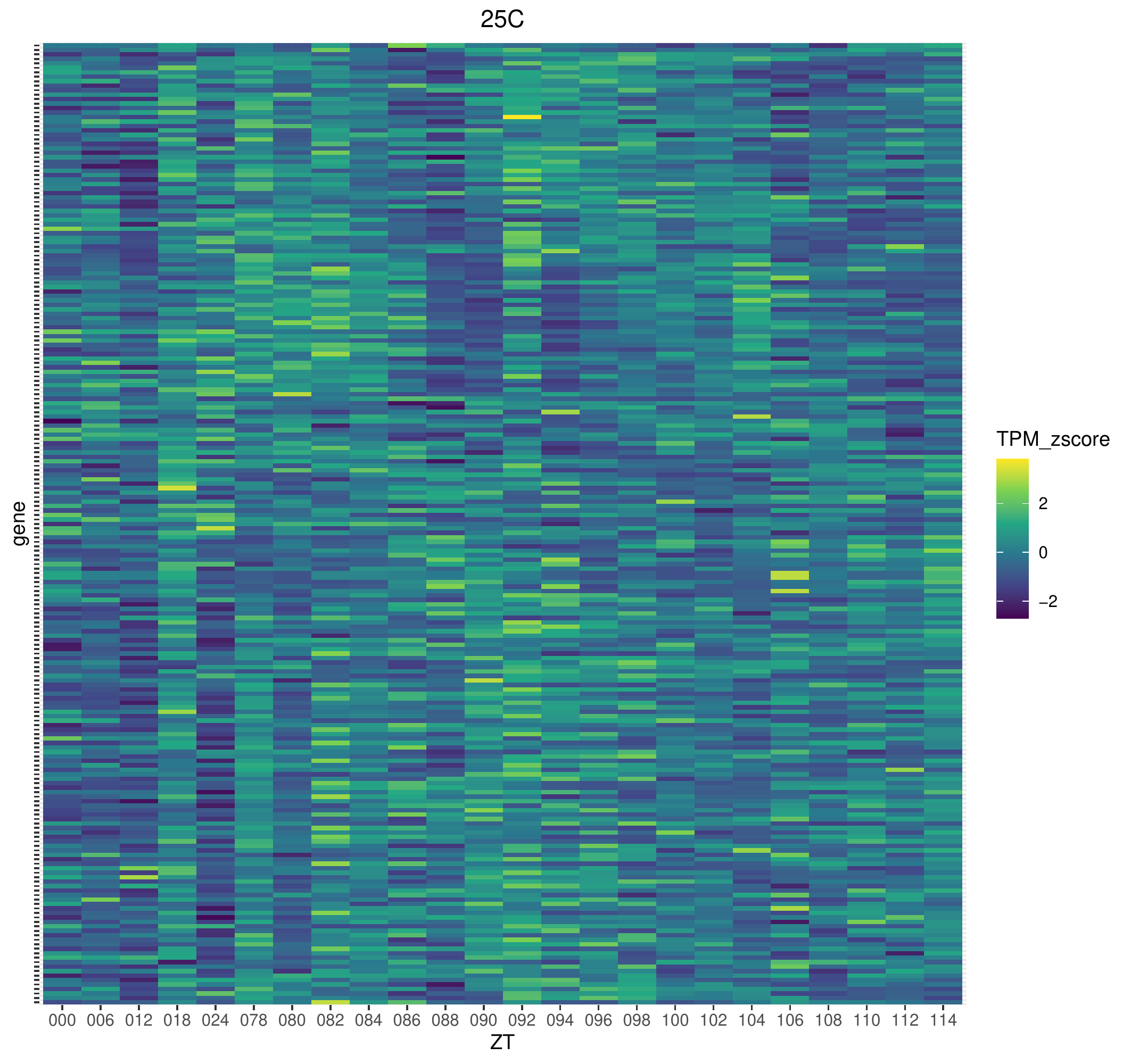}
    }
    \caption{\textbf{Heatmaps of concordant cycling genes under both temperature, ranked by phases under 18\degree{}C.} Gene expressions (TPM) are $Z$-scored within the three time windows (25\degree{}C, 18\degree{}C Day1/2, 18\degree{}C Day4/5). A large fraction of genes peaks at ZT24 in 18\degree{}C Day1 as well as around ZT96 in 18\degree{}C Day4.}
    \label{Figure.9}
\end{figure}

We plot heatmaps of concordant cycling genes under both temperatures [Figure~\ref{Figure.9}], ranked by estimated phases under 18\degree{}C. Nearly half of the cycling genes peak around ZT96 on 18\degree{}C Day4/5.  This may be because the effect LD cycle entrainment is more prominent in 18\degree{}C than 25\degree{}C, so that more cycling genes are synchronized and phase-locked. Interestingly, the 18\degree{}C limit cycle is well correlated with the dynamics during adaptation on 18\degree{}C Day1/2, such that genes peaking around ZT96 also have high expression levels at ZT24, although some continue to adapt after Day1/2.  For the remainder of genes that peak at times other than ZT96 on 18\degree{}C Day4/5, their expression levels also appear to be high at the corresponding ZT on 18\degree{}C Day1/2, although the pattern is not as recognizable as the prominent stripe at ZT24. 

\begin{figure}[H]
    \centering
    \subfigure[\textit{Clk}]{
    \includegraphics[scale=0.4]{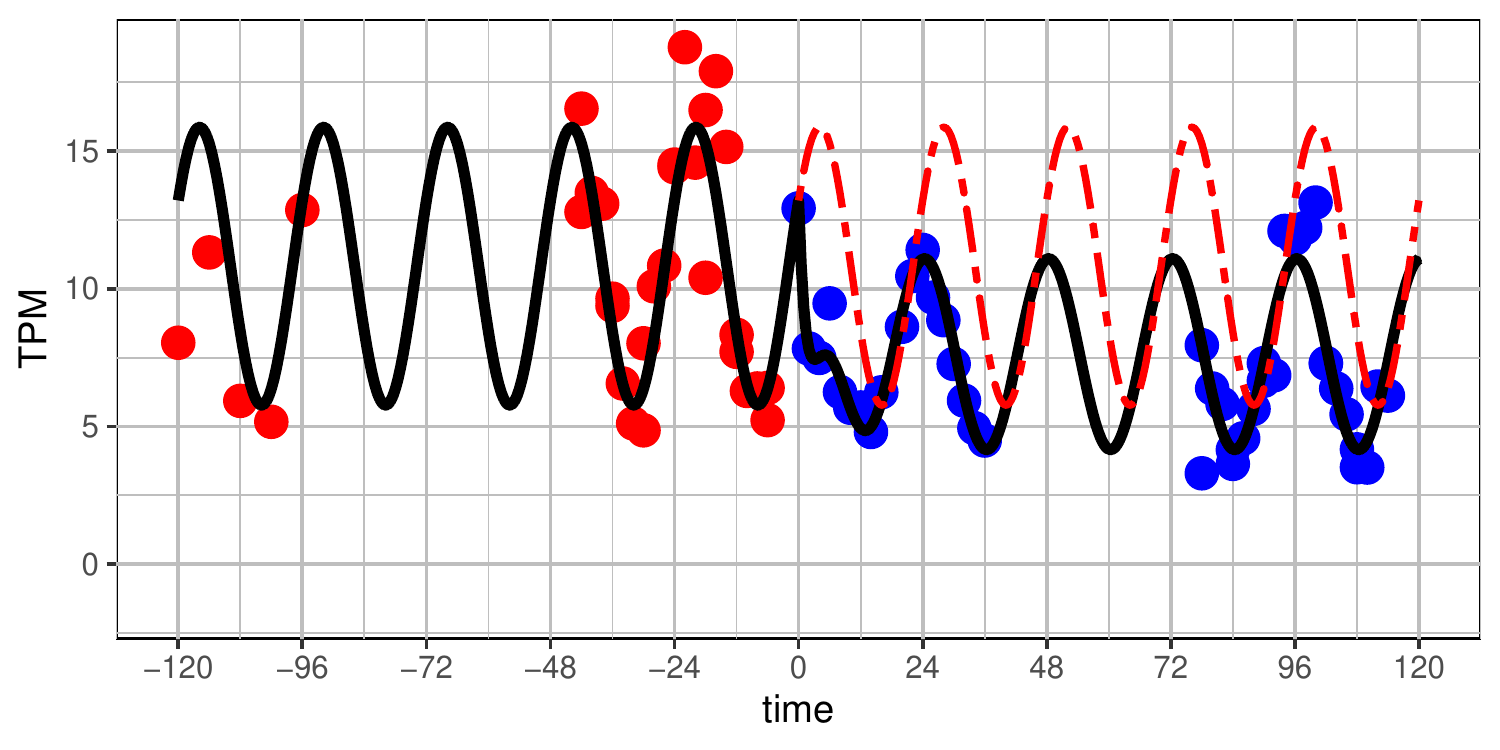}
    }
    \subfigure[\textit{vri}]{
    \includegraphics[scale=0.4]{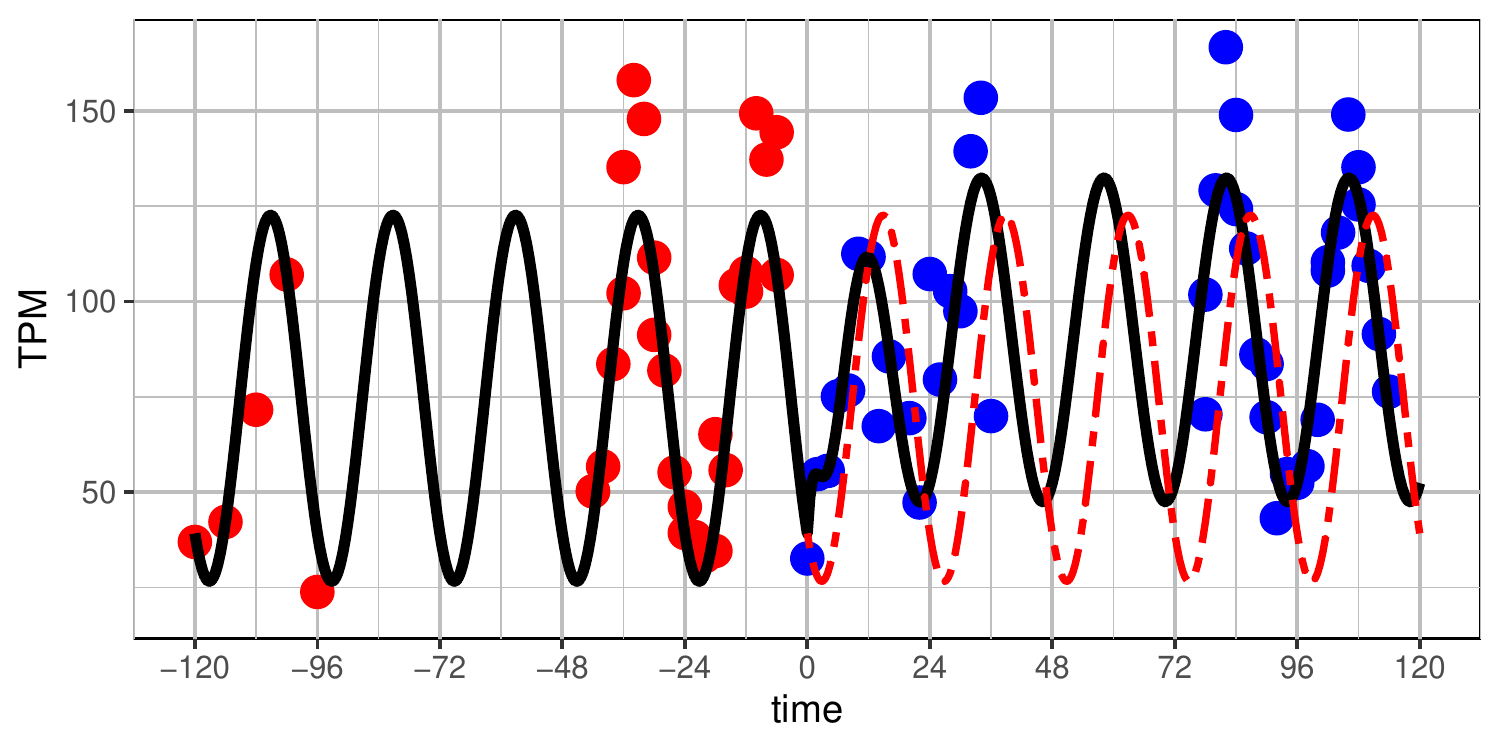}
    }
    \subfigure[\textit{per}]{
    \includegraphics[scale=0.4]{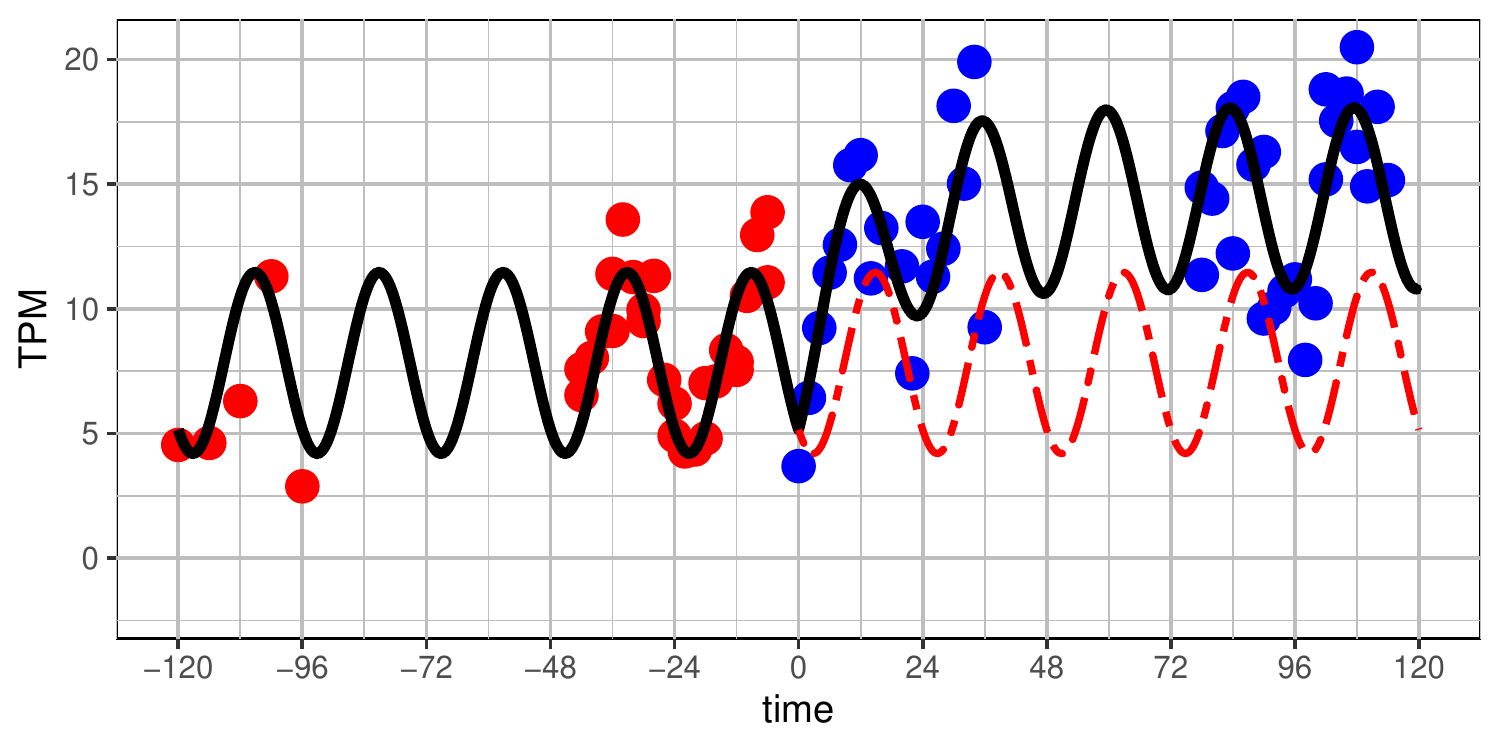}
    }
    \subfigure[\textit{tim}]{
    \includegraphics[scale=0.4]{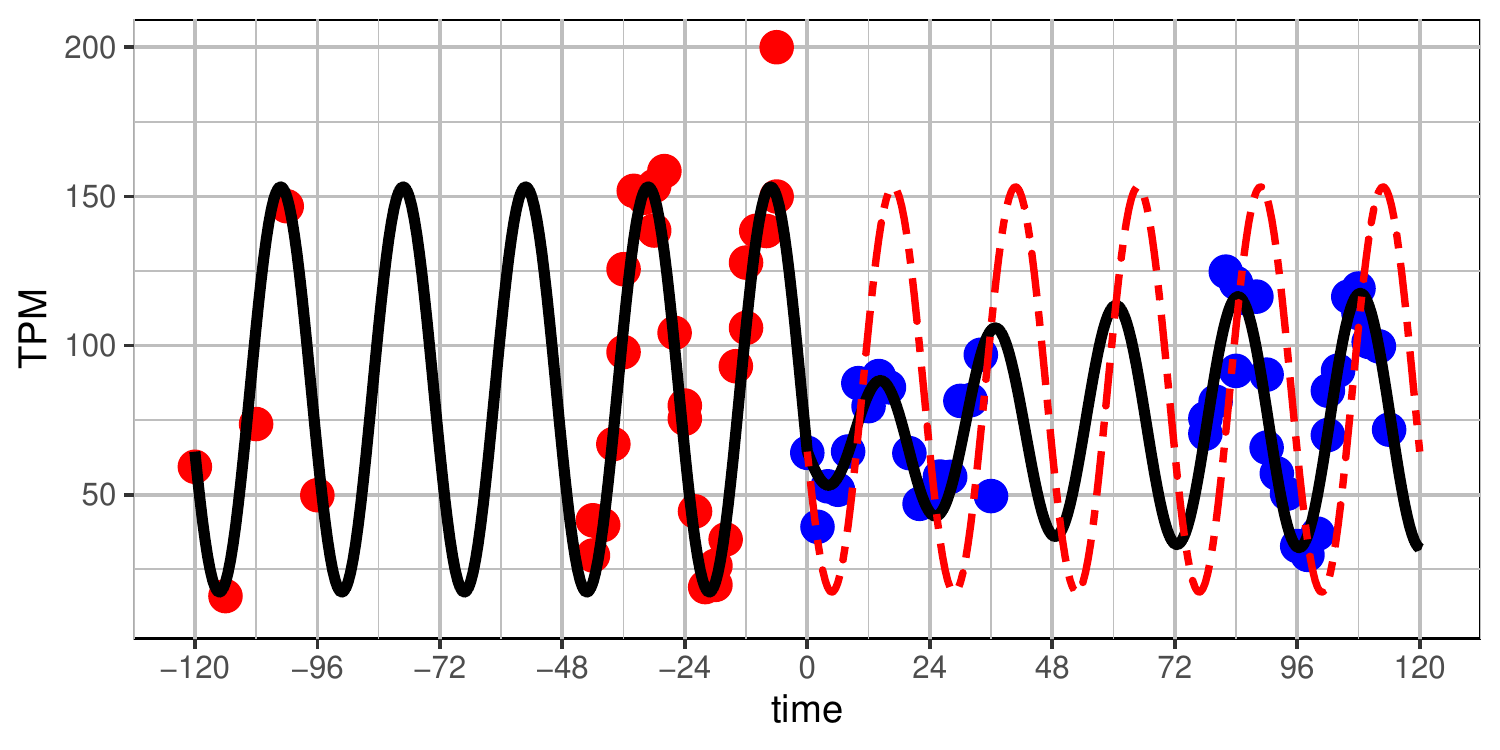}
    }
    \caption{\textbf{Gene expression and fitted results for core clock components \textit{Clk}, \textit{vri}, \textit{per}, \textit{tim}.} Each figure shows the fit of the full model, which is the same as the best model with minimum AICc. All genes are concordantly classified as Type 16 (Model 16F). \textit{Clk} has the largest adaptation rate, while \textit{tim} has the smallest rate. Black curve is the fitted model curve, and red dashed line is the extension of fitted curve under 25\degree{}C. All genes show a small phase advance given the temperature change. (a). \textit{Clk}. $\lambda=0.308hr^{-1}$, $\Delta \phi=3.8hr$. (b). \textit{vri}. $\lambda=0.306hr^{-1}$, $\Delta \phi=4.7hr$. (c). \textit{per}. $\lambda=0.073hr^{-1}$, $\Delta \phi=3.4hr$. (d). \textit{tim}. $\lambda=0.041hr^{-1}$, $\Delta \phi=4.5hr$.}
    \label{Figure.10}
\end{figure}

Core clock genes \textit{Clk}, \textit{vri}, \textit{per}, \textit{tim} are included in these concordant cycling genes and we are able to infer adaptation rates $\lambda$ for these genes [Figure~\ref{Figure.10}]. \textit{Clk} has the largest adaptation rate $\lambda=0.308$ and \textit{vri} has a slightly smaller rate $\lambda=0.306$ than \textit{Clk}. The rates for \textit{per} ($\lambda=0.073$) and \textit{tim} ($\lambda=0.041$), however, are much smaller. In response to the step temperature change, \textit{Clk}, \textit{vri} and \textit{per} adapt towards the new median baseline expression level while maintaining their oscillation amplitudes, while \textit{tim} experiences a rapid decrease in its oscillation amplitude. All four core clock genes show a slight phase advance of 3-5 hours. 

\subsection{Potential Source of Temperature Adaptation}
The observed changes in the oscillation patterns of the core clock genes may result from upstream changes in other circadian genes which regulate them. Effects of other transcription regulators or temperature-sensitive genes may also be responsible for these changes, and the four clock genes themselves may be subject to different upstream controls. In order to find potential initiators of the temperature adaptation, we select concordant genes with larger transition rates than \textit{tim} in three major GO categories: Circadian Rhythm (GO:0007623, Table~\ref{Table.3}), Response to Temperature Stimulus (GO:0009266, Table~\ref{Table.4}) and Transcription Regulator Activity (GO:0140110, SI Table~\ref{Table.5}).

\begin{figure}[H]
    \centering
    \includegraphics[scale=0.5]{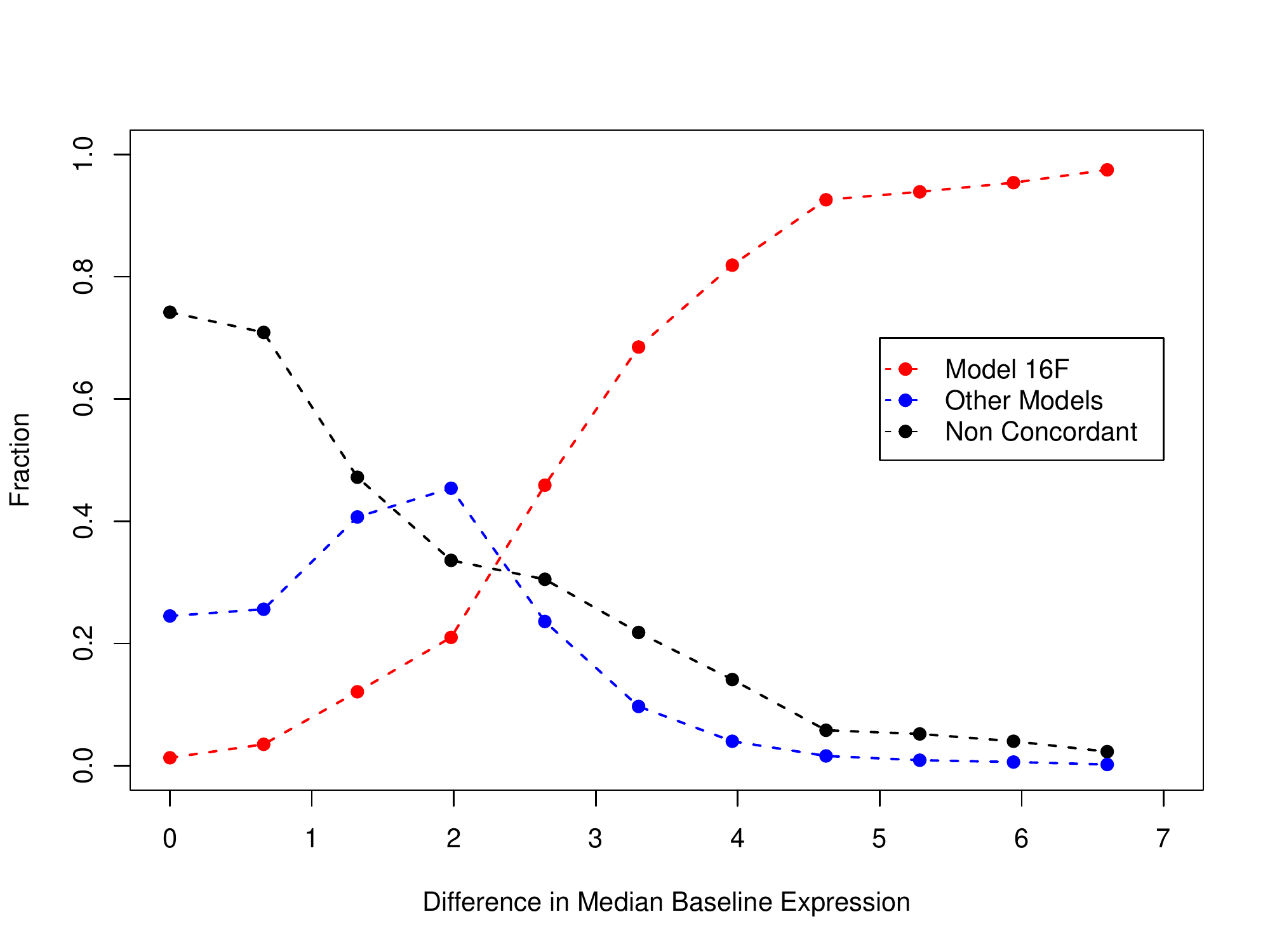}
    \caption{\textbf{Simulation results with varying difference between median baseline expressions at two temperatures.} The model curve is chosen from the parameter fit of \textit{per}, with its $A_6$ value changing from 0.11 to -6.50, such that the difference between two median baseline expressions ranges from 0 to 6.60 (difference in experimental data). Residuals are drawn from the empirical distribution of \textit{per}'s residuals, with $10^3$ trials for each $A_6$. The red curve shows the fraction of trials classified as concordant Type 16 (Model 16F).  The blue curve shows the fraction of trials being classified as other concordant types (models), and the black curve is the fraction of non-concordant trials.}
    \label{Figure.12}
\end{figure}

The majority of the listed genes with large transition rates have significantly different median baseline expressions between the two temperatures (i.e., are differentially expressed). It is possible that exponential transitions are easier to detect if one gene shows a large difference in median baseline expressions. To check this hypothesis we run simulations using the parameters from \textit{per}, which is a Model 16F curve. The $A_6$ parameter is varied so that the difference between median baseline expressions under two temperatures shrinks to zero gradually, with other eight parameters unchanged. With each new $A_6$ value synthetic data is created by adding to the model curve residuals drawn from \textit{per}'s  empirical distribution of residuals with replacement. Simulation results [Figure~\ref{Figure.12}] show that as the difference between two median baseline expressions gets smaller, a greater fraction of trials will not be concordantly classified as some type (model), and the fraction of correct classifications also decreases. This result provides one explanation why many of our selected genes are differentially expressed. 

\subsection{Mapping of Transition Rates onto Existing Gene Regulatory Networks}
\begin{figure}[H]
    \centering
    \includegraphics[scale=0.6]{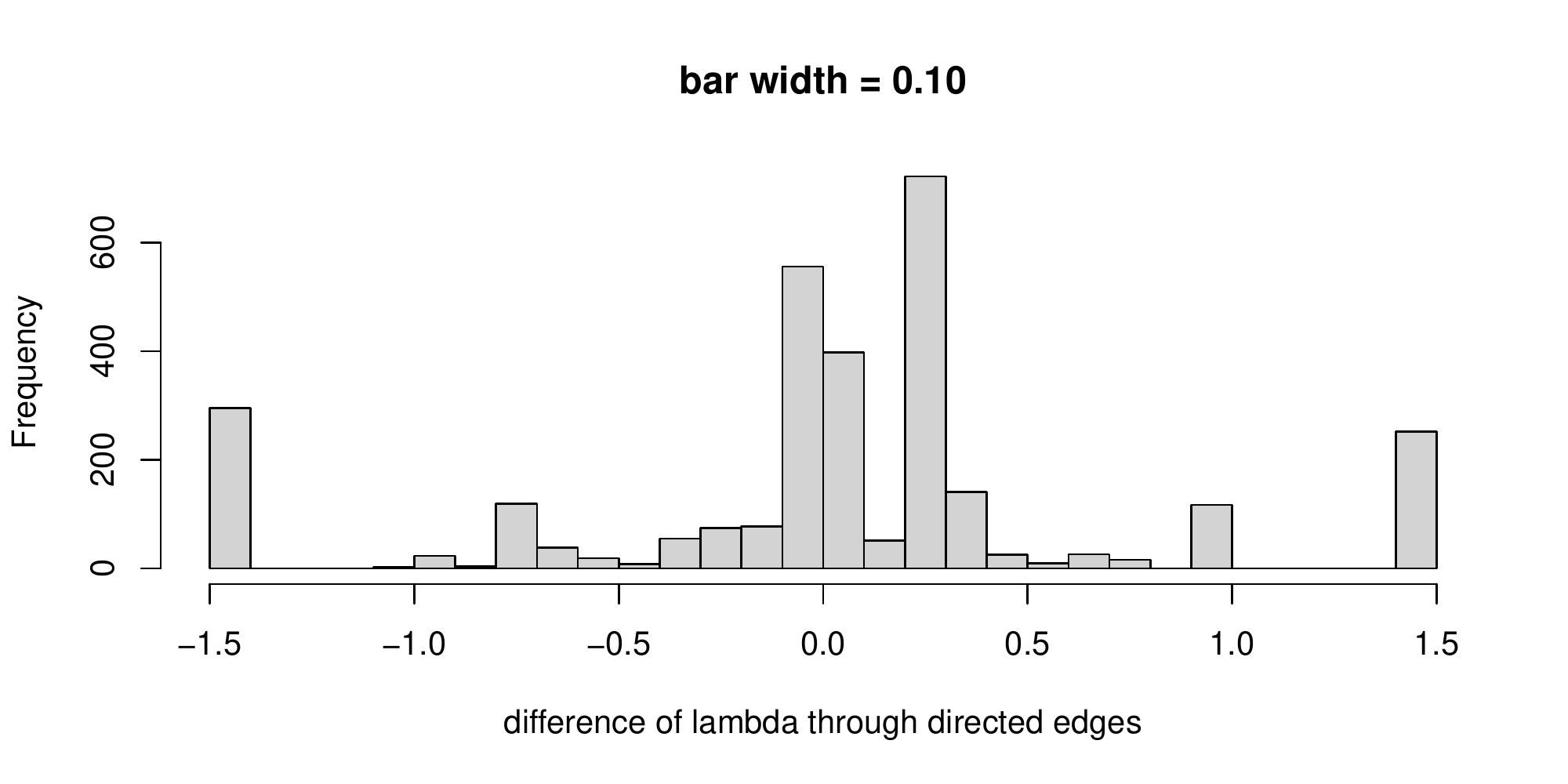}
    \caption{\textbf{Distribution of differences in adaptation rates through directed edges in the Reactome pathway, with bar width equal to 0.1.} The distribution is bimodal, with a peak centered around 0 and the other peak centered around 0.25.}
    \label{Figure.13}
\end{figure}

In addition to the GO category analysis, we also mapped our transition rate results onto existing Reactome pathways (retrieved on Jan 11 2022) \cite{gillespie2022reactome} to look for rapidly changing genes that may drive more slowly changing genes following the temperature reduction. For the directed edges connecting two genes that are concordant and pass the reproducibility criteria, the distribution of adaptation rate differences between them (terminal vertex rate minus starting vertex rate) are determined [Figure~\ref{Figure.13}, SI Figure~\ref{SIFigure.3}] (\textit{binding} relationships are excluded, as protein bindings themselves do not imply any relationships between mRNA adaptation rates).

The distribution of adaptation rate differences through these edges are bimodal. One peak is around 0 and the other is around 0.25, and they can be clearly distinguished regardless of which bar width is used for the distribution histogram. These two peaks suggest there are two scenarios of temperature-mediated control on adaptation rates: 1) if the the upstream gene is acting as the rate-limiting step, then the downstream gene should have a similar rate as the upstream gene, or 2) if the the downstream gene is acting as the rate-limiting step, then the downstream gene should have a slower rate than the upstream gene.

\begin{figure}[H]
    \centering
    \subfigure[Simulated data from existing model.]{
    \includegraphics[scale=0.4]{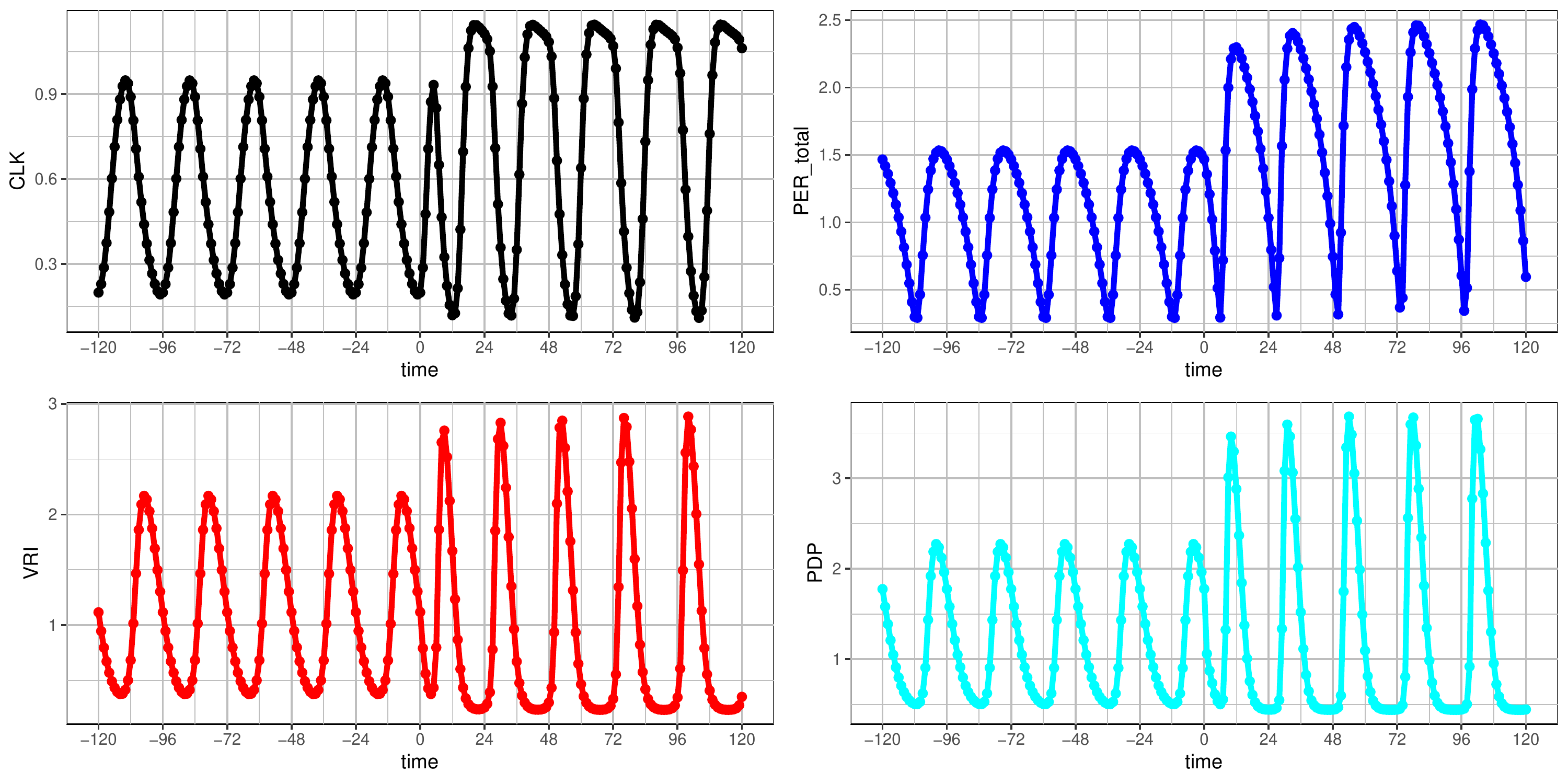}
    }
    \subfigure[Distribution of $\lambda$s.]{
    \includegraphics[scale=0.37]{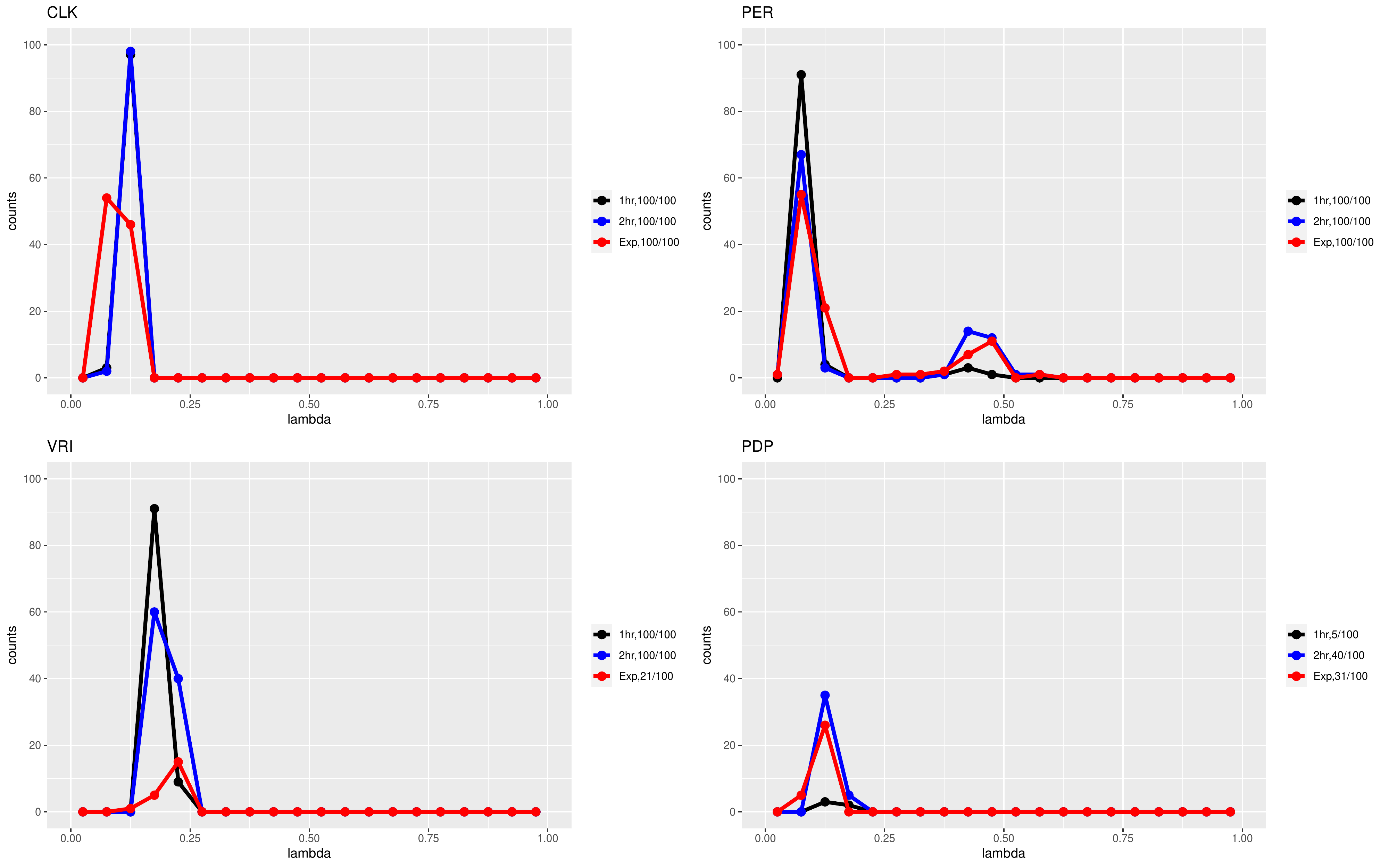}
    }
    \caption{\textbf{Simulated data and distribution of estimated $\lambda$ for all concordant trials.} Parameters change instantaneously to new values in response to temperature changes at t=0, except $v_{pyct}$ approaches its new value exponentially at rate $\lambda=0.05$. Black: sample every 1hr; blue: sample every 2hr; red: our sampling scheme.}
    \label{Figure.5}
\end{figure}

To illustrate potential temperature-mediated control mechanisms, we selected a directed subgraph including only: 1) genes (vertices) that have concordant estimates of types from the significance test and model selection, and also pass the reproducibility criteria, including genes with $\lambda=\infty$ by inference; 2) interactions (edges) that connect two concordant genes, with differences in adaptation rates larger than -0.1 but excluding \textit{binding} relationships. The tolerance of -0.1 is chosen to allow for some variation in the rate estimation, since the peak in the distribution of rate differences roughly spans [-0.1,0.1] [SI Figure~\ref{SIFigure.3}] and is consistent with the distribution of estimated rates from simulation results [Figure~\ref{Figure.5}].

\begin{figure}[H]
    \centering
    \includegraphics[scale=0.35]{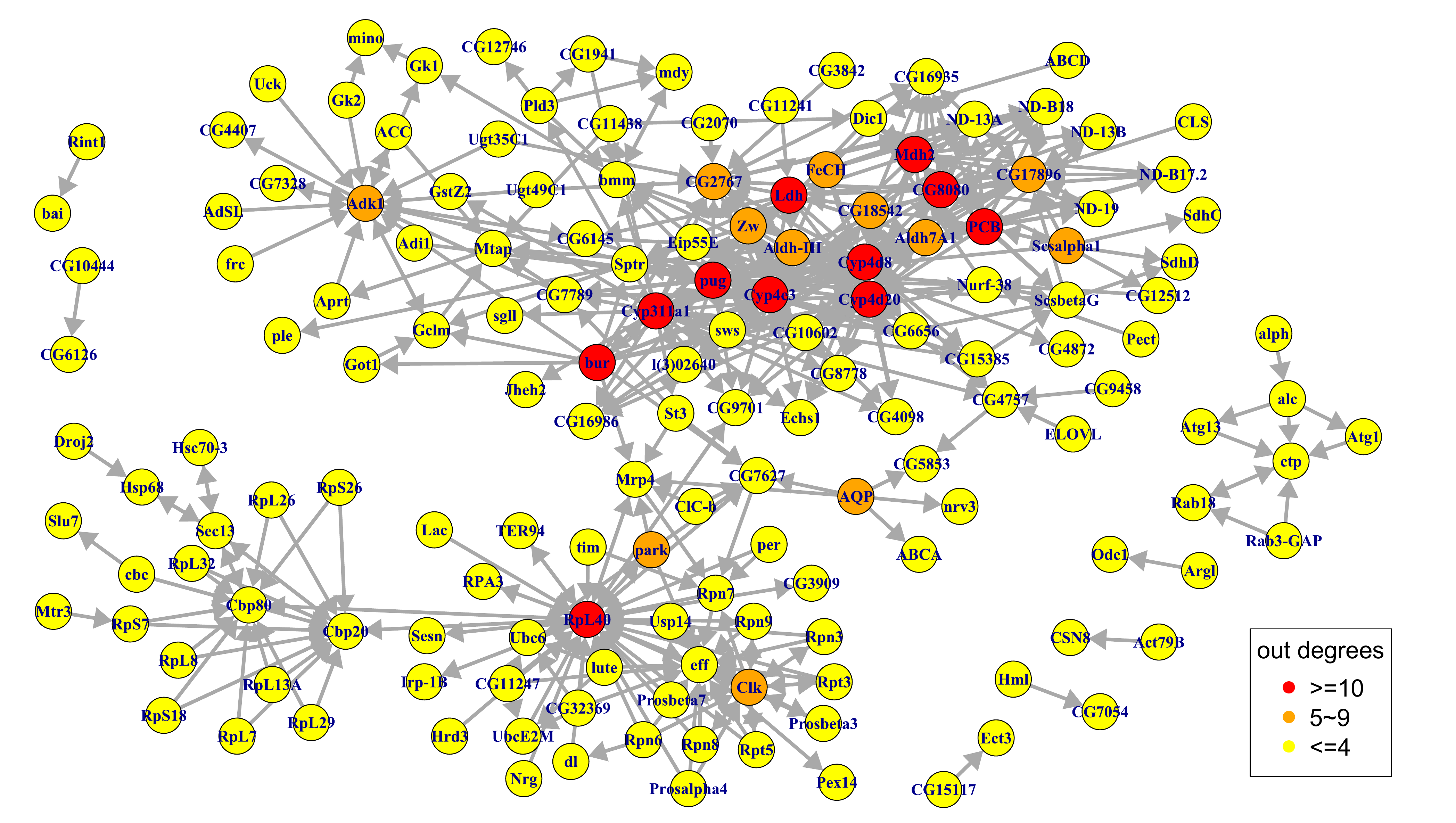}
    \caption{\textbf{Overview of subgraph of Reactome pathways showing possible control of slowly changing genes by rapidly changing genes.} Color represents out degree (red: $\geq$10, orange: 5$\sim$9, yellow: $\leq$4). The largest component in this network contains $88\%$ of the vertices and $96\%$ of the edges.}
    \label{Figure.14}
\end{figure}

The resulting subgraph [Figure~\ref{Figure.14}] has 158 vertices and 448 edges. Core clock genes \textit{Clk}, \textit{per} and \textit{tim} are present in this network, while \textit{vri} is not. A single large connected component forms the majority of this graph, containing 139 vertices and 432 edges. Various metabolic and biosynthetic processes are enriched in this network. These include proteasome-mediated ubiquitin-dependent protein catabolic process, fold enrichment=5.91, FDR=$2.86\times10^{-5}$; cytoplasmic translation, fold enrichment=7.41, FDR=$3.43\times10^{-4}$.

\begin{figure}[H]
    \centering
    \includegraphics[scale=0.48]{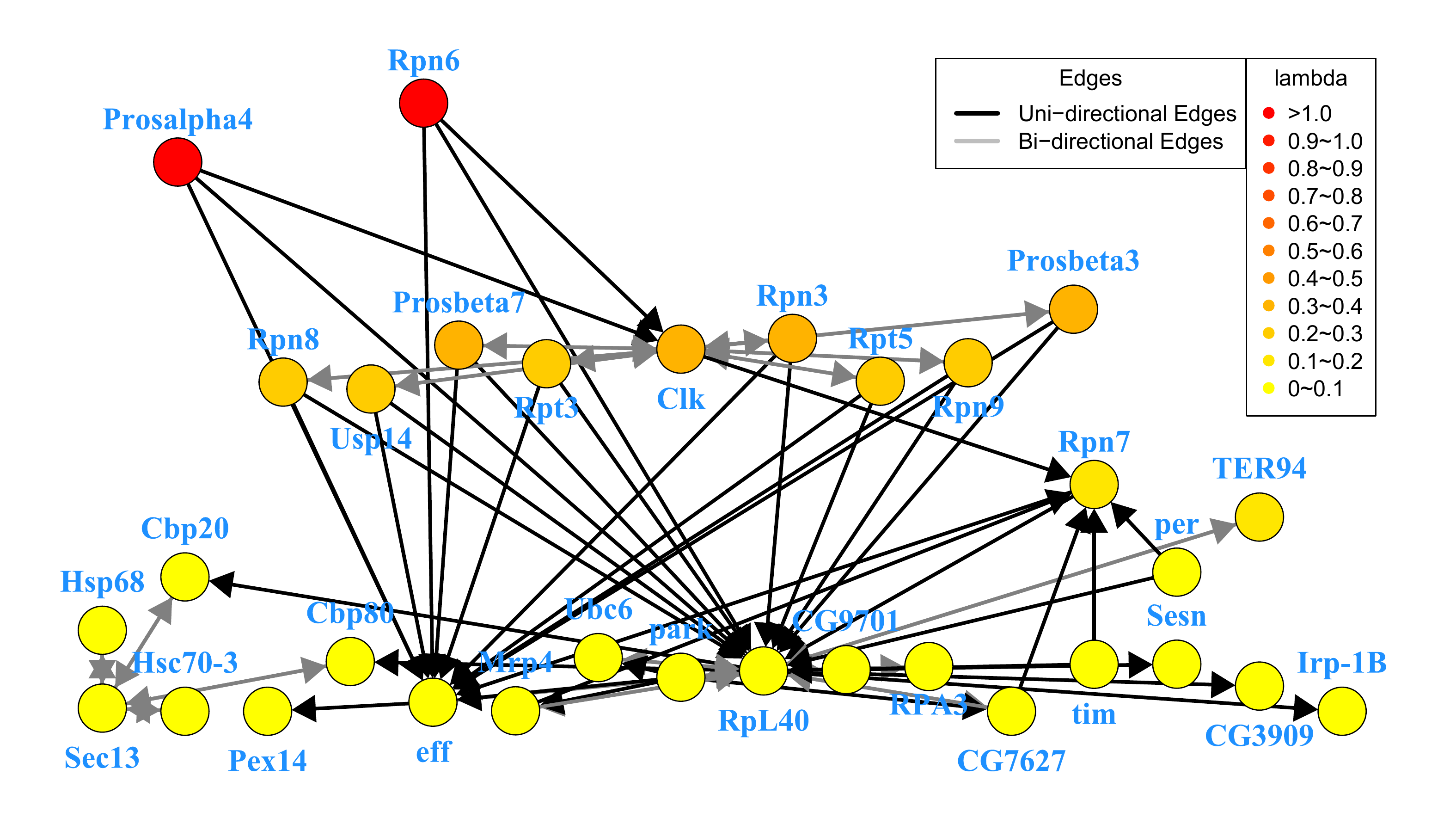}
    \caption{\textbf{Subgraph including genes connected to core clock gene \textit{Clk}, \textit{per} and \textit{tim}, colored by adaptation rates $\lambda$.} Edges are colored differently to indicate whether they are uni-directional or bi-directional. Height also represents log adaptation rates. Red: large lambda; yellow: small lambda.}
    \label{Figure.15}
\end{figure}

From this network we can extract a subgraph that is connected (either directly or indirectly) to core clock genes \textit{Clk}, \textit{per} and \textit{tim} [Figure~\ref{Figure.15}] illustrating the controls involving the core clock genes. Similar metabolic processes are also enriched in this subgraph, including proteasome-mediated ubiquitin-dependent protein catabolic process, fold enrichment=29.18, FDR=$9.82\times10^{-16}$. All GO enrichment analysis was performed on GeneOntology (http://geneontology.org/) via the PANTHER Overrepresentation Test (Released 2022/10/13).

In terms of potential regulation of temperature-mediated responses, it appears that \textit{Clk} may have a more prominent effect on controlling downstream genes than \textit{per} and \textit{tim} since it has more outgoing edges directly connecting to various proteasomes. Among all the proteasomes that are connected to \textit{Clk}, it is clear that \textit{Rpn6} and \textit{Prosalpha4} have much faster rates than \textit{Clk}, and \textit{Rpn7} has a much smaller rate than \textit{Clk}. However, since the rest of proteasomes (\textit{Prosbeta3}, \textit{Rpn3}, \textit{Prosbeta7}, \textit{Rpn9}, \textit{Rpt3}, \textit{Rpt5}, \textit{Rpn8}, \textit{Usp14}) are having very close adaptation rates comparing to \textit{Clk} and their edges are bi-directional, it is hard to distinguish whether they or \textit{Clk} should be considered to be upstream and thus might be the rate-limiting step for the temperature-mediated control. 

\begin{figure}[H]
    \centering
    \includegraphics[scale=0.48]{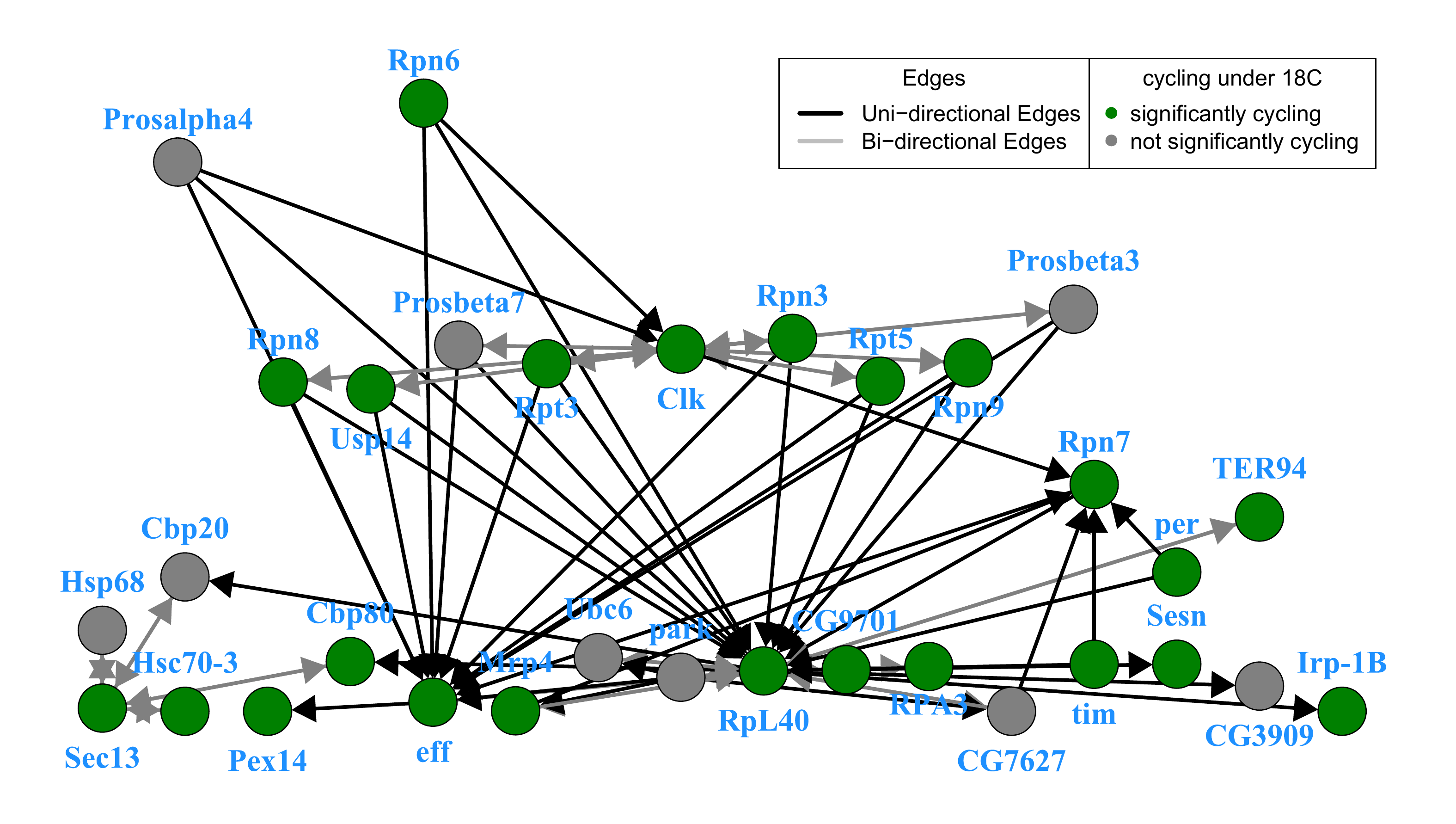}
    \caption{\textbf{Subgraph including genes connected to core clock gene \textit{Clk}, \textit{per} and \textit{tim}, colored by cycling status under 18\degree{}C.} Edges are colored differently to indicate whether they are uni-directional or bi-directional. Height also represents log adaptation rates. Green: significantly cycling under 18\degree{}C; grey: not significantly cycling under 18\degree{}C.}
    \label{Figure.16}
\end{figure}

\begin{figure}[H]
    \centering
    \includegraphics[scale=0.48]{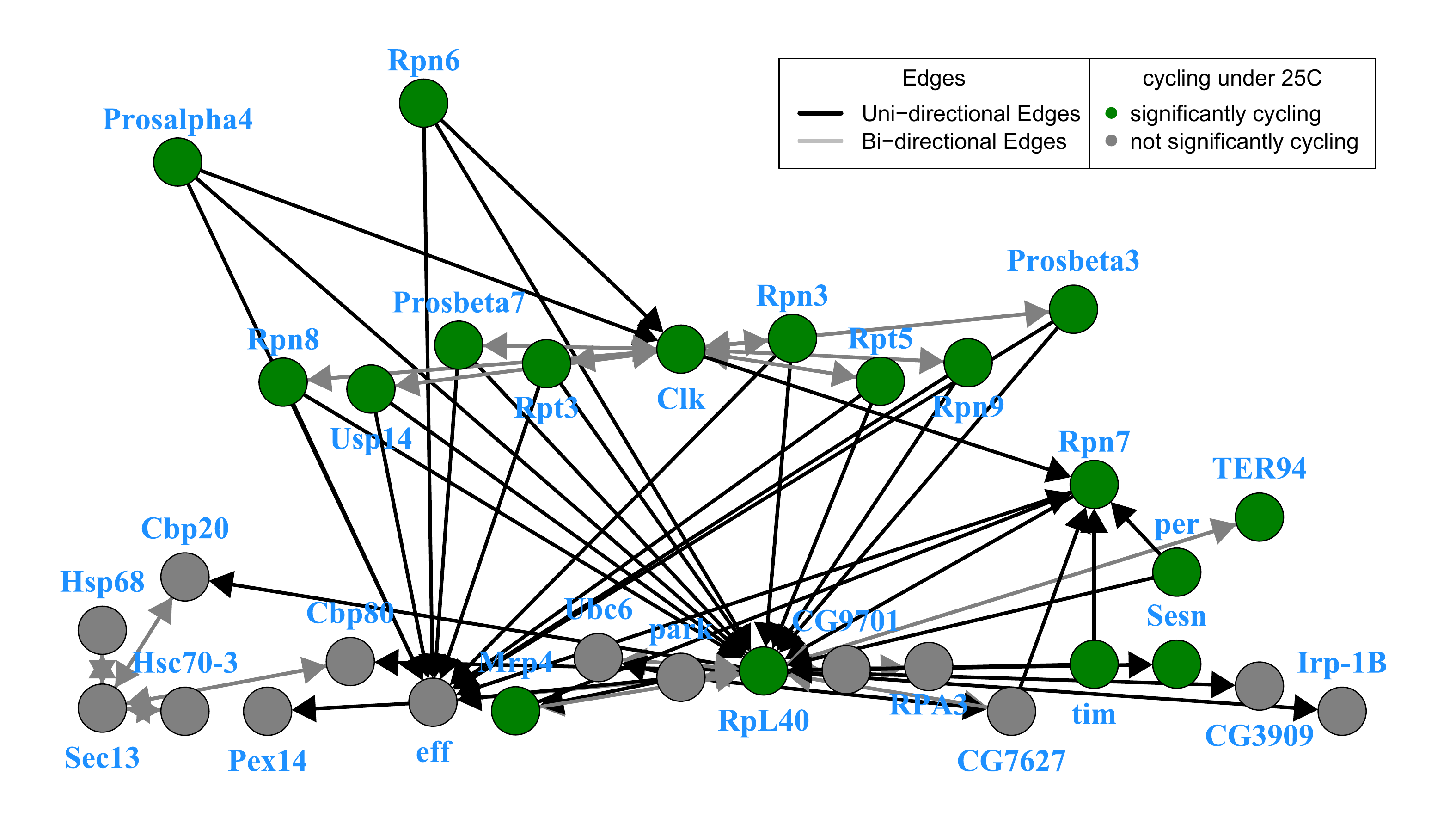}
    \caption{\textbf{Subgraph including genes connected to core clock gene \textit{Clk}, \textit{per} and \textit{tim}, colored by cycling status under 25\degree{}C.} Edges are colored differently to indicate whether they are uni-directional or bi-directional. Height also represents log adaptation rates. Green: significantly cycling under 25\degree{}C; grey: not significantly cycling under 25\degree{}C.}
    \label{Figure.17}
\end{figure}

\begin{figure}[H]
    \centering
    \includegraphics[scale=0.6]{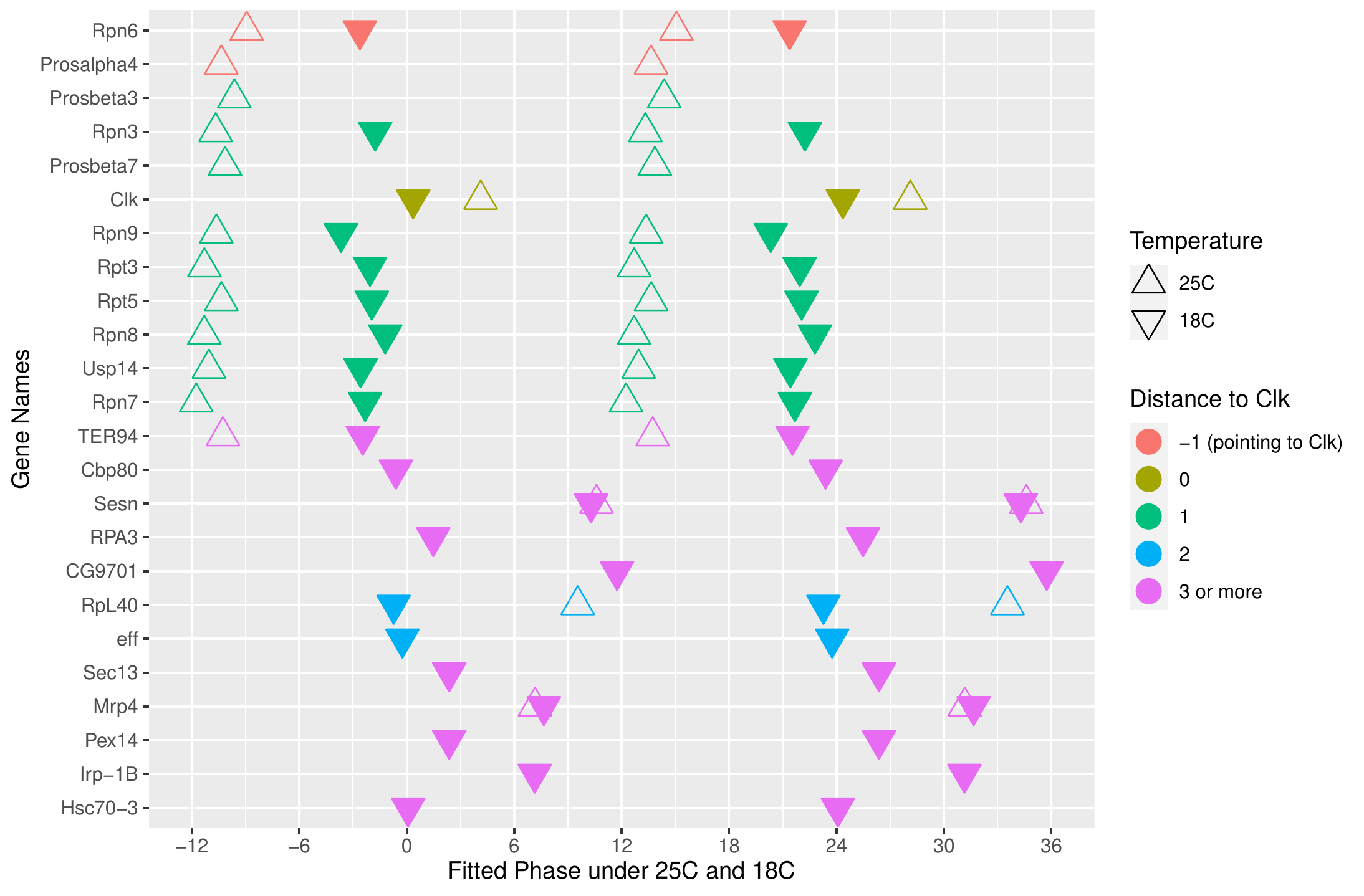}
    \caption{\textbf{Oscillation phase of genes that 
    are significantly cycling under either 18\degree{}C or 25\degree{}C and are connected to \textit{Clk}.} Genes are ordered by their adaptation rates, and colored by their distances to \textit{Clk}. A gene having distance of -1 means that it is directly pointing to \textit{Clk}. The phases are double-plotted for better visualization.}
    \label{Figure.18}
\end{figure}

\begin{figure}[H]
    \centering
    \includegraphics[scale=0.5]{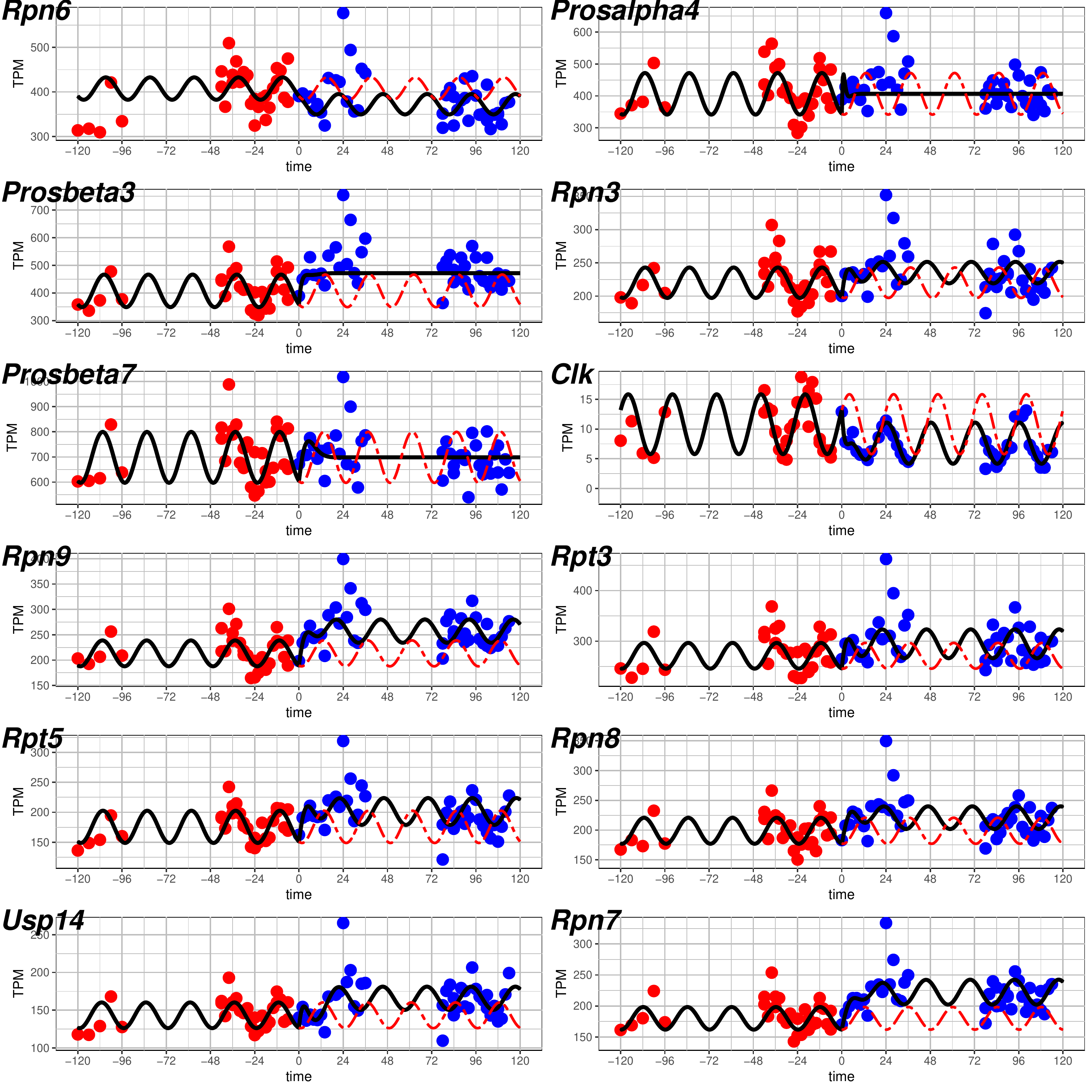}
    \caption{\textbf{Expression profiles and the best model fits for genes that are directly connected to \textit{Clk}.} Genes are ordered by their adaptation rates. Black curve is the fitted best model curve with minimum AICc, and red dashed line is the extension of fitted curve under 25\degree{}C.}
    \label{Figure.19}
\end{figure}

Interestingly, these proteasomes show significant cycling behavior under both temperatures [Figure~\ref{Figure.16}, Figure~\ref{Figure.17}], and their phases are very close to each other under both temperatures. Under 25\degree{}C \textit{Clk} is out of phase with these proteasomes, while after the transient temperature reduction to 18\degree{}C, \textit{Clk} re-synchronizes with them to have very similar oscillation phases [Figure~\ref{Figure.18}, Figure~\ref{Figure.19}]. This change of phase is observed only for these proteasomes directly connected to \textit{Clk} and for the downstream connection \textit{RpL40} (Ribosomal protein L40), and not for other genes having much larger distances to \textit{Clk}.

\newpage
\section{Discussion}
We report the first study of transient gene expression dynamics for genes under circadian control subject to a temperature step, along with a generalizable, nonparametric method to estimate the adaptation rates based on a theoretical underpinning from Floquet theory. In this study we propose an exponentially-adapting model to describe the dynamics of genes subject to periodic forcing that undergo a change in their environment and estimate their adaptation rates. To accommodate the non-\textit{i.i.d.} structure of noise and existence of outliers, we use quantile regression to fit the conditional median curve, and construct our significance test using the wild bootstrap. To find accurate and reliable estimates for adaptation rates $\lambda$, we use AICc to perform model selection and estimate adaptation rates from the minimal model only for genes that have reproducibly detected characteristics from the above two tests. 

While a previous study \cite{boothroyd2007integration} reported transcriptomic changes during a temperature thermocycle (step changes in temperature every 12hrs), the experimental designs are far from sufficient to investigate temperature adaptation and construct networks for temperature regulations. First, it usually takes longer time than 12hrs for genes to adapt to a new temperature (e.g. for \textit{per} it takes $1/\lambda\approx13.7$hrs to finish about 63\% of the adaptation). Second, dense sampling both during and after the adaptation process are needed to estimated the adaptation rates. Because those previous experiments were not designed to model adaptation rates, their sampling frequency (once every four hours) is too low to determine adaptation rates or assess early drivers of change. 

We note that the proposed method is very general, and can be applied to study data from any periodically forced dynamical system undergoing a perturbation. In a biological context, this could be used to study not only gene expression but isoforms, proteins, etc., following any environmental change (temperature, diet, etc.). While our focus here was on circadian rhythms, the method could be easily applied to study monthly or annual physiological cycles. We also envision that it may find utility outside the life sciences, e.g. for the purposes of modeling perturbations to sector rotation in economic data. 

\begin{figure}[H]
    \centering
    \subfigure[Experimental Design]{
    \includegraphics[scale=0.3]{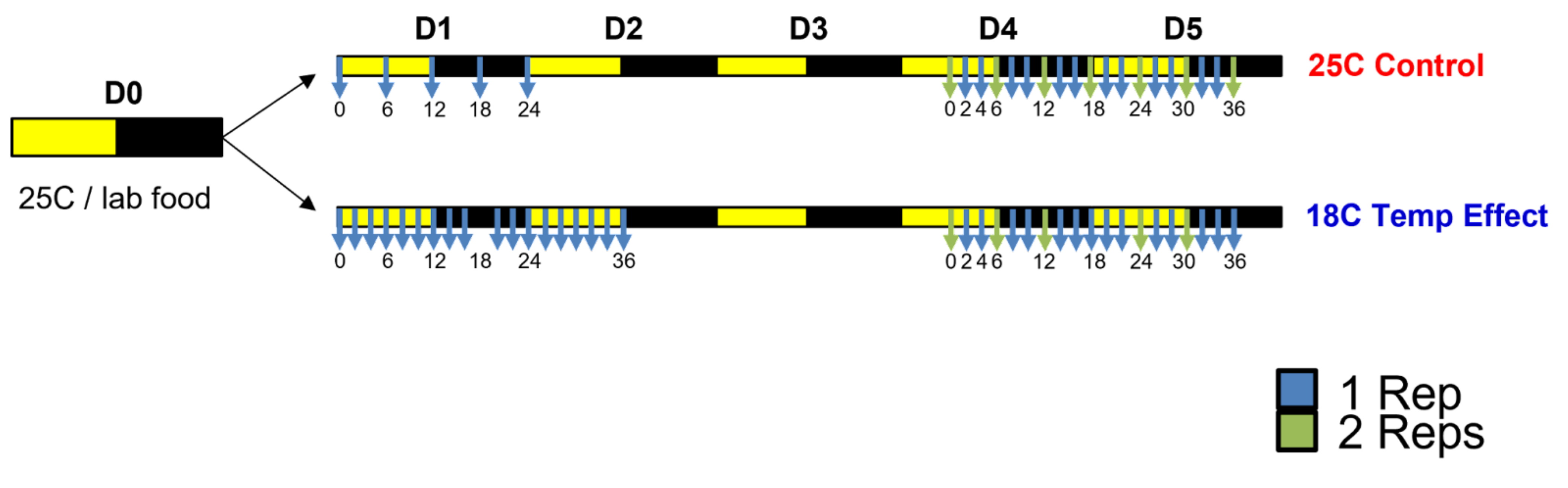}
    }
    \subfigure[Method Workflow]{
    \includegraphics[scale=0.3]{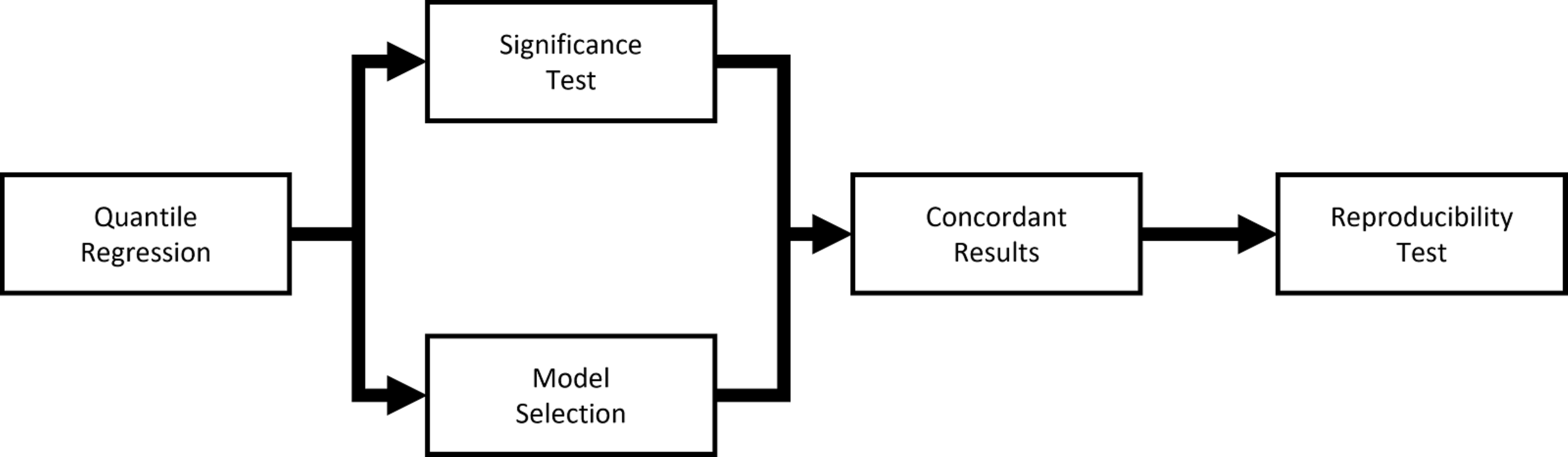}
    }
    \caption{\textbf{Design of the transient study and method flowchart.} (a). Sample data are collected on Day1/2 and Day4/5 under each temperature at illustrated time points, with a second replicate collected at intermittent time points. The first replicate at ZT18 in 18\degree{}C Day1, the second replicate at ZT24 in 18\degree{}C Day4 and the second replicate at ZT18 in 18\degree{}C Day5 do not pass quality control and are excluded from all subsequent analyses. (b). Nonlinear quantile regression is used to fit the full model curve. Significance test and model selection are performed in parallel, and we only estimate adaptation rates for genes with concordant results from the above two methods. Additional reproducibility testing using a previously collected dataset, which filters out any differentially-cycling genes between two data sets under either temperature, is performed for more robust inference on the dynamics of gene expressions.}
    \label{Figure.1}
\end{figure}

Simulation studies show that by combining a significance test and model selection we can reduce the chance of inferring incorrect characteristics and accurately estimate adaptation rates for each gene, at the cost of not being able to give estimates for the remaining non-concordant genes.   This could be because the noise level impairs the interpretability of the characteristics tested by our method, or the transient dynamics during the adaptation stage is not resolved sufficiently with our sampling scheme.  A limitation of the experimental data presented in this paper is the gap in sampling between day 2 and day 4 [Figure~\ref{Figure.1}], with no observations beyond day 5. To ensure interpretable results, we place a lower bound $\lambda=0.03$ on $\lambda$, such that all exponential terms would at least decrease to $e^{-0.03*96}\approx5.6\%$ of their initial values, consistent to our hypothesis that genes would reach steady limit cycles at Day5. To make improvements on the experiment design, simulations from an existing model reveals that sampling every 1hr with 2 replicates that covers the whole 5 days would be a good choice for future experiments. 

\begin{figure}[H]
    \centering
    \subfigure[]{
    \includegraphics[scale=0.5]{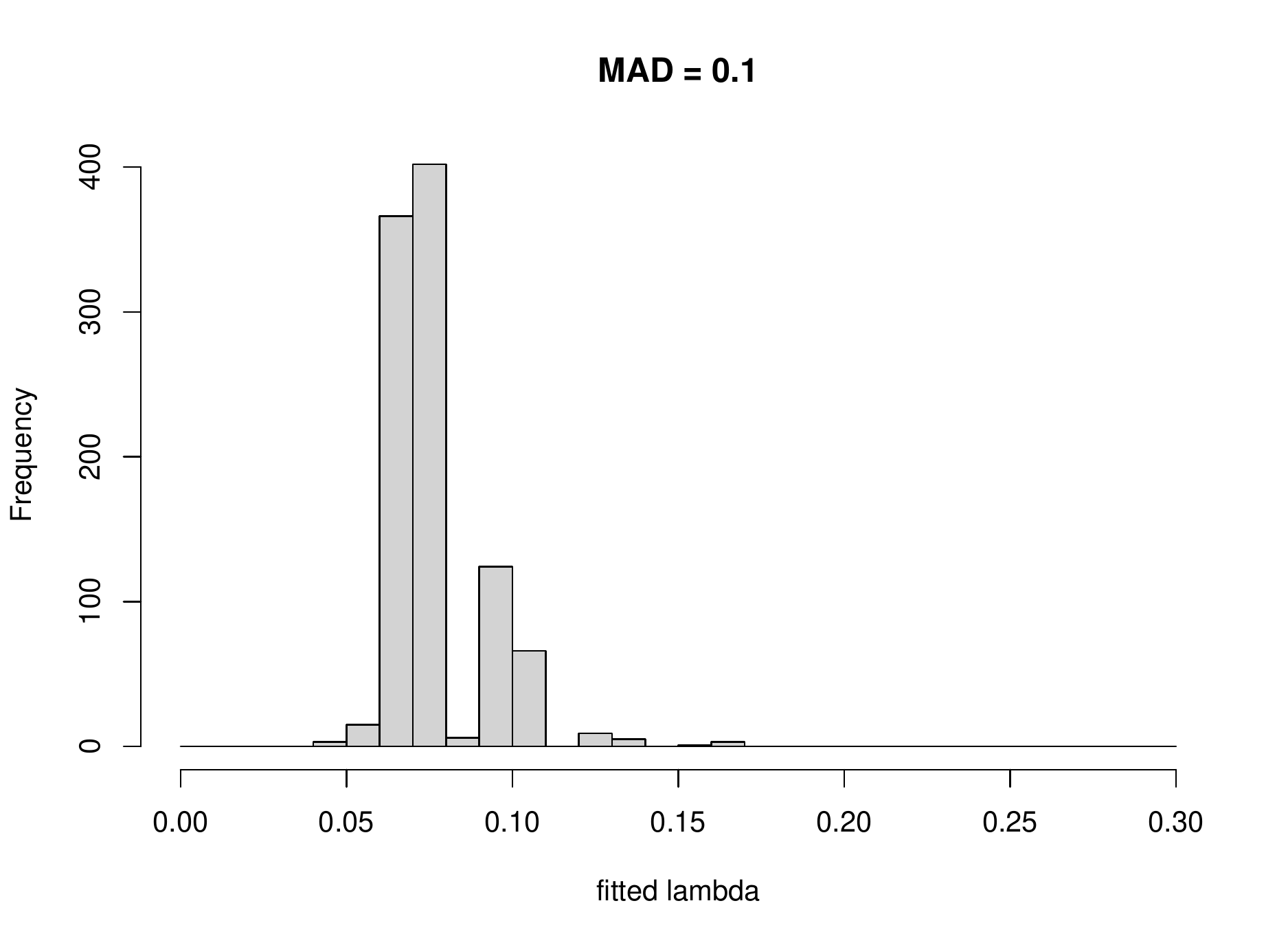}
    }
    \subfigure[]{
    \includegraphics[scale=0.5]{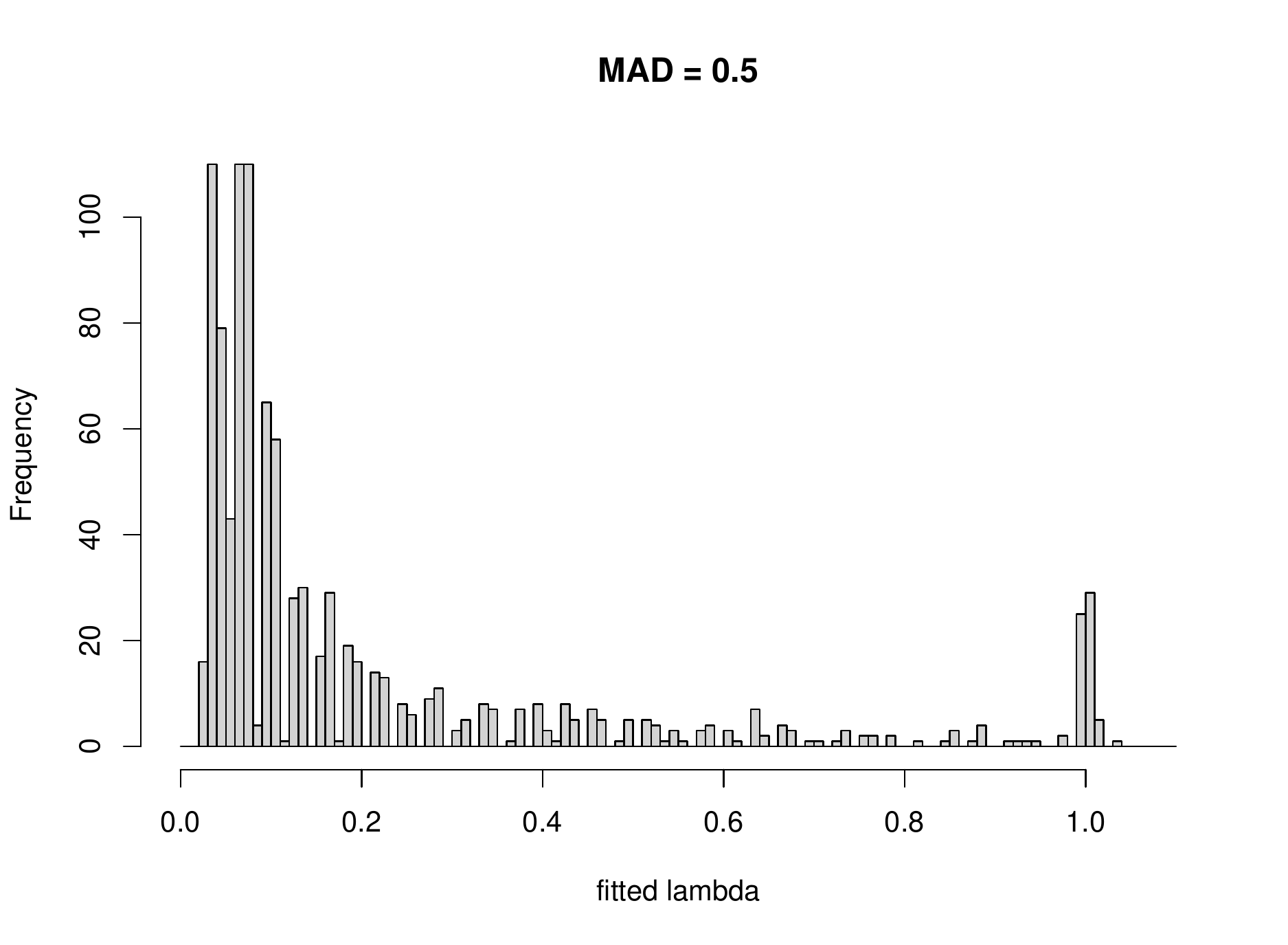}
    }
    \caption{\textbf{Simple simulation when the null model includes two different exponents.} The null model is $f(t;\lambda_1,\lambda_2)=3$ for $t<0$, and $f(t;\lambda_1,\lambda_2)=1+e^{-\lambda_1 t}+e^{-\lambda_2 t}$ for $t\geq0$. Synthetic data are made by drawing residuals from a Laplace distribution with $\mu=0$ and different $b$. $\lambda_1=0.4$, $\lambda_2=0.04$. The two plots illustrate the distribution of fitted $\lambda$ in the context of small and large noise. (a). $b=10$. MAD$=\frac{1}{b}=0.1$. (b). $b=2$. MAD$=\frac{1}{b}=0.5$.}
    \label{Figure.11}
\end{figure}

One can also expand our model to include additional rates for the purpose of their own studies, since Floquet theory shows the dynamics is not restricted to a single real Floquet exponent $\lambda$. Simple simulation [Figure~\ref{Figure.11}] shows that when the null model has two different exponents, our model with one exponent tends to fit a rate $\lambda$ that is in between the two but much closer to the slow rate.  This suggests that only including one rate in our model is reasonable, because we are looking for possible control pathways for temperature regulation and it is the slow exponent from the upstream gene that would control the slow exponents in the downstream gene if such control exists. 

The experimental data validate our assumption that residuals may have different variances (or MADs) given changes in external environments. Therefore, we believe it is important to take residual heteroskedasticity into consideration when researchers use parametric models to conduct studies across different conditions (e.g., different temperatures, sexes, light settings). While in this study we do not assume any relationships between residual MAD and gene expressions, we do observe moderate correlations between residual MAD and fitted median gene expression level and fitted oscillation amplitude. Thus in future studies with environmental changes it may be reasonable to assume the residuals have such properties (e.g., residuals increase with average expression level). 

Our results point to the importance of \textit{Clk} in adaptation to lower temperatures. In particular, the observation that many genes shift to an earlier phase in 18\degree{}C appears to be driven in large part by an early shift in \textit{Clk}. \textit{Clk} has a more rapid adaptation rate than \textit{vri}, \textit{per} and \textit{tim}, and it possibly regulates more genes than \textit{per} and \textit{tim} in terms of temperature-mediated responses. In the circadian GO category, there are no concordant genes that have a more rapid adaptation rate than \textit{Clk}, which suggests that the upstream regulators of \textit{Clk} may not be only responsible for the circadian clock. These regulators may be involved in general processes, for example, transcription and degradation processes. We hypothesize that proteasomes may play an important role in temperature adaptation of the circadian system. Several proteasomes can regulate \textit{Clk} and subsequently regulate one ribosomal protein \textit{RpL40}. To confirm these regulations in temperature adaptation, future studies could perform knockouts of these proteasomes that either regulates or be regulated by \textit{Clk} under the same experimental setting as this temperature perturbation study, and investigate any changes in dynamics of gene expressions. 

\newpage
\section{Acknowledgement}
This work was supported by NSF grant DMS-1764421 and Simons Foundation grant 597491.

\newpage

\bibliographystyle{unsrt} 
\bibliography{source.bib} 

\newpage
\section{Tables and Figures}
\subsection{Table.1}
\begin{table}[ht]
    \centering
    \begin{tblr}{|c|c|c|}
        \hline
        Characteristic & High Temperature & Low Temperature \\
        \hline\hline
        Median Baseline Expression & $A_1$ & $A_1-A_4-A_6-A_7$ \\
        \hline
        Oscillation Pattern & $(A_2,A_3)$ & $(A_2+A_7,A_3+A_8)$ \\
        \hline
        Oscillation Amplitude & $\sqrt{\strut A_2^2+A_3^2}$ &$\sqrt{\strut (A_2+A_7)^2+(A_3+A_8)^2}$ \\
        \hline
        Oscillation Phase & $\arctan({A_3}/{A_2})$ & $\arctan({[A_3+A_8]}/{[A_2+A_7]})$\\
        \hline
    \end{tblr}
    \caption{\textbf{List of Tested Characteristics in mRNA expression.} Differences in these characteristics between the two temperatures is also tested, using the difference between corresponding terms.}
    \label{Table.1}
\end{table}

\newpage
\subsection{Table.2}
\begin{table}[ht!]
    \centering
    \begin{tblr}{|c|c|c|c|c|c|}
        \hline\hline
        Model Type & 25CYC & 18CYC & DIFF CYC & DIFF EXP & Theoretically Valid\\
        \hline\hline
        \cellcolor[HTML]{332288}Type 01 & - & - & - & - & Valid\\
        \hline
        \cellcolor[HTML]{88CCEE}Type 02 & - & - & - & Significant & Valid\\
        \hline
        Type 03 & - & - & Significant & - & -\\
        \hline
        Type 04 & - & - & Significant & Significant & -\\
        \hline
        Type 05 & - & Significant & - & - & -\\
        \hline
        Type 06 & - & Significant & - & Significant & -\\
        \hline
        \cellcolor[HTML]{44AA99}Type 07 & - & Significant & Significant & - & Valid\\
        \hline
        \cellcolor[HTML]{117733}Type 08 & - & Significant & Significant & Significant & Valid\\
        \hline
        Type 09 & Significant & - & - & - & -\\
        \hline
        Type 10 & Significant & - & - & Significant & -\\
        \hline
        \cellcolor[HTML]{999933}Type 11 & Significant & - & Significant & - & Valid\\
        \hline
        \cellcolor[HTML]{DDCC77}Type 12 & Significant & - & Significant & Significant & Valid\\
        \hline
        \cellcolor[HTML]{CC6677}Type 13 & Significant & Significant & - & - & Valid\\
        \hline
        \cellcolor[HTML]{882255}Type 14 & Significant & Significant & - & Significant & Valid\\
        \hline
        \cellcolor[HTML]{AA4499}Type 15 & Significant & Significant & Significant & - & Valid\\
        \hline
        \cellcolor[HTML]{DDDDDD}Type 16 & Significant & Significant & Significant & Significant & Valid\\
        \hline\hline
    \end{tblr}
    \includegraphics[scale=0.6]{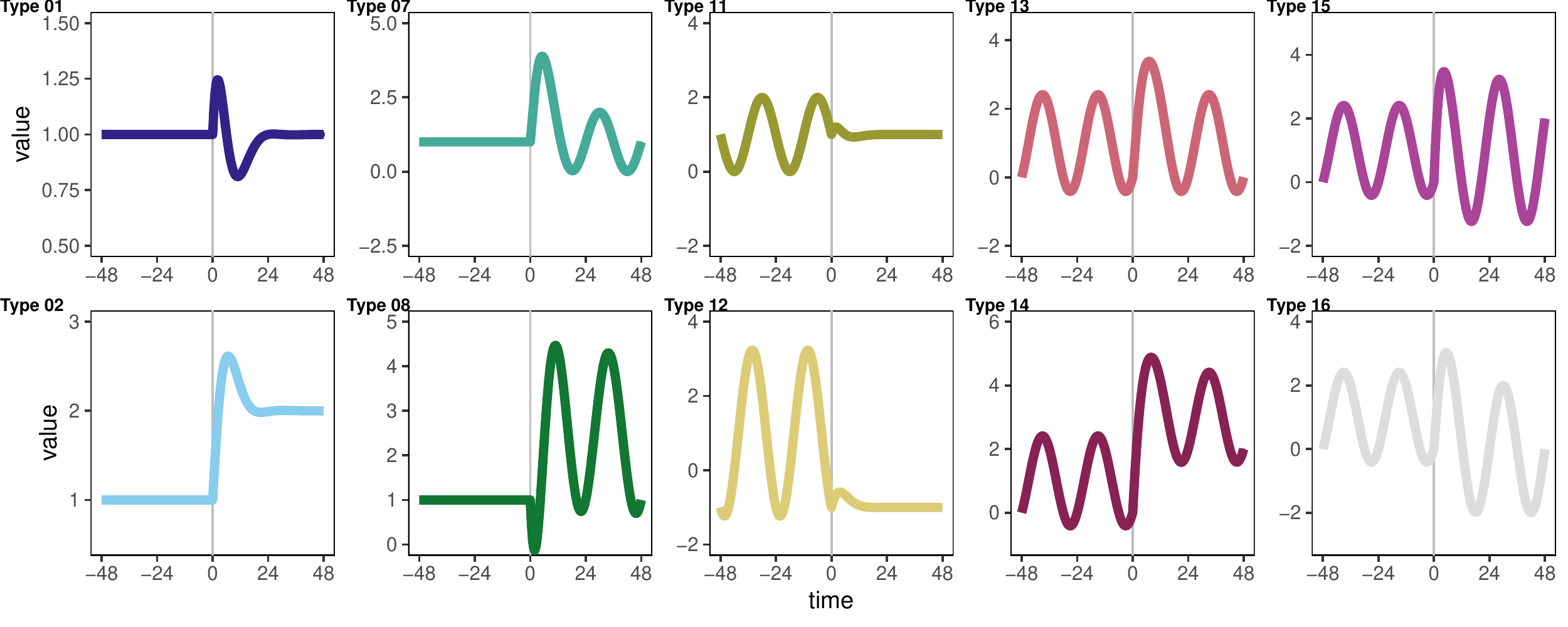}
    \caption{\textbf{Classification of all gene expression dynamics into 16 types depending on 4 features.} 10 out of 16 types are theoretically valid, and a sketch of each type (with transients) is shown here. 25CYC: cycling at high temperature; 18CYC: cycling at low temperature; DIFF CYC: different oscillation patterns across different temperatures (differentially cycling); DIFF EXP: different median baseline expression across different temperatures (differentially expressed).}
    \label{Table.2}
\end{table}

\newpage
\subsection{Table.3}
\begin{table}[!htbp]
    \centering
    \begin{tblr}{|c|c|c|c|c|c|}
    \hline\hline
	Gene ID & Gene Name & Model & Transition Rate & $\Delta AICc$ & $2^{nd}$ Best Model\\ \hline\hline
	FBgn0023076 & Clk & 16F & 0.308 & 11.98 & 16P\\ \hline
	FBgn0016076 & vri & 16F & 0.306 & 10.55 & 15F\\ \hline
	FBgn0003068 & per & 16F & 0.073 & 14.91 & 14F\\ \hline
	FBgn0014396 & tim & 16F & 0.041 & 9.237 & 15F\\ \hline\hline
    \end{tblr}
    \caption{\textbf{Concordant genes with larger transition rates than \textit{tim} in GO category Circadian rhythm (GO:0007623).} Model represents the candidate model with the minimum AICc, as well as the classified type from significance test since these genes are concordant. Adaptation rates are rounded to three decimal places. There is no concordant gene with $\lambda=\infty$ by inference in this GO category.}
    \label{Table.3}
\end{table}

\newpage
\subsection{Table.4}
\begin{table}[htbp]
    \centering
    \begin{tabular}{|c|c|c|c|c|c|}
    \hline\hline
	Gene ID & Gene Name & Model & Transition Rate & $\Delta AICc$ & $2^{nd}$ Best Model\\ \hline\hline
	FBgn0000229 & bsk & 02F & 0.492 & 2.849 & 12F\\ \hline
	FBgn0023076 & Clk & 16F & 0.308 & 11.98 & 16P\\ \hline
	FBgn0035110 & thoc7 & 08F & 0.190 & 2.755 & 07F \\ \hline
	FBgn0005626 & ple & 12F & 0.130 & 4.360 & 11F\\ \hline
	FBgn0038145 & Droj2 & 14F & 0.130 & 3.796 & 08F\\ \hline
	FBgn0003068 & per & 16F & 0.073 & 14.91 & 14F\\ \hline\hline
    \end{tabular}
    \caption{\textbf{Concordant genes with larger transition rates than \textit{tim} in GO category Response to Temperature Stimulus (GO:0009266).} Model represents the candidate model with the minimum AICc, as well as the classified type from significance test since these genes are concordant. Transition rates are rounded to three decimal places. There is no concordant gene with $\lambda=\infty$ by inference in this GO category.}
    \label{Table.4}
\end{table}

\newpage
\subsection{Table.5}
\begin{table}[htbp]
    \centering
    \begin{tabular}{|c|c|c|c|c|c|}
    \hline\hline
	Gene ID & Gene Name & Model & Transition Rate & $\Delta AICc$ & $2^{nd}$ Best Model\\ \hline\hline
	FBgn0036318	& Wbp2 & 02P & $\infty$ & 2.656 & 12P\\ \hline
    FBgn0260632	& dl & 02P & $\infty$ & 2.169 & 02F\\ \hline
	FBgn0013263 & Trl & 08F & 0.579 & 4.211 & 16F\\ \hline
	FBgn0000520 & dwg & 07F & 0.426 & 2.113 & 01F\\ \hline
	FBgn0037120	& CG11247 & 08F & 0.366 & 2.520 & 16F\\ \hline
	FBgn0023076 & Clk & 16F & 0.308 & 11.98 & 16P\\ \hline
	FBgn0016076 & vri & 16F & 0.306 & 10.55 & 15F\\ \hline
	FBgn0003512 & Sry-delta & 02F & 0.277 & 2.004 & 08F\\ \hline
	FBgn0262656	& Myc & 14F & 0.252 & 2.718 & 08F\\ \hline
	FBgn0085432 & pan & 08F & 0.218 & 2.894 & 16F\\ \hline
	FBgn0260401	& MED9 & 02F & 0.135 & 2.726 & 08P\\ \hline
	FBgn0034534 & maf-S & 14F & 0.101 & 3.516 & 16F\\  \hline
	FBgn0036581 & MED10 & 02F & 0.100 & 2.014 & 14F\\ \hline
	FBgn0003068 & per & 16F & 0.073 & 14.91 & 14F\\ \hline
	FBgn0036804	& Sgf11 & 08F & 0.071 & 3.787 & 14F\\ \hline\hline
    \end{tabular}
    \caption{\textbf{Concordant genes with larger transition rates than \textit{tim} in GO category Transcription Regulator Activity (GO:0140110).} Model represents the candidate model with the minimum AICc, as well as the classified type from significance test since these genes are concordant. Transition rates are rounded to three decimal places. This table includes concordant genes with $\lambda=\infty$ by inference.}
    \label{Table.5}
\end{table}

\newpage
\subsection{Algorithm.1}
\RestyleAlgo{boxruled}
\LinesNumbered
\begin{algorithm}[ht]
  \caption{Model Fitting Procedure\label{Algorithm.1}}
  Choose a sequence of $\lambda_j$ from a coarse grid $\Lambda_1$. \\
  For every $\lambda_j$, fit the model with fixed $\lambda=\lambda_j$ using linear quantile regression. Record all parameter estimates $A_1^j,...,A_8^j$ and objective function value $Z_j$. \\
  Select index $j=\text{argmin}_j Z_j$. \\
  If $\lambda_k=1$, repeat step 2 and step 3 with a finer grid $\Lambda_2$. \\
  Randomly perturb $A_1^k,...,A_8^k,\lambda_k$ by $\pm 1\%$. Use them as initial guess for the L-BFGS-B method to fit the nonlinear model, with constraints on $\lambda \in \Lambda_3$.
\end{algorithm}

\newpage
\subsection{Figure.1}

\newpage
\subsection{Figure.2}

\newpage
\subsection{Figure.3}

\newpage
\subsection{Figure.4}

\newpage
\subsection{Figure.5}

\newpage
\subsection{Figure.6}

\newpage
\subsection{Figure.7}

\newpage
\subsection{Figure.8}

\newpage
\subsection{Figure.9}

\newpage
\subsection{Figure.10}

\newpage
\subsection{Figure.11}

\newpage
\subsection{Figure.12}

\newpage
\subsection{Figure.13}

\newpage
\subsection{Figure.14}

\newpage
\subsection{Figure.15}

\newpage
\subsection{Figure.16}

\newpage
\subsection{Figure.17}

\newpage
\subsection{Figure.18}

\newpage
\subsection{Figure.19}

\newpage
\newgeometry{left = 1.4cm, right = 1.4cm, top = 2.5cm, bottom = 3cm}
\subsection{Table.\protect\ref{SITable.1}}
\begin{table}[!h]
    \centering
    \begin{tabular}{|c|c|c|c|}
        \hline\hline
        Type & Features & FULL MODEL & PARTIAL MODEL \\
        \hline\hline
        \cellcolor[HTML]{332288}\textcolor{white}{Type 01} & -----,-----,-----,----- & $F(t_i;A_1,0,0,A_4,A_5,-A_4,0,0,\lambda)$ & $F(t_i;A_1,0,0,0,0,0,0,0,0)$ \\
        \hline
        \cellcolor[HTML]{88CCEE}Type 02 & -----,-----,-----,DE & $F(t_i;A_1,0,0,A_4,A_5,A_6,0,0,\lambda)$ & $F(t_i;A_1,0,0,0,0,0,0,0,0) + C_1\mathbbm{1}_{t \geq 0}$ \\
        \hline
        \cellcolor[HTML]{44AA99}Type 07 & -----,LC,DC,----- & $F(t_i;A_1,0,0,A_4,A_5,A_6,-A_4-A_6,A_8,\lambda)$ & $F(t_i;A_1,0,0,0,0,0,0,A_8,0)$ \\
        \hline
        \cellcolor[HTML]{117733}Type 08 & -----,LC,DC,DE & $F(t_i;A_1,0,0,A_4,A_5,A_6,A_7,A_8,\lambda)$ & $F(t_i;A_1,0,0,0,0,0,A_7,A_8,0)$ \\
        \hline
        \cellcolor[HTML]{999933}Type 11 & HC,-----,DC,----- & $F(t_i;A_1,A_2,A_3,A_4,A_5,A_2-A_4,-A_2,-A_3,\lambda)$ & $F(t_i;A_1,0,A_3,0,0,0,0,-A_3,0)$ \\
        \hline
        \cellcolor[HTML]{DDCC77}Type 12 & HC,-----,DC,DE & $F(t_i;A_1,A_2,A_3,A_4,A_5,A_6,-A_2,-A_3,\lambda)$ & $F(t_i;A_1,A_2,A_3,0,0,0,-A_2,-A_3,0)$ \\
        \hline
        \cellcolor[HTML]{CC6677}Type 13 & HC,LC,-----,----- & $F(t_i;A_1,A_2,A_3,A_4,A_5,-A_4,0,0,\lambda)$ & $F(t_i;A_1,A_2,A_3,0,0,0,0,0,0)$ \\
        \hline
        \cellcolor[HTML]{882255}Type 14 & HC,LC,-----,DE & $F(t_i;A_1,A_2,A_3,A_4,A_5,A_6,0,0,\lambda)$ & $F(t_i;A_1,A_2,A_3,0,0,0,0,0,0) + C_1\mathbbm{1}_{t \geq 0}$ \\
        \hline
        \cellcolor[HTML]{AA4499}Type 15 & HC,LC,DC,----- & $F(t_i;A_1,A_2,A_3,A_4,A_5,A_6,-A_4-A_6,A_8,\lambda)$ & $F(t_i;A_1,A_2,A_3,0,0,0,0,A_8,0)$ \\
        \hline
        \cellcolor[HTML]{DDDDDD}Type 16 & HC,LC,DC,DE & $F(t_i;A_1,A_2,A_3,A_4,A_5,A_6,A_7,A_8,\lambda)$ & $F(t_i;A_1,A_2,A_3,0,0,0,A_7,A_8,0)$ \\
        \hline\hline
    \end{tabular}
    \includegraphics[scale=0.6]{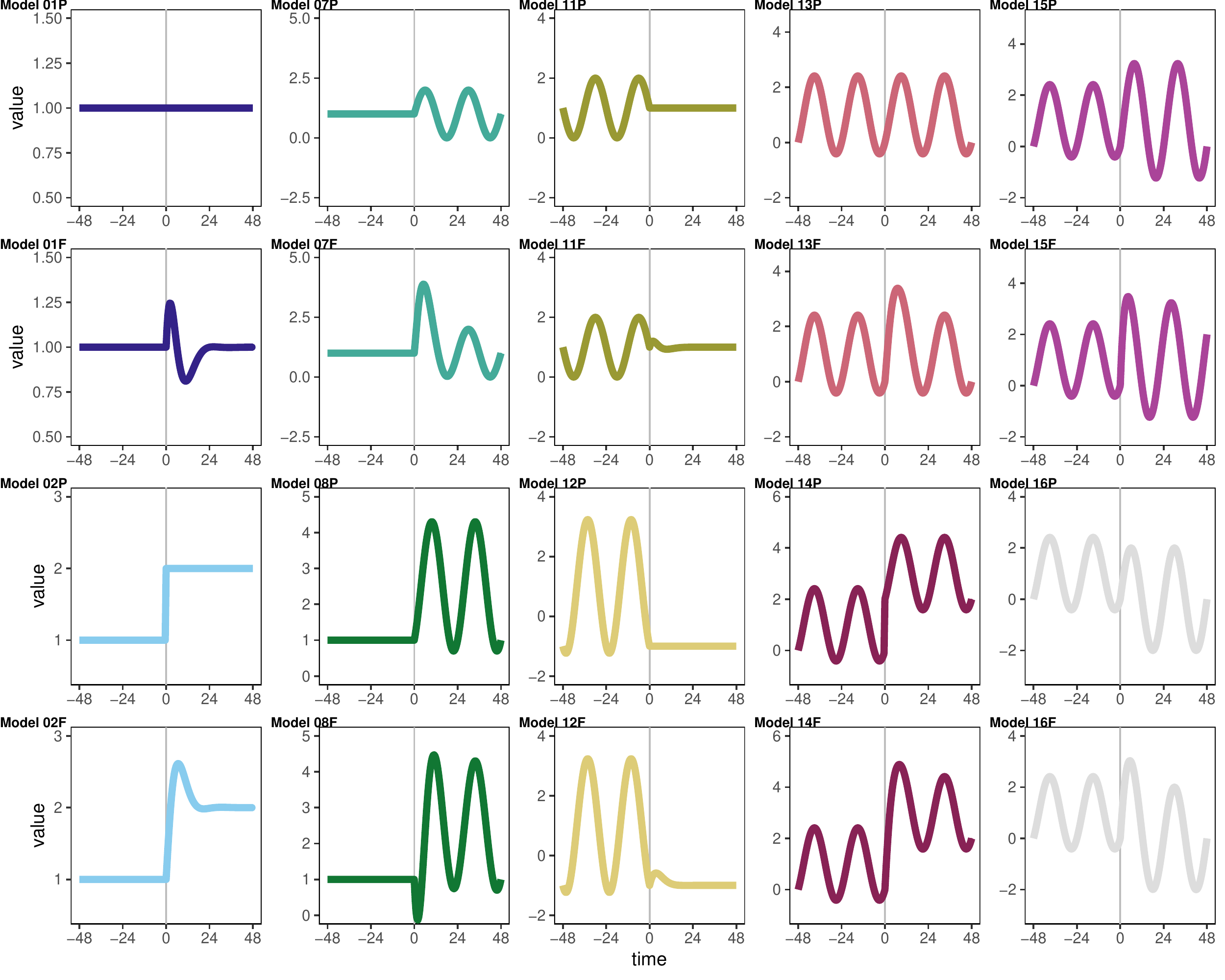}
    \caption{\textbf{Detailed candidate models for each classified possible types.} $F(\cdot)$ is the full model shown in [Equation.1]. FULL MODEL: models containing exponential terms, and constrained by characteristics in each type; PARTIAL MODEL: models excluding exponential terms, and constrained by characteristics in each type; HC: cycling at high temperature; LC: cycling at low temperature; DC: differentially cycling; DE: differentially expressed. A sketch of each model is shown here. The sketches of full models are identical to sketched in Table~\ref{Table.2}.}
    \label{SITable.1}
\end{table}
\restoregeometry

\begin{figure}[H]
    \centering
    \subfigure[]{
    \includegraphics[scale=0.5]{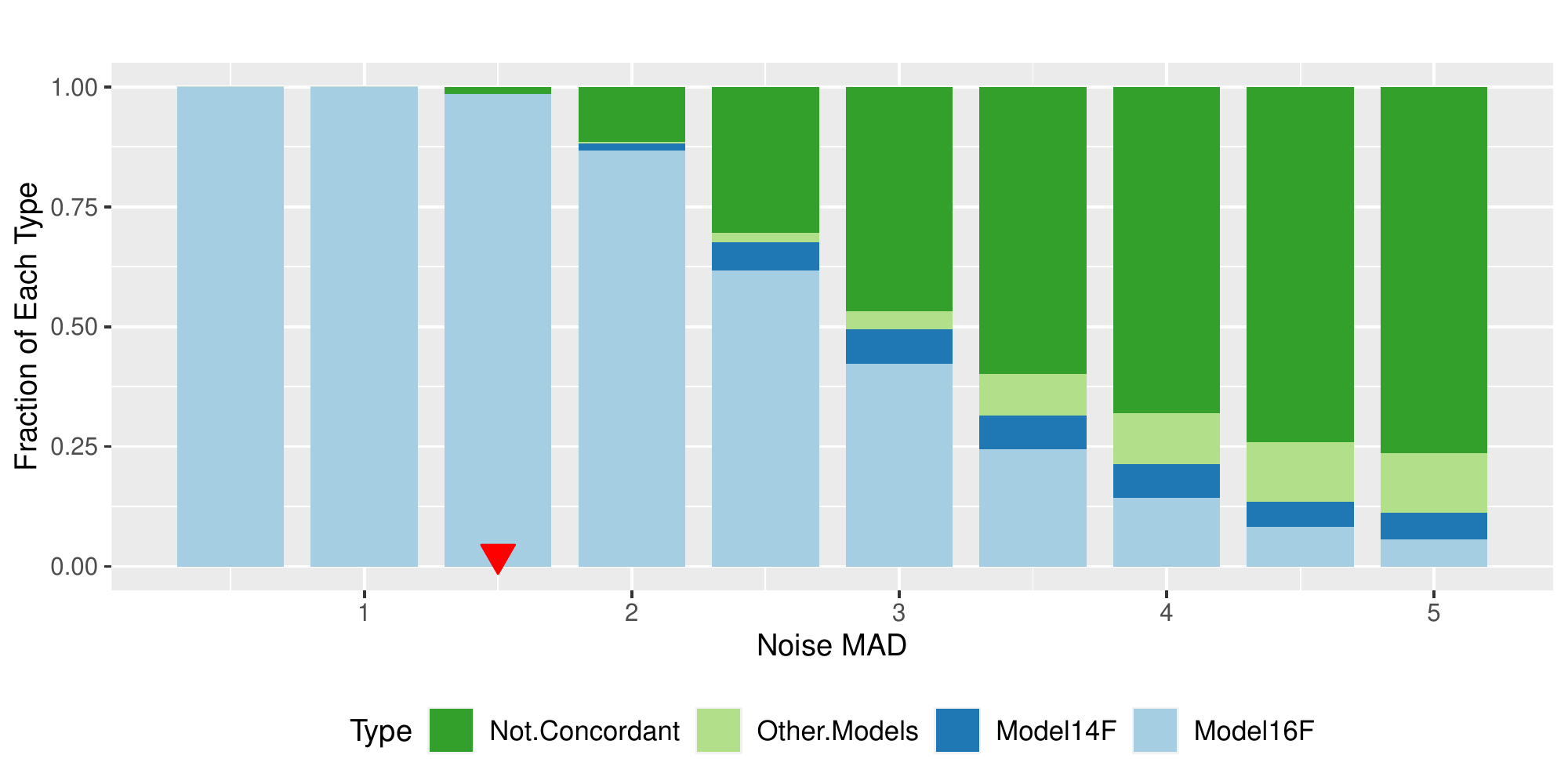}
    }
    \subfigure[]{
    \includegraphics[scale=0.5]{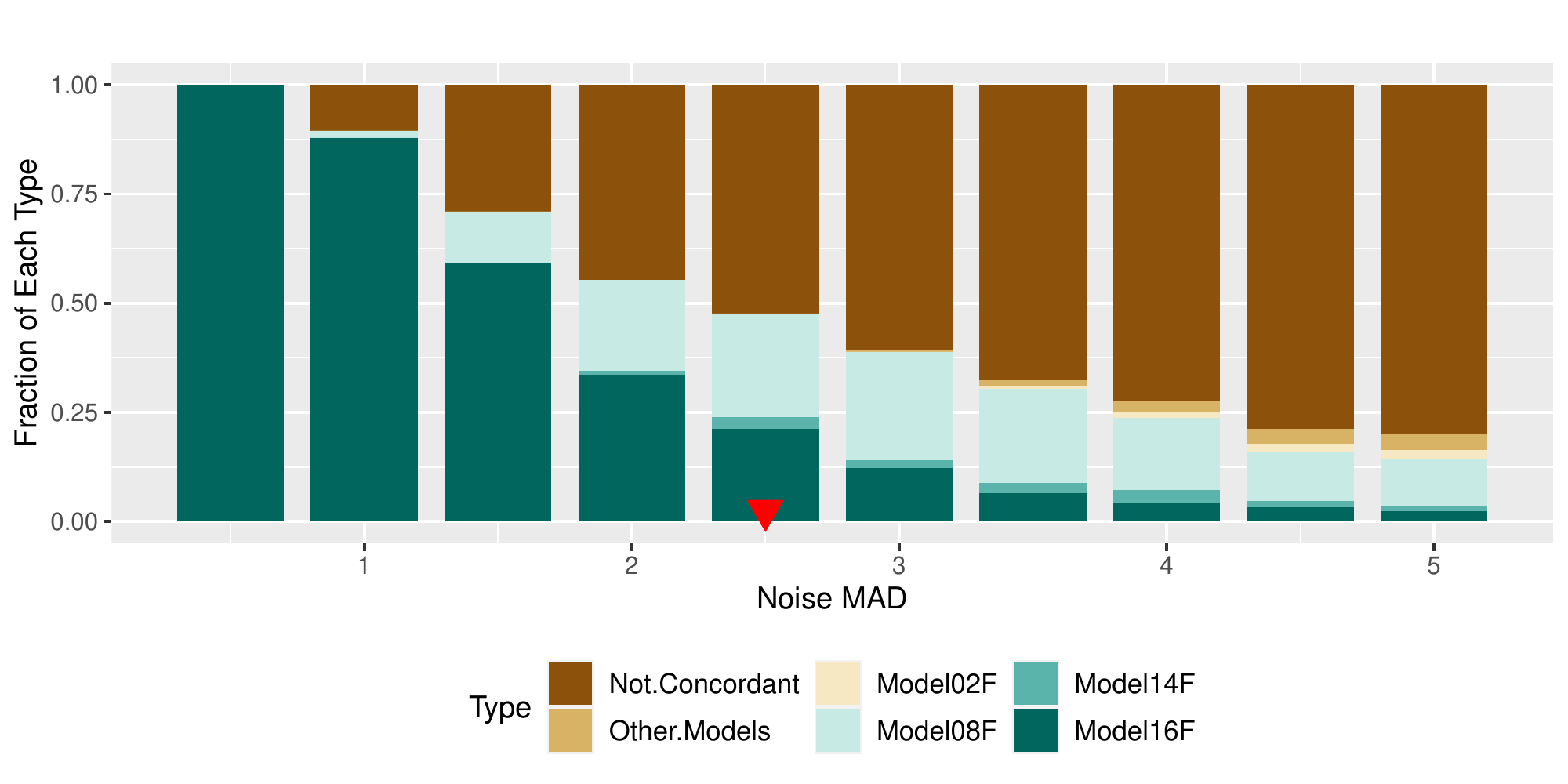}
    }
    \caption{\textbf{Classifications by significance test and model selection, with model parameters from two curves (fitted full model curves from \textit{per} and \textit{Hsf}).} Simulations are performed with parameters (rounded to two decimal places) $A_1=7.83, A_2=-2.68, A_3=-2.46, A_4=0.82, A_5=1.08, A_6=-6.50, A_7=-0.93, A_8=2.97, \lambda=0.07$ (fitted curve for \textit{per}) in (a) and $A_1=30.26, A_2=-0.95, A_3=0.60, A_4=865.46, A_5=124.71, A_6=-872.82, A_7=1.58, A_8=2.47, \lambda=0.98$ (fitted curve for \textit{Hsf}) in (b). Residuals are drawn from the Laplace distribution with different mean average deviations (MAD), with $n=10^3$ trials for each MAD. Red triangle indicates the closest bar of residual MAD in the experimental data of \textit{per} (1.590) and \textit{Hsf} (2.730). Distributions of different classified types among all concordant trials are shown with different colors along with non-concordant trials.}
    \label{SIFigure.1}
\end{figure}

\newpage
\subsection{Figure.\protect\ref{SIFigure.1}}

\begin{figure}[H]
    \centerline{\includegraphics[width=0.75\textwidth]{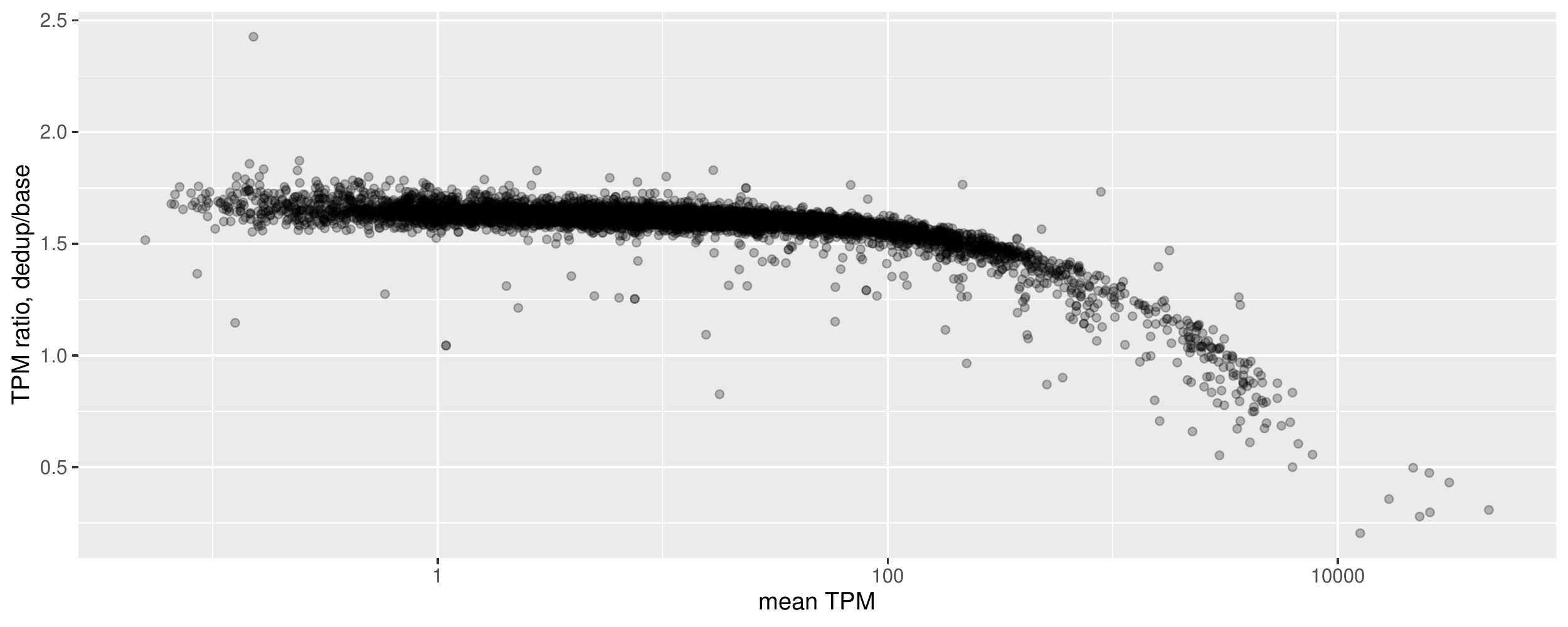}} 
    \caption{Plot of the TPM ratio (de-duplicated/base values) as a function of base TPM values.   De-duplication significantly decreases the number of reads with large TPM values (e.g., $>$ 100), and as the result the TPM values of genes with smaller TPM values are increased by approximately a factor of 1.6-1.7.}
    \label{SIFigure.2}
\end{figure}

\newpage
\subsection{Figure.\protect\ref{SIFigure.2}}

\begin{figure}[H]
    \centering
    \subfigure[]{
    \includegraphics[scale=0.6]{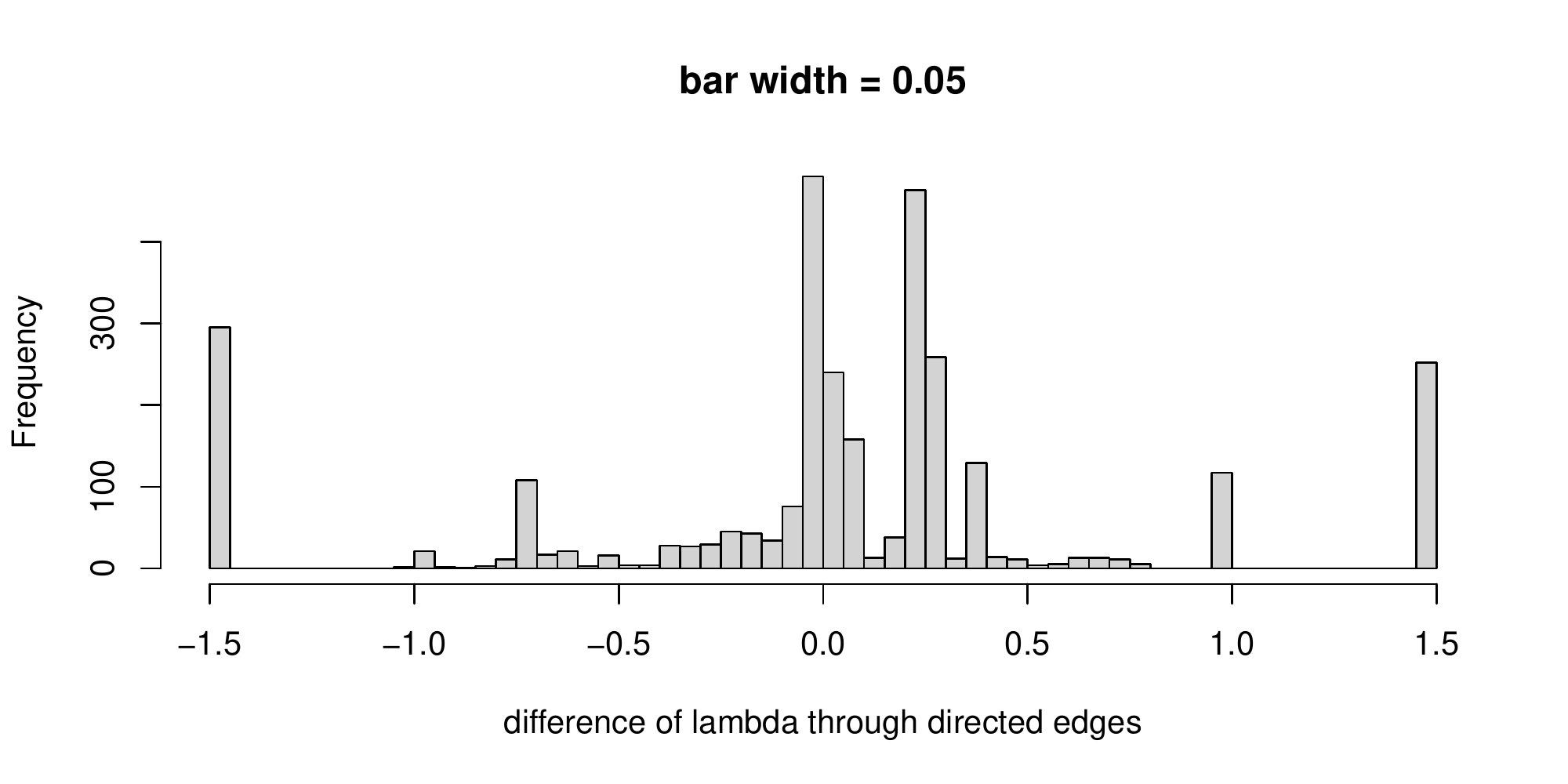}
    }
    \subfigure[]{
    \includegraphics[scale=0.6]{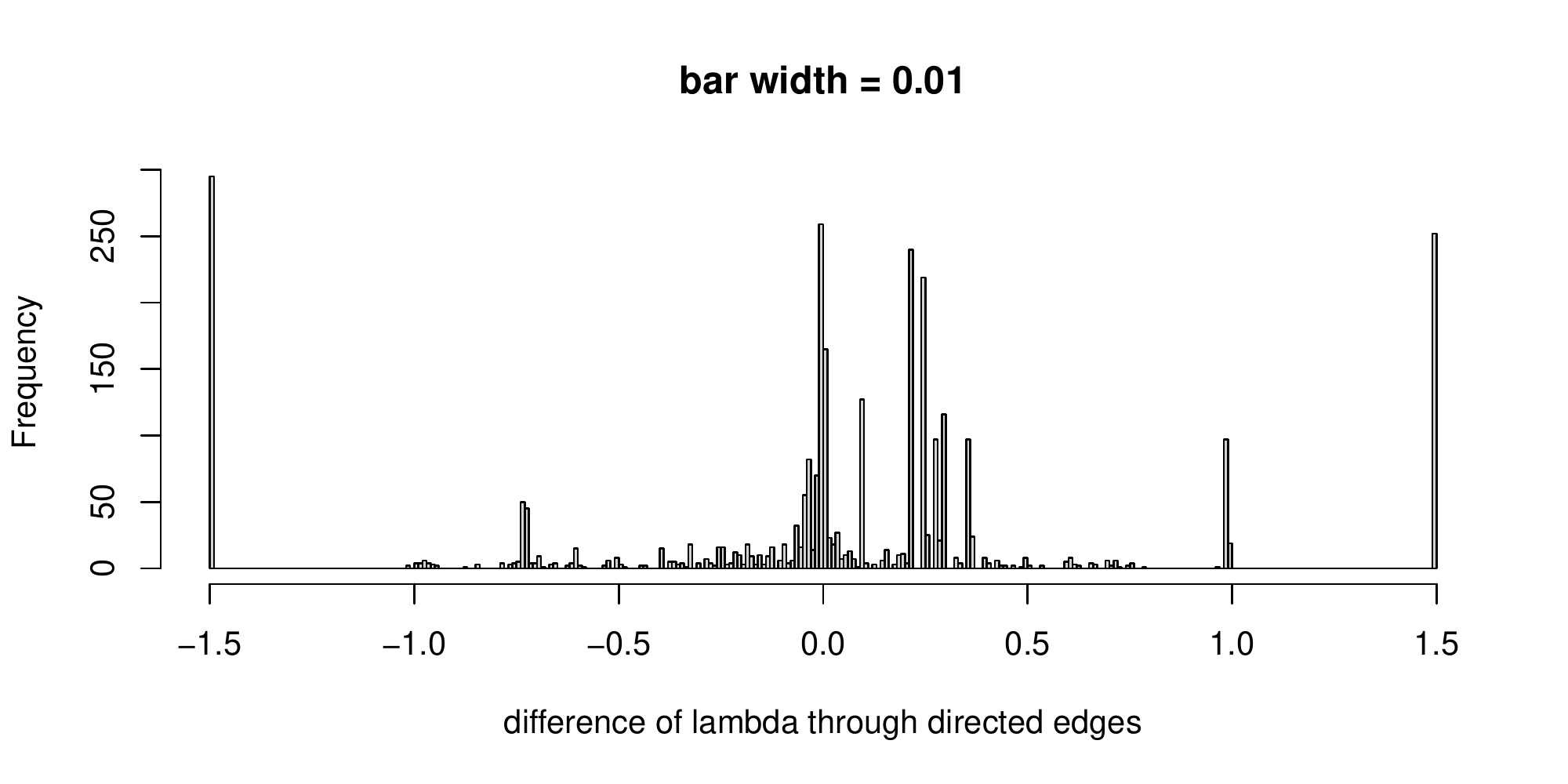}
    }    
    \caption{\textbf{Distribution of differences in adaptation rates through directed edges in the Reactome pathway, with bar width equal to 0.05 and 0.01.} The peak in the center of the distribution roughly spans [-0.1,0.1] regardless of which resolution the distribution is plotted with.}
    \label{SIFigure.3}
\end{figure}

\newpage
\subsection{Figure.\protect\ref{SIFigure.3}}

\clearpage
\appendix
\section{Supplemental Information}
\label{s:SI}
\vspace*{0.5\baselineskip}

{\bf Floquet theory}

Consider a general system 
\[
    \frac{d\vec{x}}{dt} = F(\vec{x},t)
\]
where $F(\vec{x},t)$ is $T$-periodic in its second argument.   If this equation has a $T$-periodic solution $\vec{x}_0(t)$, we linearize by assuming $\vec{x} = \vec{x}_0 + \vec{y}$, where $|\vec{y}|\ll |\vec{x}_0|$ is small, to get
\[
    \frac{d\vec{y}}{dt} = \frac{\partial F}{\partial \vec{x}}(\vec{x}_0(t),t)\vec{y} \equiv A(t)\vec{y}\,.
\]
Here $A(t)$ is $T$-periodic, and we also need an initial condition $\vec{y}(0)=\vec{y}_0$.  The fundamental solution of this equation is the matrix that solves
\[
     \frac{dY}{dt} = A(t)Y \quad\mbox{with}\quad Y(0)=I\,.
\]
Since $Y(t{+}T)$ is also a solution of this equation and $Y(t)$ is non-singular for all time if $\mbox{Tr}(A)$ is well-behaved, Floquet theory \cite{Chicone2000,Hale1971,Stoker1950} shows that
\[
    Y(t) = P(t) e^{-tR}
\]
for some $T$-periodic $P(t)$ and a constant matrix $R$. 
Assuming that $R$ can be diagonalized, i.e., $U^{\dag}RU = \Lambda$ or $R=U\Lambda U^{\dag}$, then $e^{-tR} = U e^{-t\Lambda} U^{\dag}$, and
\[
    \vec{y}(t) = P(t) U e^{-t\Lambda} U^{\dag} \vec{y}_0 = P^*(t) e^{-t\Lambda} \vec{c}\,.
\]    
This is, equivalently, 
\begin{align*}
    \vec{y}(t) & = \left[ \vec{p}_1^{\;*}(t)\ \vec{p}_2^{\;*}(t)\ \ldots\ \vec{p}_n^{\;*}(t)\right]
                 \left[\begin{array}{c c c c} 
                             e^{-t\lambda_1} & 0 & \cdots & 0 \\
                             0 & e^{-t\lambda_2} &  \cdots & 0 \\
                             \vdots & \vdots & \ddots & \vdots \\
                             0 & \cdots & 0 & e^{-t\lambda_n} \end{array}\right]
                  \left[\begin{array}{c} c_1 \\[1ex] \vdots \\[1ex] c_n \end{array}\right]  \\[0.5ex]
           & =   c_1 e^{-t\lambda_1} \vec{p}_1^{\;*}(t) + c_2 e^{-t\lambda_2} \vec{p}_2^{\;*}(t) + \cdots + c_n e^{-t\lambda_n} \vec{p}_n^{\;*}(t)  \,,         
\end{align*}   
i.e., the solution is a sum of terms with each being the product of an exponential and a periodic function. 

\end{document}